\documentclass[12pt]{amsart}
\usepackage{amssymb}
\usepackage{amsmath}
\usepackage{amsfonts}
\usepackage{mathrsfs}
\usepackage{graphicx}
\usepackage{color}
\usepackage[onehalfspacing]{setspace}
\usepackage{caption}
\usepackage{natbib}
\usepackage{longtable}
\usepackage{placeins}
\usepackage{enumerate}
\usepackage{tabularx}
\usepackage[charter,cal=cmcal]{mathdesign}
\usepackage[colorlinks=true,citecolor=blue,urlcolor=blue,pdfpagemode=UseNone,pdfstartview=FitH]{hyperref}
\usepackage{apptools}
\usepackage{bibunits}
\makeatletter
\def\section{\@startsection{section}{1}
	\z@{1.0\linespacing\@plus\linespacing}{.8\linespacing}{\Large}}

\def\subsection{\@startsection{subsection}{2}
	\z@{.8\linespacing\@plus.7\linespacing}{.7\linespacing}{\large}}

\def\subsubsection{\@startsection{subsubsection}{3}
	\z@{.5\linespacing\@plus.7\linespacing}{-.5em}{\normalfont\bfseries}}
\makeatother

\numberwithin{equation}{section}
 
\newtheorem{theorem}{Theorem}[section]
\newtheorem{lemma}{Lemma}[section]

\newtheorem{proposition}{Proposition}[section]
\theoremstyle{definition}
\newtheorem{definition}{Definition}[section]

\theoremstyle{definition}
\newtheorem{assumption}{Assumption}[section]

\theoremstyle{definition}

\DeclareTextFontCommand{\bi}{%
	\fontseries\bfdefault 
	\itshape
}

\newcommand{\CI}{\perp\negthinspace\negthinspace\negthinspace\perp}

\DeclareTextFontCommand{\bi}{%
	\fontseries\bfdefault 
	\itshape
}

\title{}
\begin{document}
	\vspace*{5ex minus 1ex}
	\begin{center}
		\Large \textsc{Measuring Diffusion over a Large Network}
		\bigskip
	\end{center}
	
	\date{%
		\today%
	}
	
	\vspace*{7ex minus 1ex}
	\begin{center}
		
		Xiaoqi He and Kyungchul Song\\
		\textit{Central University of Finance and Economics and University of British Columbia}
		
	\end{center}
\begin{bibunit}[econometrica]
\begin{abstract}
{\footnotesize This paper introduces a measure of the diffusion of binary outcomes over a large, sparse network, when the diffusion is observed in two time periods. The measure captures the aggregated spillover effect of the state-switches in the initial period on their neighbors' outcomes in the second period. This paper introduces a causal network that captures the causal connections among the cross-sectional units over the two periods. It shows that when the researcher's observed network contains the causal network as a subgraph, the measure of diffusion is identified as a simple, spatio-temporal dependence measure of observed outcomes. When the observed network does not satisfy this condition, but the spillover effect is nonnegative, the spatio-temporal dependence measure serves as a lower bound for diffusion. Using this, a lower confidence bound for diffusion is proposed and its asymptotic validity is established. The Monte Carlo simulation studies demonstrate the finite sample stability of the inference across a range of network configurations. The paper applies the method to data on Indian villages to measure the diffusion of microfinancing decisions over households' social networks.}\bigskip

{\footnotesize \ }

\bigskip

\end{abstract}
\thanks{We are grateful to Stephane Bonhomme, Arun Chandrasekhar, Yoosoon Chang, Wei Li, Vadim Marmer, Azeem Shaikh and Jeff Wooldridge for insightful comments and discussions, and thank the seminar participants at Brown University, a Harvard/MIT joint seminar, Indiana University, McMaster University, Michigan State University, Northwestern University, Rice University, Texas A\&M, University of Chicago, and the University of British Columbia. We thank Ning Ding, John Zhou and Clemens Possnig for assistance at the early stage of this research. We thank Arun Chandrasekhar for providing part of the data that we used for our empirical application. We also thank a co-editor and three anonymous referees for valuable comments. All errors are ours. Xiaoqi He acknowledges financial support from the National Natural Science Foundation of China under grant 72173007. Kyungchul Song acknowledges financial support from the Social Sciences and Humanities Research Council of Canada. Corresponding address: Kyungchul Song, Vancouver School of Economics, University of British Columbia, 6000 Iona Drive, Vancouver, BC, V6T 1L4, Canada. Email address: kysong@mail.ubc.ca.}
\maketitle

\pagebreak

\section{Introduction}

The phenomenon of diffusion over a network, where a person's change of state leads to another change in her neighborhood, has been of interest in epidemiology, economics, sociology, and marketing research.\footnote{To name but a few examples, see, e.g., \cite{Coleman/Katz/Menzel:57:Sociometry}, \cite{Conley/Udry:01:AJAE} and \cite{Conley/Udry:10:AER} for diffusion of technology and innovation; \cite{Banerjee/Chandrasekhar/Duflo/Jackson:13:Science}) for diffusion of microfinancing decisions over villages; \cite{Leskovec/Adamic/Huberman:07:ACM} for diffusion of product recommendations; and \cite{deMatos/Ferreira/Krackhardt:14:MISQ} for peer influence on iPhone purchases among students.} Examples include diffusion of disease, technology, innovation, or product recommendations over various social, industrial, or community networks. The main purpose of this paper is to propose a measure of the diffusion of state changes over a large network as a causal parameter and to develop statistical inference on the measure.

There are challenges in developing such an empirical measure of diffusion. First, the diffusion of actions tends to occur constantly in real time, whereas the researcher makes multiple snapshot observations during the diffusion process. Second, diffusion can begin locally around a fraction of units and spread over the entire cross-section of units over time. This propagation can be fast over a short period, even if the contact network is sparse, and induces extensive cross-sectional dependence of outcomes, which can hamper asymptotic inference. Third, the researcher rarely observes the true underlying contact network accurately, but often observes only its proxy. In such a situation, the underlying contact network, not the proxy, determines the cross-sectional dependence structure of the observed outcomes. Without knowing this structure, developing statistical inference on diffusion is a nontrivial task.

This paper considers a generalized diffusion model over a large latent contact network, where the researcher observes the states of people twice, first in the initial period and then in the second period. The researcher also observes a network over them, where the observed network does not necessarily coincide with or even ``approximate'' the contact network. In this setting, we introduce an aggregate measure of diffusion and provide asymptotic inference on the measure. Here we explain several key features of our proposal.

First, we adopt the Neyman-Rubin potential outcome approach in the program evaluations literature (\cite{Imbens/Rubin:15:CausalInference}) and measure diffusion as the expected increase in the number of people until the second observation when one additional randomly chosen person switches the state in the initial period. We call this measure the \textit{Average Diffusion at the Margin (ADM)} in the paper. The potential outcome approach in the paper regards the initial switches in the first period as the ``treatment'', and the states observed in the second period as the ``outcomes''. Our approach requires that we can accurately estimate the probability of the initial switch. Thus, the paper's framework fits the situation where many initial switches are triggered, for example, by targeted advertisements based on their observed characteristics and networks. It does not apply to a situation where diffusion begins with a single switch or a few ``seeds''.

The potential outcome approach in this paper has both weaknesses and strengths. The weakness is that its focus is rather narrow, which is on the measurement of the average causal effect of the initial state-switches on neighbors' outcomes by a later time. The approach does not aim to recover from data any aspects of the causal mechanism between the initial state-switches and subsequent outcomes. Learning such aspects from data can be useful in generating predictions in a counterfactual scenario about the diffusion process. The strength of the potential outcome approach is that, as we show later, we can adopt a generalized diffusion model that encompasses a wide range of diffusion models used in the literature as special cases.

Second, we explicitly consider that the researcher's observation of diffusion can be asynchronous with the actual process of diffusion in real time. More specifically, we assume that the researcher observes the states in the initial period and in a later period, without observing the actual process of diffusion between these periods. Although our setting appears similar to the two-period network interference models studied in the literature (e.g., \cite{Aronow/Samii:17:AAS}), it has distinctive features. First, we distinguish between the contact network and the network that governs the causal relations between the observed states. For the latter network, we introduce the notion of the \textit{causal graph} which assigns an edge between two people if and only if we can trace one back to the other along a walk in the contact network. Second, we allow the unobserved heterogeneity in the potential outcomes to be cross-sectionally dependent with a dependence structure shaped by the contact network.

Third, we allow the observed network to be different from the contact network. It is well known that observing a large contact network accurately is difficult in practice (see, e.g.,  \cite{Breza/Chandrasekhar/Tahbaz-Salehi:18:WP} and \cite{Banerjee/Chandrasekhar/Duflo/Jackson:19:ReStud} for references and discussion.) We often observe only a proxy network measured through various survey questions. In dealing with the discrepancy between the contact network and the observed network, our development is divided into two cases. First, we consider a situation where the observed network contains the causal graph as a subgraph - which we call the \textit{subgraph hypothesis} in this paper. In this case, while the observed network does not capture the details of the causal relations among the cross-sectional units, it encodes conditional independence among the observed outcomes.\footnote{It suffices for the observed graph to capture the dependence structure, not the details on the causal relations, because the researcher rarely obtains accurate information on the direction of influence between two people from survey data in practice. For example, when there is a recorded edge from one farmer to another whenever the former borrows money from the latter, it is not obvious which direction between the two will be the correct direction of influence in studying the diffusion of agricultural technology. In such a case, our framework allows the researcher simply to take the undirected version of the observed graph (by turning all the directed edges to undirected ones) and use it as the observed graph.} We show that this is sufficient to identify the ADM as a simple spatio-temporal dependence measure between the actions in the observed graph. Second, we consider the case where the subgraph hypothesis fails, but the spillover effect of each person on her neighbors is nonnegative. In this case, we show that the spatio-temporal dependence measure still serves as a lower bound for the ADM. Later in the paper, we provide a method for directional testing of the subgraph hypothesis.

Fourth, the researcher using our framework can be entirely agnostic about the cross-sectional dependence structure of the covariates or the relationship between the covariates and the contact network. Our framework allows a wide class of network formation models for the contact network. This flexibility is due to our defining the ADM as a quantity that is conditional on the contact network. When the contact network is formed homophilously, that is, people with similar characteristics are more likely to become neighbors with each other, the covariates exhibit cross-sectional dependence conditional on the contact network. However, the researcher rarely knows the cross-sectional dependence structure of the covariates because they do not observe the contact network. The unknown cross-sectional dependence structure poses a challenge in developing statistical inference on the ADM. To address this challenge, we follow the approach of \cite{Kuersteiner/Prucha:13:JOE} in linear panel models. Thus, we treat the covariates, contact network, and observed graph as part of \textit{common shocks} so that the asymptotic validity of the inference is not affected by the details of the way the contact network or the observed network is formed or by the way the covariates are cross-sectionally dependent.

The literature has shown that similarity in the observed actions among the cross-sectional units may stem from the similarity in their attributes, and this similarity can produce what seems like diffusion, even when there is no diffusion of outcomes over the network. We call this \textit{spurious diffusion}.\footnote{Spurious diffusion arises especially when the network is formed with homophily, so that correlated actions among neighbors primarily come from shared characteristics. An early recognition of this was made by \cite{Manski:93:Restud} in a linear-in-means model when he distinguished between correlated effects and endogenous effects. For more recent literature recognizing this issue, see \cite{Aral/Muchnick/Sundararajana:09:PNAS} and \cite{Shalizi/Thomas:11:SMR}.} The severity of spurious diffusion depends on the cross-sectional dependence structure of omitted attributes. To control for this effect of similarity in attributes, it is important to measure diffusion after controlling for the covariates. The ADM does precisely this. Our identification result shows that the ADM is identified as a spatial-temporal measure among the residuals after \textit{controlling for the covariates}.

To illustrate the usefulness of our measure, we apply it to estimate the diffusion of microfinancing decisions over various measures of social networks using data on Indian villages used by \cite{Banerjee/Chandrasekhar/Duflo/Jackson:13:Science}. We consider two definitions of the initial state-switch. First, we define a person's initial state-switch to be an event that the person is a ``leader'', which is determined based on the expected connectedness in the social network, such as teachers, shopkeepers, and saving group leaders. According to the experiment design of \cite{Banerjee/Chandrasekhar/Duflo/Jackson:13:Science}, the leaders are those who first learned about the microfinancing program through private meetings with credit officers. Hence the initial state-switch by a person is equivalent to the person's having access to the information on the micro-financing decisions in the initial period. The second definition of the initial state-switch is an event in which the person was a leader and participated in the program. We call those people ``leader-adopters''.

Our first study focuses on the causal effect of ``leader-adopters'' on other households' participation in micro-financing along the network. For the observed networks, our directional tests find no evidence against the subgraph hypothesis. We find that the estimated ADM is positive with statistical significance at 5\%. We also estimate the ADM by redefining the initial triggers to be the ``leaders'' instead. In this case, the estimated ADM turns out to be statistically insignificant even at 10\%. Our results demonstrate the importance of being precise about the initial triggers in the study of diffusion.\medskip

\noindent \textbf{Related Literature}\smallskip

Diffusion of disease, information, and technology, has received a great deal of attention in the literature. The traditional literature on the diffusion of disease uses a parsimonious model in which every individual is assumed to meet every other person randomly, and the dynamics of the aggregate measure of diffusion are characterized by a few parameters such as the infection rate and meeting rate. (See \cite{Newman:10:Networks}, chapter 17, for a review of the models and literature.) This approach extends to various approaches of mean-field approximation. (See \cite{Jackson/Rogers:07:AER}, \cite{Jackson/Yariv:07:AER} and \cite{Young:09:AER}, among others.)

In the literature of statistics, computer science, and economics, a great deal of attention has been paid to the diffusion of information or actions, and to its interaction with network structures (see \cite{DeMarzo/Vayanos/Zwiebel:03:QJE},  \cite{Leskovec/Adamic/Huberman:07:ACM}, \cite{Golub/Jackson:10:AEJ}, \cite{Campbell:13:AER}, and \cite{Banerjee/Chandrasekhar/Duflo/Jackson:13:Science}, to name but a few). A focus from the policy perspective is often the problem of optimal seeding, that is, the problem of finding a group of people who can be seeded with information to diffuse it to the whole population most effectively. For example, motivated by marketing research, \cite{Kempe/Kleinberg/Tardos:03:KDD} provide approximation guarantees for the problem of finding the influence maximizing subset of nodes in threshold and cascade diffusion models. Recent contributions in the economics literature include \cite{Banerjee/Chandrasekhar/Duflo/Jackson:19:ReStud}, \cite{Akbarpour/Malladi/Saberi:20:WP}, and \cite{Beaman/BenYishay/Magruder/Mobarak:21:AER}. Furthermore, a stream of literature on economic theory studies strategic interactions in the diffusion process. For example, see \cite{Morris:00:ReStud}, \cite{Jackson/Yariv:07:AER} and \cite{Sadler:20:AER} and references therein.

The potential outcome approach we adopt in this paper is closely related to the literature on network interference in program evaluations. Since treatment always precedes the observed outcomes, a program evaluation set-up can be viewed as a two-period panel environment, where the first period decisions are the treatments and the second period decisions are the outcomes. Recent research in this area includes \cite{vanderLaan:14:JCI}, \cite{Aronow/Samii:17:AAS}, \cite{Athey/Eckles/Imbens:18:JASA}, and \cite{Leung:20:ReStat}. Using a dynamic structural model, \cite{vanderLaan:14:JCI} proposes a method of causal inference with network interference. In his setting, the cross-sectional dependence of potential outcomes arises solely from overlapping treatment exposures.  \cite{Aronow/Samii:17:AAS} distinguish treatment assignment (controlled by the random experiment design) and treatment exposure (not controlled by the design). They develop asymptotically valid inference for average treatment effects for network interference, under the assumption that the exposure map is known yet flexible. \cite{Athey/Eckles/Imbens:18:JASA} consider various hypothesis testing problems on treatment effects with network interference. They propose various tests that are exact in finite samples, assuming randomized experiments on the network effects. \cite{Leung:20:ReStat} studies identification of causal parameters under the assumption that the potential outcomes do not depend on the labels of the nodes in the network. He develops asymptotic inference under weak assumptions that account for the randomness of the underlying graphs. \cite{Leung:22:ECTA} proposes asymptotic inference on treatment spillover effects which permits a treated unit to have an impact on all other units, with the impact becoming weaker as the other units are farther away from the treated unit in a network.

The literature on social interactions and networks has recognized that recovering the underlying network accurately is difficult in practice (see \cite{Banerjee/Chandrasekhar/Duflo/Jackson:19:ReStud} for a discussion and references.) \cite{Choi:17:JASA} proposes methods of inference for the average treatment effect under network interference, assuming that the individual treatment effects are monotone, but without requiring any information on the network. \cite{dePaula/Rasul/Souza:20:WP} propose identification and estimation of the network structure in a linear social interactions model with panel data. \cite{Lewbel/Qu/Tang:21:WP} consider a linear social interactions model with many disjoint groups of people, where each group is connected under a social network that is not observed. They provide identification analysis and develop estimation methods. \cite{Zhang:20:WP} studies identification and estimation of spillover effects with mismeasured networks, employing the measurement error approach of \cite{Hu:08:JOE}. \cite{Li/Sussman/Kolaczyk:21:arXiv} analyze the bias and variance of causal effect estimators with mismeasured networks in the framework of \cite{Aronow/Samii:17:AAS} and propose a method of moments estimator that reduces the bias.

This paper is organized as follows. Section 2 introduces a generalized diffusion model and the ADM as a causal parameter from the potential outcome perspective. The section provides identification results, and discusses spurious diffusion. Section 3 presents an inference method under the subgraph hypothesis and establishes its asymptotic validity. The section also presents a directional test of the subgraph hypothesis. Section 4 develops inference for the case where the subgraph hypothesis fails. Section 5 discusses the Monte Carlo simulation results. Section 6 presents an empirical application that uses the microfinancing data from \cite{Banerjee/Chandrasekhar/Duflo/Jackson:13:Science}. Section 7 concludes. The Supplemental Note provides the mathematical proofs of the results established in the paper.

\section{Measuring Diffusion over a Network}
\subsection{A Generalized Diffusion Model}
\label{subsec: a generl diff model}
Suppose that a set $N$ of people are connected along a simple directed graph called a \bi{contact network} $G_{\mathsf{ctt}} = (N,E_{\mathsf{ctt}})$ (where subscript ``$\mathsf{ctt}$'' is a mnemonic for ``contact''). Edge set $E_{\mathsf{ctt}}$ consists of edges $ij$, where edge $ij$ represents person $i$  being exposed to the influence of person $j$. Each person $i \in N$ at each time $t$ has two possible states, the default state $0$ that everyone is endowed with in the beginning, and the state $1$ to which a person may switch. When person $i$ has not yet switched her state, she is exposed to the influence of the previous state-switches of people in her neighborhood (denoted by $N_{\mathsf{ctt}}(i)$) in the contact network.

We assume that once the state is switched, it is \textit{irreversible}. Hence, each person is allowed to switch her state at most once during the observation periods. For example, the binary state-switch may represent an investment in a technology which cannot be reversed during the observed periods. Another example is the decision to abandon a technology permanently. The irreversibility of the state-switch is satisfied trivially by definition when the focus of interest is the diffusion of a \textit{first-time} event, such as the first-time purchase of a product of a new brand.

More formally, initial state-switches, $A_{j,0} \in \{0,1\}$, are realized for each person $j$ at time $t=0$, and then the diffusion of state-switches arises subsequently along discrete time $t \in \{1,2,...\}$. The state-switch of person $i$ at time $t$, denoted by $A_{i,t} \in \{0,1\}$, is a function of $(A_{j,t-1})_{j \in N_{\mathsf{ctt}}(i)}$ and her own state vector $U_{i,t} \in \mathbf{R}^d$: for $t = 1,2,3,...$,
\begin{align}
	\label{actual outcome}
	A_{i,t} = \left\{ \begin{array}{ll}
		\rho_{i,t}\left((A_{j,t-1})_{j \in N_{\mathsf{ctt}}(i)},U_{i,t};G_{\mathsf{ctt}}\right), & \text{ if } A_{i,s} = 0, \text{ for all } s = 0,1,...,t-1\\
		0, & \text{ otherwise},
	\end{array}
	\right.
\end{align}
for some map $\rho_{i,t}$, which may depend on the contact network $G_{\mathsf{ctt}}$. If $A_{i,s} = 1$ for some $s=0,1,...,t-1$, we automatically have $A_{i,t} = 0$, by the irreversibility of the state-switch, that is, person $i$ cannot switch her state, once she has switched it previously. For example, in the context of technology adoption in agriculture, when a farmer $i$ adopted a new technology at time $t-1$ and kept using the technology at time $t$, we record the state-switches as $A_{i,t-1} = 1$ and $A_{i,t} = 0$. The indicator $A_{i,t-1}=1$ does \textit{not} represent the state of using the technology, but the \textit{switch} of the state. Hence what triggers the state-switches in our model is neighbors' previous \textit{state-switches}, not their states. Our diffusion model excludes the possibility of state-switches depending on the persistence of a certain state of a neighbor. At each time $t$, we call each node $i$ a \textit{\textbf{switcher}} at $t$ if $A_{i,t} = 1$, and a \textit{\textbf{stayer}} at $t$ if $A_{i,t} = 0$.

In the literature, diffusion models that accommodate network information can be divided into two categories: threshold diffusion models and cascade diffusion models (\cite{Leskovec/Adamic/Huberman:07:ACM}). In a threshold diffusion model, an individual adopts an action if there are enough people in the person's neighborhood who adopt the same action. This adoption behavior can be due to pressure for conformity. A special case of such a model is a linear threshold model of diffusion (\cite{Granovetter:78:ASJ})\footnote{See \cite{Acemoglu/Ozdaglar/Yildiz:11:IEEE} and \cite{Banerjee/Chandrasekhar/Duflo/Jackson:13:Science} for more recent applications of this model.}
\begin{align}
	\label{linear threshold}
	\rho_{i,t}\left((A_{j,t-1})_{j \in N_{\mathsf{ctt}}(i)},U_{i,t};G_{\mathsf{ctt}}\right) = 1\left\{ \sum_{j \in N_{\mathsf{ctt}}(i)} p_{ij} A_{j,t-1} \ge U_{i,t} \right\},
\end{align}
where $p_{ij}$ is a nonnegative weight that person $i$ assigns to person $j$. In the linear threshold model (\ref{linear threshold}) with $p_{ij} = 1/|N_{\mathsf{ctt}}(i)|$, person $i$ switches the state if she has not yet done so and a large enough fraction of her contact neighbors have switched their states previously.

A cascade model assumes that for each person, her neighbors take turns influencing her action. The model specifies the probability that a neighbor successfully influences a person. Its special case is an independent cascade model, where the probability of a person's successful influence does not depend on the prior history of successes or failures by her neighbors (\cite{Goldenberg/Libai/Muller:01:ML}). However, as noted by \cite{Kempe/Kleinberg/Tardos:03:KDD}, the threshold model and the cascade model, once generalized, can be shown to be equivalent. Our diffusion model in (\ref{actual outcome}) encompasses this generalized version.

The generality of our diffusion model allows the view that the map $\rho_{i,t}$ in (\ref{actual outcome}) arises from an equilibrium in a strategic game. Consider the diffusion game introduced by \cite{Sadler:20:AER}. In this game, at each period $t$, a player is said to \textit{get exposed}, if one of her neighbors in the contact network switched the state in the previous period. At the beginning of the diffusion process, each player $i$ learns about her private value $U_i$ and the number of her neighbors. Then, she decides whether to be a \textit{potential adopter} or not, by maximizing her expected payoff. When the player becomes a potential adopter, she switches her state once she gets exposed, and this state-switch becomes irreversible. Player $i$'s strategy is represented by map $\sigma_i$, which assigns $0$ or $1$ to $(u_i,d_i)$, so that $\sigma(u_i,d_i) =1$ indicates that player $i$ with private value $U_i = u_i$ and $|N_{\mathsf{ctt}}(i)| = d_i$ decides to be a potential adopter. At the end of the diffusion process, each player receives a payoff that depends on the private value, the number of her contact neighbors, and the total number of switchers in her contact neighborhood.

Once strategy $\sigma_i$ is assumed to arise as a perfect Bayesian equilibrium as in \cite{Sadler:20:AER}, each player $i$'s actual state-switch is determined as in (\ref{actual outcome}) with
\begin{align*}
	\rho_{i,t}\left((A_{j,t-1})_{j \in N_{\mathsf{ctt}}(i)},U_i;G_{\mathsf{ctt}}\right) = \sigma_i\left(U_i,|N_{\mathsf{ctt}}(i)| \right)1\left\{\sum_{j \in N_{\mathsf{ctt}}(i)} A_{j,t-1} >0 \right\}.
\end{align*}
When player $i$ is a potential adopter (i.e., $\sigma_i(U_i,|N_{\mathsf{ctt}}(i)|) = 1$), she switches her state only when she has never switched the state before and gets exposed, that is, $\sum_{j \in N_{\mathsf{ctt}}(i)} A_{j,t-1} >0$.\footnote{Since our generalized diffusion model allows the unobserved heterogeneity $U_{i,t}$ to be time-varying, its time-invariance in this game is accommodated as a special case.}  In our context, we assume that we observe diffusion over a single large network. This means that if there are multiple equilibria in this diffusion game, only one equilibrium is involved in generating the data. Our causal inference then focuses on this equilibrium.

Despite its generality, our diffusion model excludes some models of diffusion that are used in the literature. For example, it excludes the model considered in \cite{Jackson/Yariv:07:AER}. They use mean-field approximation, where each person's decision to switch the state depends on the previous actions of her contact neighbors only through the probability of her neighbors' actions in the previous period. In such a model, decisions about state-switches are realized through random meetings, rather than through exposure to neighbors' influence on a given network.

\subsection{The Researcher's Observation}
\label{subsec: the researcher's obs}
\begin{figure}[t]
	\begin{center}
		\includegraphics[scale=0.45]{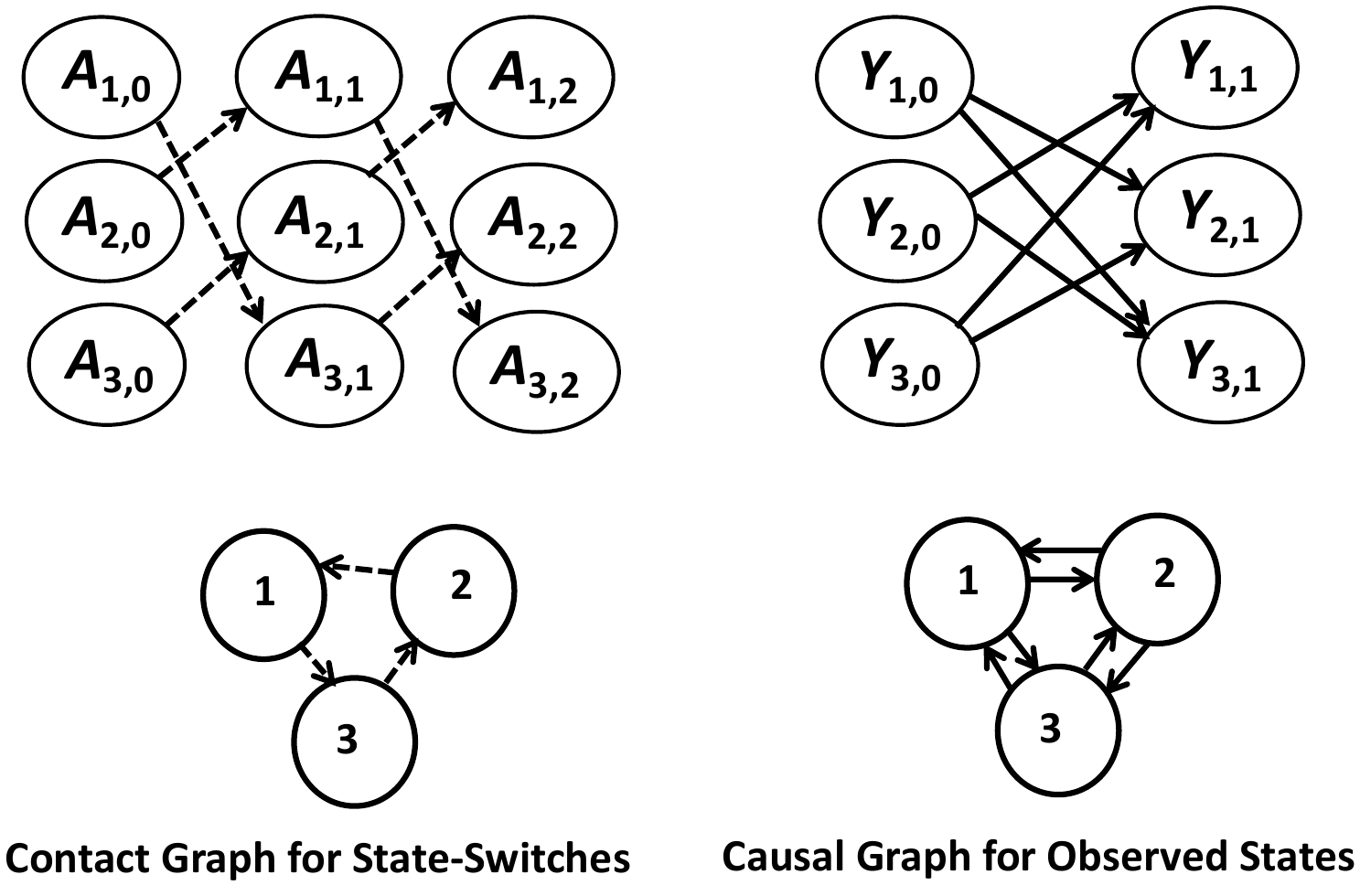}
		\caption{Illustration of a Contact Network and Its Causal Graph: \footnotesize The figure illustrates a contact network with three nodes, $N = \{1,2,3\}$, its causal graph at the second observation time being $t_1 = 2$. The edge set of the contact network is given by $\{12, 23, 31\}$. For example, the influence on $A_{1,2}$ is traced back to $A_{3,0}$ and that of $A_{1,1}$ is traced back to $A_{2,0}$.  Hence the causal in-neighborhood of node $1$ is $\{2,3\}$. Reasoning similarly with nodes $2$ and $3$, we obtain the causal graph between $(Y_0,Y_1)$ as one with the edge set $\{12,21,23,32,31,13\}$.}
		\label{fig:Causal Graph2}
	\end{center}
\end{figure}
Throughout the paper, we assume that the researcher observes the state of each person at two points in time: at time $t=0$ and at time $t=t_1$. We denote person $i$'s state at time 0 as $Y_{i,0}$ and her state at time $t = t_1$ as $Y_{i,1}$. We can relate the observed states to the state-switches as follows:
\begin{align}
	\label{obs outcome2}
	Y_{i,0} = A_{i,0}, \text{ and } Y_{i,1}  = \sum_{t=0}^{t_1} A_{i,t}.
\end{align}
Since each person can switch the state at most once, $Y_{i,1}$ takes a value of either 1 or 0, depending on whether person $i$ has switched the state at some time $t = 0,1,...,t_1$ or not. In other words, $Y_{i,1} = 1$ if and only if person $i$ has switched the state by time $t_1$.

The causal relations between $Y_0 = (Y_{j,0})_{j=1}^n$ and $Y_1 = (Y_{i,1})_{i=1}^n$ are determined by the model (\ref{actual outcome}) of diffusion. To express these causal relations, we define a graph $G_{\mathsf{cau}} = (N,E_{\mathsf{cau}})$, where $ij \in E_{\mathsf{cau}}$ if and only if $i$ and $j$ are on a walk from $j$ to $i$ of length less than or equal to $t_1$ in the contact network $G_{\mathsf{ctt}}$.\footnote{Person $i$ at time $t_1$ can be traced back to person $j$ at time $0$ through the contact network, that is, there exist $i_1,i_2,...,i_{t_1-1}, i_{t_1} = i$ such that $j \in N_{\mathsf{ctt}}(i_1)$ and $i_{s-1} \in N_{\mathsf{ctt}}(i_s)$, for $s=2,...,t_1$. The same person can appear repeatedly in the sequence.} We call the graph $G_{\mathsf{cau}}$ the \bi{causal graph} of $(Y_0,Y_1)$, because, when $ij \notin E_{\mathsf{cau}}$, the state $Y_{i,1}$ does not depend on $Y_{j,0}$ and hence there is no causal effect of $Y_{j,0}$ on $Y_{i,1}$. (See Figure \ref{fig:Causal Graph2} for an illustration of a contact network and its causal graph when $t_1 = 2$.)

In practice, the researcher rarely observes the causal graph $G_{\mathsf{cau}}$ perfectly. Throughout the paper, we assume that the researcher does not observe the causal graph $G_{\mathsf{cau}}$, but there is an \bi{observed graph} $G_{\mathsf{obs}} = (N,E_{\mathsf{obs}})$ which is a simple, directed or undirected graph on $N$. In this paper, we require that the causal graph and the observed graph are sparse, as stated in Assumption \ref{assump: network formation} later. On the relationship between the two graphs, we consider two scenarios: one in which the causal graph is a subgraph of the observed graph, and the other in which a monotone-treatment-effect type condition is satisfied. Section \ref{subsec:Identification of the ADM} provides more details on this.

To summarize, the researcher observes $(Y_1,Y_0,X,G_{\mathsf{obs}})$, where $X = (X_i)_{i \in N}$ is a profile of covariate vectors. We regard graphs $G_{\mathsf{ctt}}$ and $G_{\mathsf{obs}}$ as stochastic. Since we allow the covariates to be arbitrarily dependent across people, each covariate vector $X_i$ may include group specific characteristics, such as the population size of person $i$'s village, or it may include the size of person $i$'s neighborhood or the average of some covariates over person $i$'s neighbors in $G_{\mathsf{obs}}$. Let $\mathcal{C}$ be a vector of common shocks which represents part of the environment shared by all the agents and yet unobserved by the researcher. We take $\mathcal{F}$ to be the $\sigma$-field generated by $(G_{\mathsf{ctt}},G_{\mathsf{obs}},X, \mathcal{C})$. The $\sigma$-field $\mathcal{F}$ represents the environment in which the diffusion of state-switches among people begins.

\subsection{Average Diffusion at the Margin}
\label{subsec:ADM}
We adopt a potential outcome approach to introduce a measure of diffusion as a causal parameter. Our measure is inspired by the weighted average treatment effects in the program evaluation literature (\cite{Hirano/Imbens/Ridder:03:Eca}). In the standard setting with no spillover effects, the treatment of a person affects the person only. However, in our case with spillover effects, the ``treatment'' by a person, which corresponds to her initial state-switch from 0 to 1, potentially influences multiple people's outcomes. Thus, we first define the individual influence of each person's initial state-switch as the expected increase in the number of switchers due to the person's initial switch, and then take the measure of diffusion to be a weighted average of the individual influences.

We begin by defining a potential outcome for each person $i$, when the initial state-switch of a person $j$ in her neighborhood in $G_{\mathsf{cau}}$ is counterfactually \textit{fixed} to be $d \in \{0,1\}$. To define the potential outcome, we first relate $Y_{i,1}$ to her neighbors' initial state-switches by iterating (\ref{actual outcome}) recursively up to $t_1$ so that for some map $\tilde \rho_i$, we can write
\begin{align}
	\label{actual outcome2}
	Y_{i,1} = \tilde \rho_i\left((Y_{k,0})_{k \in \overline N_{\mathsf{cau}}(i)},V_i ; G_\mathsf{ctt} \right),
\end{align}
where $V_{i}$ denotes the vector of $U_{k,s}$'s with $k \in \overline N_{\mathsf{cau}}(i)$ and $s \le t_1$, and
\begin{align*}
	\overline N_{\mathsf{cau}}(i) = N_{\mathsf{cau}}(i) \cup \{i\},
\end{align*}
with $N_{\mathsf{cau}}(i) = \{j \in N: ij \in E_{\mathsf{cau}}\}$, that is, the in-neighborhood of $i$ in $G_{\mathsf{cau}}$. For $k,j \in N$ and $d \in \{0,1\}$, we define
\begin{align*}
	Y_{k,0}^{[d,j]} = \left\{\begin{array}{ll}
		d, & \text{ if } k = j \\
		Y_{k,0}, & \text{ if } k \ne j,			
	\end{array}
	\right.
\end{align*}
so that $Y_{k,0}^{[d,j]}$ is the same as $Y_{k,0}$ except that it is fixed to be $d$ when $k = j$. The potential outcome for person $i$ with person $j$'s initial state-switch fixed at the ``treatment status'' $d$ is defined to be
\begin{align*}
	Y_{ij}^*(d) = \left\{\begin{array}{ll}
		\tilde \rho_i\left((Y_{k,0}^{[d,j]})_{k \in \overline N_{\mathsf{cau}}(i)},V_i ; G_\mathsf{ctt} \right), & \text{ if } j \in \overline N_{\mathsf{cau}}(i) \\
		Y_{i,1}, & \text{ if } j \notin \overline N_{\mathsf{cau}}(i).			
	\end{array}
	\right.
\end{align*}

Using the potential outcomes $Y_{ij}^*(d)$, we introduce a measure of diffusion as follows. First, for $j=1,...,n$, we define
\begin{align*}
	\Delta_j = \sum_{i \in N} \mathbf{E}[Y_{ij}^*(1) - Y_{ij}^*(0) \mid \mathcal{F}],
\end{align*}
where we recall that $\mathcal{F}$ is the $\sigma$-field of $(G_{\mathsf{ctt}}, G_{\mathsf{obs}},X, \mathcal{C})$, which reflects the environment at the start of diffusion. Then $\Delta_j$ represents the expected increase in the number of switchers by time $t_1$ just due to one person $j$'s being a switcher at $t=0$, when other people, say, $k$, choose  $Y_{k,0}$, at time $t=0$.\footnote{The conditional expectation given $\mathcal{F}$ involves the joint distribution of other people's initial state-switches $Y_{k,0}$, because $Y_{ij}^*(d)$ depends on $Y_{k,0}$'s, $k \in N_{\mathsf{cau}}(i) \setminus\{j\}$.} Hence $\Delta_j$ measures the individual influence of the initial state-switch by person $j$ on the outcomes of other people by $t_1$.

Our measure of diffusion is taken to be a weighted average of the individual influences $\Delta_j$ over $j \in N$. More specifically, we define
\begin{align*}
	\mathsf{ADM} = \sum_{j \in N} w_j \Delta_j,
\end{align*}
where $w_j \ge 0$ are weights such that $\sum_{j \in N} w_j = 1$. We call this quantity the \bi{Average Diffusion at the Margin (ADM)}.
 The ADM measures the expected increase in the number of switchers \textit{just due to} one additional person's being an initial switcher when the person is randomly selected with probability $w_j$, while the rest of the people, say, $k$, switch their states according to $Y_{k,0}$ in the initial period.

We recommend using equal weights $(w_j = 1/n)$ in most cases. However, it may be of interest to consider non-equal weights in some cases, especially to compare the spillover effect between two demographic groups. For example, the setup could be that for each $j=1,...,n$,
\begin{align*}
	w_{j,\mathsf{M}} &= \frac{1}{n_{\mathsf{M}}} 1\{j \text{ is male }\} \text{ and }\\
	w_{j,\mathsf{F}} &= \frac{1}{n_{\mathsf{F}}} 1\{j \text{ is female }\},
\end{align*}
where $n_{\mathsf{M}}$ and $n_{\mathsf{F}}$ denote the number of males and females, respectively. We define the weight vectors $w_{\mathsf{M}} = (w_{1,\mathsf{M}},..., w_{n,\mathsf{M}})$ and $w_{\mathsf{F}} = (w_{1,\mathsf{F}},..., w_{n,\mathsf{F}})$. Then, by comparing the two ADM's with weights $w_{\mathsf{M}}$ and $w_{\mathsf{F}}$, we can see which gender group causes more spillover effects in terms of the ADM.

The ADM tends to decrease in the second observation period $t_1$ if $t_1$ is large enough. Since there are other initial switchers (other than the additional initial switcher $j$) and state-switches are irreversible, most people who are susceptible to a switch may have already switched their states after a long while, leaving only a few people susceptible to a switch, regardless of whether the additional person $j$ initially switched her state or not. Therefore, when $t_1$ is large, the ADM tends to be close to zero. This means that the ADM with large $t_1$ may not be an effective measure of the impact of people's initial switch of the state on others.\footnote{This is analogous to a situation where a biologist tries to measure the impact of a chemical on the proliferation rate of bacteria by comparing two petri dishes of bacteria, one treated with the chemical and the other is the control. When measuring the impact of the chemical, the timing of measurement matters; the impact cannot be measured well when the measurement is made after the bacteria have fully multiplied in both dishes. (We thank Jungsun Ghil for helping us with this analogy.)}

Unlike a typical two-period network interference setting (e.g., \cite{Rosenbaum:07:JASA}, \cite{Aronow/Samii:17:AAS}, \cite{vanderLaan:14:JCI}, and \cite{Leung:20:ReStat}), we explicitly consider that there may have been progress in diffusion between the two observation periods. In this case, we cannot assume that the unobserved heterogeneity $V_{i}$ in (\ref{actual outcome2}) is cross-sectionally independent conditional on $\mathcal{F}$. To see this, suppose that $t_1 = 2$, and
\begin{align}
	\label{eq32}
	i_1' \in N_{\mathsf{ctt}}(i_1), i_2' \in N_{\mathsf{ctt}}(i_2),  \text{ and } i'' \in N_{\mathsf{ctt}}(i_1') \cap N_{\mathsf{ctt}}(i_2'),
\end{align}
for some $i_1', i_2', i'' \in N$. From model (\ref{actual outcome}), we have
\begin{align*}
	A_{i,1} &= \rho_{i,1}\left( (A_{j,0})_{j \in N_{\mathsf{ctt}}(i)}, U_{i,1}; G_{\mathsf{ctt}} \right), \text{ for } t = 1, \text{ and }\\
	A_{i,2} &= \rho_{i,2}\left( (A_{j,1})_{j \in N_{\mathsf{ctt}}(i)}, U_{i,2}; G_{\mathsf{ctt}}\right), \text{ for } t = 2.
\end{align*}
By composing maps $\rho_{i_1,2}$ and $\rho_{j,1}$, $j \in N_{\mathsf{ctt}}(i_1)$, we find that
\begin{align*}
	Y_{i_1,1} = \tilde \rho_{i_1}\left( (Y_{j,0})_{j \in \overline N_{\mathsf{cau}}(i_1)}, V_{i_1}; G_{\mathsf{ctt}} \right),
\end{align*}
for some map $\tilde \rho_{i_1}$, where $V_{i_1}$ is a vector that contains $U_{i'',0}$ by (\ref{eq32}). Arguing similarly for $i_2$, we obtain $Y_{i_2,1} = \tilde \rho_{i_2}\left( (Y_{j,0})_{j \in \overline N_{\mathsf{cau}}(i_2)}, V_{i_2}; G_{\mathsf{ctt}} \right)$, for some map $\tilde \rho_{i_2}$, where again, $V_{i_2}$ contains $U_{i'',0}$. Since both $V_{i_1}$ and $V_{i_2}$ contain $U_{i'',0}$, they cannot be assumed to be independent.

\subsection{Identification of the ADM}
\label{subsec:Identification of the ADM}

For the initial state-switches and unobserved heterogeneity in (\ref{actual outcome}), we introduce the following assumption.

\begin{assumption}
	\label{assump: cond indep}
	$(U_j,Y_{j,0})$'s are conditionally independent across $j$'s given $\mathcal{F}$, where $U_j = (U_{j,1},...,U_{j,t_1})$.
\end{assumption}	

Assumption \ref{assump: cond indep} is satisfied if each individual's decision to switch the state depends only on $\mathcal{F}$ and idiosyncratic shocks that are cross-sectionally independent conditional on $\mathcal{F}$. For the $V_{i}$'s in (\ref{actual outcome2}), the assumption allows $V_{i_1}$ and $V_{i_2}$ to be cross-sectionally dependent given $\mathcal{F}$ as long as the in-neighborhoods of $i_1$ and $i_2$ in causal graph $G_{\mathsf{cau}}$ overlap.

As a causal parameter, the ADM is generally not identified if the initial actions $Y_{j,0}$ are allowed to be correlated with potential outcomes conditional on observables. We introduce an analogue of the ``unconfoundedness condition'' used in the program evaluation literature. (See \cite{Imbens/Wooldridge:09:JEL} for an explanation of this condition.)

\begin{assumption}
	\label{assump: unconfounded}
	(i) For all $i,j \in N$ such that $i \ne j$, $(Y_{ij}^*(1),Y_{ij}^*(0),G_{\mathsf{ctt}}, \mathcal{C})$ is conditionally independent of $Y_{j,0}$ given $G_{\mathsf{obs}},X$.

    (ii) For some $c \in (0,1/2)$, $c \le \mathbf{E}[Y_{j,0}\mid G_{\mathsf{obs}},X] \le 1 - c$, for all $j \in N$ and $n \ge 1$.
\end{assumption}

Assumption \ref{assump: unconfounded}(i) is satisfied if each person's action $Y_{j,0}$ at time $t=0$ arises only as a function of $(G_{\mathsf{obs}},X)$ and some random events that are independent of all other components. Assumption \ref{assump: unconfounded}(ii) is a version of the overlap condition, where the ``propensity score'' $\mathbf{E}[Y_{j,0}\mid G_{\mathsf{obs}},X]$ is required to be bounded away from zero uniformly over $j \in N$ and over $n$. This overlap condition is not plausible, for example, if nearly all the agents are initial switchers, or there are only a small number of initial switchers.

For example, suppose that people's initial decisions are triggered by targeted advertisements to which they are exposed, and these advertisements are made based only on their observed characteristics, their network positions in the contact network, and some random, independent shocks. Then, the correlation among their initial decisions is fully explained by the covariates and network positions. In this case, the conditional independence conditions in Assumptions \ref{assump: cond indep} and \ref{assump: unconfounded}(i) are plausible. However, these conditions are not satisfied, for example, if a person's initial action $Y_{j,0}$ is based on her observation of other people's characteristics that are not observed by the researcher.

We allow for unobserved common shock $\mathcal{C}$ yet in a limited way. While we permit the common shock $\mathcal{C}$ to be arbitrarily correlated with the potential outcomes, covariates, and both the contact and observed graphs, we require that it is conditionally independent of initial state-switches $Y_{j,0}$ given $G_{\mathsf{obs}}$ and $X$. In other words, if the common shock affects the initial state-switches, it does so only through affecting $G_{\mathsf{obs}}$ and $X$. When this is not the case, Assumption \ref{assump: unconfounded} can fail. For example, suppose that $Y_{j,0}$'s are generated as follows:
\begin{align*}
	Y_{j,0} = 1\{X_j'\gamma_0 + V_0 \ge \varepsilon_j \},
\end{align*}
where $V_0$ is an unobserved random variable that is part of the common shock $\mathcal{C}$, and the $\varepsilon_j$'s are unobserved random variables that are independent of $X$ and $V_0$. Furthermore, we assume that $V_0$ is not a function of $X$ or $G_{\mathsf{obs}}$. Thus, there is an unobserved common shock that influences people's initial state-switches and is not captured by the observed covariates or the observed graph. In this case, the common shock $\mathcal{C}$ fails to be conditionally independent of $Y_{j,0}$ given $G_{\mathsf{obs}}$ and $X$, violating Assumption \ref{assump: unconfounded}.

To describe the condition for the observed graph, it is convenient to introduce the following notion.

\begin{definition}
	\label{def: subgraph hypothesis}
	We say that the observed graph $G_{\mathsf{obs}}$ satisfies the \bi{subgraph hypothesis}, if it contains the causal graph $G_{\mathsf{cau}}$ as a subgraph.
\end{definition}

If the observed graph satisfies the subgraph hypothesis, it captures all the potential dependencies among observed state-switches that are relevant in measuring the diffusion.\footnote{In the Supplemental Note, we introduce weaker conditions based on what we call the \textit{Dependency Causal Graph (DCG)} which captures this notion more precisely. By simply taking $G_{\mathsf{obs}}$ to be a complete graph, that is, a graph with an edge for every pair $i,j \in N$, $i \ne j$, we can fulfill the subgraph hypothesis automatically. However, such an option is excluded by a sparsity condition in Assumption \ref{assump: network formation}. We exclude such a graph primarily because many observed edges may induce extensive dependence among observed outcomes in a way that invalidates the asymptotic inference we develop later.} The subgraph hypothesis does not require that the observed graph should ``approximate'' the causal graph $G_{\mathsf{cau}}$ in any sense.  The hypothesis is substantially weaker than the assumption for the networks used in the linear-in-means models in the literature (e.g., \cite{Manski:93:Restud} and \cite{Bramoulle/Djebbari/Fortin:09:JOE}). In these models, the observed network that is used to define the endogenous peer effects should be equal to, or at least ``approximate'' the true network; it is not enough to assume that the observed network contains the true network as a subgraph. (See \cite{dePaula/Rasul/Souza:20:WP} and \cite{Lewbel/Qu/Tang:21:WP} for approaches that do not require information on the networks in the data in the context of linear interactions models.) In Section \ref{subsec: directional test}, we develop a directional test of the subgraph hypothesis.

We define the following:
\begin{align}
\label{CG}
	\mathsf{C} =  \sum_{j \in N} \sum_{i \in N: ij \in E_{\mathsf{obs}}}  \frac{\text{Cov}(Y_{j,0},Y_{i,1}\mid \mathcal{F})w_j}{\text{Var}(Y_{j,0}\mid \mathcal{F})},
\end{align}
where $E_{\mathsf{obs}}$ denotes the edge set of $G_{\mathsf{obs}}$. The quantity $\mathsf{C}$ captures the spatio-temporal dependence of states between the two observation periods. It can be viewed as a measure of the empirical relevance of an observed network $G_{\mathsf{obs}}$ in explaining the \textit{residual} cross-sectional covariance pattern between $Y_1$ and $Y_0$ after ``controlling for covariates and network characteristics''. We present our identification result as follows.

\begin{lemma}
	\label{lemm: identification}
	Suppose that Assumptions \ref{assump: cond indep} and \ref{assump: unconfounded} hold.
	
	(i) If the observed graph $G_{\mathsf{obs}}$ satisfies the subgraph hypothesis, then
	\begin{align*}
	\mathsf{ADM} = \mathsf{C}.
	\end{align*}
	
	(ii) If for each $ij \in E_{\mathsf{cau}}$,
	\begin{align}
		\label{MDC0}
		   \mathbf{E}\left[Y_{ij}^*(1) - Y_{ij}^*(0) \mid \mathcal{F} \right] \ge 0,
	\end{align}
     then
     \begin{align*}
     	\mathsf{ADM} \ge \mathsf{C}.
     \end{align*}
\end{lemma}
 \medskip

Lemma \ref{lemm: identification} shows that when the observed graph satisfies the subgraph hypothesis, the ADM is identified as a weighted sum of covariances under Assumption \ref{assump: unconfounded}. When Assumption \ref{assump: unconfounded} is in doubt, the inference result on $\mathsf{C}$ is interpreted in non-causal terms, only as a measure of covariation between the outcomes in one period with their neighboring outcomes in the next period. The quantity $\mathsf{C}$ is related to the population version of the well-known inverse probability weighted estimator in the program evaluation literature:
\begin{align}
	\label{C2}
	\mathsf{C} =  \sum_{j \in N} \sum_{i \in N: ij \in E_{\mathsf{obs}}}  \mathbf{E}\left[\frac{Y_{j,0} Y_{i,1}}{\mu_{j,0}} - \frac{(1-Y_{j,0}) Y_{i,1}}{1-\mu_{j,0}} \mid \mathcal{F}\right] w_j,
\end{align}
where $\mu_{j,0} = \mathbf{E}[Y_{j,0} \mid \mathcal{F}]$. Here we can see that $Y_{j,0}$ plays the role of the treatment indicator and $\mu_{j,0}$ that of the propensity score in a program evaluation setting.

When the observed graph does not satisfy the subgraph hypothesis, but condition (\ref{MDC0}) is satisfied, $\mathsf{C}$ is a lower bound for the ADM. Condition (\ref{MDC0}) is a variant of the monotone treatment effect assumption used by \cite{Manski:97:Eca} and \cite{Choi:17:JASA}. The condition requires that the spillover effect be nonnegative in the sense that when one of person $i$'s in-neighbors in the causal graph becomes a switcher at time $t=0$, this should not make it less likely that person $i$ becomes a switcher by time $t_1$.

The intuition behind the result of Lemma \ref{lemm: identification}(ii) is as follows. By Assumptions \ref{assump: cond indep} and \ref{assump: unconfounded}, the conditional covariance between $Y_{j,0}$ and $Y_{i,1}$ given $\mathcal{F}$ is non-zero only if $ij \in E_{\mathsf{cau}} \cap E_{\mathsf{obs}}$. (For all $ij \in E_{\mathsf{obs}}\setminus E_{\mathsf{cau}}$, $Y_{j,0}$ and $Y_{i,1}$ have zero conditional covariance.) Furthermore, the terms in the sum in $\mathsf{C}$ do not include those $j$ such that $ij \in E_{\mathsf{cau}} \setminus E_{\mathsf{obs}}$, which influence $i$ non-negatively by (\ref{MDC0}) and contribute to the ADM. Since $\mathsf{C}$ does not include these nonnegative effects, it is a lower bound for the ADM.

\subsection{Spurious Diffusion}
In the literature on diffusion or network interference, it has been noted that when the networks are formed homophilously, that is, people with similar characteristics becoming neighbors with each other, the observed correlation of actions between people who are adjacent in the network may not represent the spillover effect.\footnote{For example, \cite{Aral/Muchnick/Sundararajana:09:PNAS} develop methods to distinguish between influence-based contagion and homophily-based diffusion. They find that the peer influence is generally overestimated when the homophily effect is ignored. \cite{Shalizi/Thomas:11:SMR} point out challenges arising from the confounding of social contagion, homophily, and the influence of individual traits.} For example, suppose that $Y_{i,1}$ represents the indicator of the $i$-th student purchasing a new smartphone, and $G_\mathsf{obs}$ represents the friendship network of the students. Suppose further that the friendship network exhibits homophily along the income of the student's parents, and that a student from a high-income household is more likely to purchase a new smartphone than a student from a low-income household. In this case, even if there is no diffusion of new smartphone purchases along a friendship network, the estimated diffusion without controlling for heterogeneity in income may be significantly different from zero. When the measured diffusion is not zero only because of failings to condition on certain covariates, we refer to this phenomenon as \textit{\textbf{spurious diffusion}}.

The primary reason for spurious diffusion under omitted covariates is that the unconfoundedness condition in Assumption \ref{assump: unconfounded} may fail when we omit covariates. However, the way the consequence of omitting covariates arises in the context of diffusion over a network is different from the standard setting of causal inference. In the standard setting of causal inference with no spillover effects, omitting covariates that affect both the potential outcomes and the treatment variable can cause asymptotic bias in the estimated causal effect. In the context of diffusion over a network, however, the treatment variable ($Y_{j,0}$) and the outcome variable $(Y_{i,1})$ do not belong to the same cross-sectional unit, that is, $i \ne j$. Hence, as we show below, spurious diffusion due to omitted variables arises only when the covariates exhibit cross-sectional dependence conditional on the contact and observed graphs.

To formalize this observation, let $\mathbb{S} = \{1,...,p\}$, where $p$ denotes the dimension of $X_i$. For some $S \subset \mathbb{S}$, we let the vector $X_i$ be decomposed into two subvectors $X_{i,S}$ and $X_{i,-S}$, where $X_{i,S}$ denotes the subvector of $X_i$ having its entry indices restricted to $S$, and $X_{i,-S}$ the subvector of $X_i$ whose entries are not in $X_{i,S}$. Suppose further that the researcher observes only $X_{-S} = (X_{i,-S})_{i \in N}$. Let $\mathsf{C}_{-S}$ be the same as $\mathsf{C}$, except that $X_{-S}$ replaces $X$, that is, the covariates in $X_S$ are omitted. Without observing $X_{S}$, we can recover $\mathsf{C}_{-S}$, but not $\mathsf{C}$, from data. The following proposition shows that when there is no cross-sectional dependence among the $(U_i,Y_{i,0},X_i)$'s given $(G_{\mathsf{obs}},G_{\mathsf{ctt}}, \mathcal{C})$, there is no spurious diffusion.

\begin{proposition}
	\label{prop: spurious diffusion}
	Suppose that there is no diffusion in the sense that state-switches, $A_{i,t}$, are generated as follows: for $t=1,2,3,...$,
	\begin{align}
		\label{actual outcome22}
		A_{i,t} = \left\{ \begin{array}{ll}
			\rho_{i,t}\left(U_{i,t};G_{\mathsf{ctt}}\right), & \text{ if } A_{i,s} = 0, \text{ for all } s = 0,1,...,t-1\\
			0, & \text{ otherwise}.
		\end{array}
		\right.
	\end{align}
Suppose further that $(U_i,Y_{i,0},X_i)$'s are conditionally independent across $i$'s given $(G_{\mathsf{obs}}, G_{\mathsf{ctt}}, \mathcal{C})$.

    Then there is no spurious diffusion from omitting $X_S$, that is,
	\begin{align*}
		\mathsf{C}_{-S} = 0.
	\end{align*}
\end{proposition}	

The requirement that the $(U_i,Y_{i,0},X_i)$'s are conditionally independent given $(G_{\mathsf{obs}}, G_{\mathsf{ctt}}, \mathcal{C})$ is stronger than Assumption \ref{assump: cond indep}, as it includes the cross-sectional conditional independence of the $X_i$'s given $(G_{\mathsf{obs}}, G_{\mathsf{ctt}}, \mathcal{C})$. The proposition shows the role of cross-sectional dependence of covariates in creating spurious diffusion.\footnote{See \cite{Song:22:ET} for a decomposition method that can be used in empirically investigating the role of covariates in creating spurious diffusion.}

Cross-sectional dependence of the covariates is not merely a theoretical concern. If the contact network is formed homophilously, that is, with edges formed between people with similar observed characteristics, then, conditional on the two people being neighbors, their observed characteristics are correlated. Hence, when the contact network is formed homophilously, it creates cross-sectional dependence conditional on the network, and omitting part of the covariates can create spurious diffusion.

\section{Inference on Diffusion under the Subgraph Hypothesis}

\subsection{Estimation of the ADM}
We saw that when $G_{\mathsf{obs}}$ satisfies the subgraph hypothesis, the ADM is identified as $\mathsf{C}$ which was defined in (\ref{CG}). Here we consider the estimation of $\mathsf{C}$. First, recall the definition: $\mu_{j,0} = \mathbf{E}[Y_{j,0}\mid \mathcal{F}]$. By Assumption \ref{assump: unconfounded}(i), we have
\begin{align*}
	\mu_{j,0} = \mathbf{E}[Y_{j,0}\mid G_{\mathsf{obs}}, X].
\end{align*}
We let $\sigma_{j,0}^2 = \mu_{j,0}(1 - \mu_{j,0})$ and rewrite
\begin{align*}
	\mathsf{C} = \sum_{j \in N} \frac{w_j}{\sigma_{j,0}^2} \sum_{i \in N: ij \in E_{\mathsf{obs}}}\mathbf{E}\left[(Y_{i,1} - c_i)(Y_{j,0} - \mu_{j,0}) \mid \mathcal{F} \right],
\end{align*}
where $c_i$ is any random variable that is $\mathcal{F}$-measurable, taking values in $[0,1]$. Later, we will provide specifications of $\mu_{j,0}$ and $c_i$ which we can estimate consistently.\footnote{It would be ideal if we could find a specification of $\mathbf{E}[Y_{i,1} \mid \mathcal{F}]$ that is consistently estimable and use it in place of $c_i$. However, it is difficult to find such a specification that is compatible with the generalized diffusion model in (\ref{actual outcome}) because $Y_{i,1}$ arises through the diffusion process over a complex, unobserved contact network.} Once we have consistent estimators $\hat c_i$ and $\hat \mu_{j,0}$ of $c_i$ and $\mu_{j,0}$, we construct a sample analogue estimator of $\mathsf{C}$ as follows:
\begin{align*}
	\mathsf{\hat C} = \sum_{j \in N} \frac{w_j}{\hat \sigma_{j,0}^2} \sum_{i \in N: ij \in E_{\mathsf{obs}}}(Y_{i,1} - \hat c_i)(Y_{j,0} - \hat \mu_{j,0}),
\end{align*}
where $\hat \sigma_{j,0}^2 = \hat \mu_{j,0} (1 - \hat \mu_{j,0})$. In the next subsection, we develop inference on $\mathsf{C}$ using this estimator.

Alternatively, we could define $\mathsf{\hat C}$ simply by setting $\hat c_i = 0$, and prove that this alternative estimator is consistent for $\mathsf{C}$. Then this becomes the sample analogue of (\ref{C2}):
\begin{align*}
	\sum_{j \in N} \sum_{i \in N: ij \in E_{\mathsf{obs}}}  \left(\frac{Y_{j,0} Y_{i,1}}{\hat \mu_{j,0}} - \frac{(1-Y_{j,0}) Y_{i,1}}{1-\hat \mu_{j,0}} \right) w_j.
\end{align*}
However, our unreported simulation studies show that the choice of $c_i$ and $\hat c_i$ that we propose later (see (\ref{ci spec}) and (\ref{hat ci}) below) leads to substantially shorter confidence intervals than this alternative estimator. Furthermore, the use of $c_i$ and $\hat c_i$ that we propose does not complicate the asymptotic inference procedure much because the estimation error in $\hat c_i$ does not influence the asymptotic variance of estimator $\mathsf{\hat C}$ after an appropriate scale-location normalization.

\subsection{Confidence Intervals for $\mathsf{C} $}

The ADM, as identified as $\mathsf{C}$ under Lemma \ref{lemm: identification}(i), measures the diffusion among a finite set of people, $N$, and naturally depends on the sample size $n$. To construct a confidence interval for $\mathsf{C}$, we employ asymptotic approximation as the size of the contact network $n$ grows to infinity. In other words, we aim to construct a confidence interval whose coverage probability is bounded from below by a given level $1 - \alpha$ up to an error that goes to zero as $n \rightarrow \infty$.

We first provide an overview of the main idea of constructing confidence intervals. Details on the estimators used here will follow in Section \ref{subsec: Example of Conditional Mean Spec}, where we present a concrete example of the conditional mean specification for $Y_{j,0}$ and a choice of $c_i$. Define
\begin{align}
	\label{a_i}
	a_i = \sum_{j\in N: ij \in E_{\mathsf{obs}}} \frac{ n w_j (Y_{j,0} - \mu_{j,0})}{\sigma_{j,0}^2} \text{ and } \hat a_i = \sum_{j\in N: ij \in E_{\mathsf{obs}}} \frac{ n w_j (Y_{j,0} - \hat \mu_{j,0})}{\hat \sigma_{j,0}^2},
\end{align}
where estimators $\hat \mu_{j,0}$ and $\hat \sigma_{j,0}^2$ are those used in the construction of $\mathsf{\hat C}$. Then, under the conditions stated below, we can show the asymptotic linear representation:
\begin{align}
	\label{AL}
	\sqrt{n}(\mathsf{\hat C} - \mathsf{C} ) = \frac{1}{\sqrt{n}}\sum_{i \in N}(q_i - \mathbf{E}[q_i\mid \mathcal{F}]) + o_P(1),
\end{align}
as $n \rightarrow \infty$, where
\begin{align*}
	q_i = (Y_{i,1} - c_i) a_i + \psi_i,
\end{align*}
and $\psi_i$ is a random variable that appears in the following representation:
\begin{align}
	\label{asymp lin hat ai}
	\frac{1}{\sqrt{n}}\sum_{i \in N} (Y_{i,1} - c_i) (\hat a_i - a_i) = \frac{1}{\sqrt{n}}\sum_{i \in N} (\psi_i - \mathbf{E}[\psi_i \mid \mathcal{F}])  + o_P(1).
\end{align}
Representation (\ref{AL}) suggests the construction of confidence intervals as follows.

Using an estimator $\hat \psi_i$ of $\psi_i$, we define\footnote{As noted by \cite{Kojevnikov:21:WP}, $\hat{\sigma}^2$ is always positive semidefinite.}
\begin{align*}
	\hat{\sigma}^2 = \frac{1}{n}\sum_{i_1,i_2 \in N: \overline N_{\mathsf{obs}}(i_1) \cap \overline N_{\mathsf{obs}}(i_2) \ne \varnothing} (\hat q_{i_1} -
	\hat h_{i_1})(\hat q_{i_2} -\hat h_{i_2}),
\end{align*}
where $\overline N_{\mathsf{obs}}(i) = N_{\mathsf{obs}}(i) \cup \{i\}$ with $N_{\mathsf{obs}}(i)=\left\{j \in N: ij \in E_{\mathsf{obs}}\right\}$,
\begin{align*}
	\hat q_i = (Y_{i,1} - \hat c_i) \hat a_i + \hat \psi_i, \text{ and } \hat h_i = \hat \lambda' X_i,
\end{align*}
and
\begin{align*}
	\hat \lambda = \left( \frac{1}{n} \sum_{i \in N} X_i X_i'\right)^{-1} \frac{1}{n}\sum_{i \in N} X_i  \hat q_i.
\end{align*}
Then, the $(1 - \alpha)$-level confidence interval for $\mathsf{C} $ is given by
\begin{align*}
	\mathbb{C}_{1-\alpha} = \left[ \mathsf{\hat C} -\frac{ z_{1-\alpha/2}\hat{\sigma}}{\sqrt{n}},\ \mathsf{\hat C} +\frac{ z_{1-\alpha/2}\hat{\sigma}}{\sqrt{n}}\right],
\end{align*}
where $z_{1-\alpha/2}$ is the $1-(\alpha/2)$ percentile of $N(0,1)$.

To see the motivation behind this method of constructing confidence intervals, note first that we would ideally like to estimate the following quantity consistently:
\begin{align}
	\label{sigma 2}
	\sigma^2 = \text{Var}\left(\frac{1}{\sqrt{n}}\sum_{i \in N} q_i \mid \mathcal{F} \right).
\end{align}
However, since $G_{\mathsf{ctt}}$ is not observed, we cannot estimate $\mathbf{E}[q_i\mid \mathcal{F}]$ consistently without making further assumptions. Such assumptions will have implications for the underlying process of diffusion over the contact network. Instead, we consider the following linear projection
\begin{align}
	\label{h i tau}
	h_i = \lambda' X_i,
\end{align}
where $X_i$ includes an intercept term, and
\begin{align*}
	\lambda = \left( \frac{1}{n} \sum_{i \in N} X_i X_i'\right)^{-1} \frac{1}{n}\sum_{i \in N} X_i \mathbf{E}\left[ q_i \mid \mathcal{F} \right].
\end{align*}
While this use of linear projection makes the inference conservative, it makes the procedure relatively simple and seems to work well in our simulation study.\footnote{Instead of using a linear projection on $X_i$, one may also use $\Phi(X_i)$, where $\Phi: \mathbf{R}^{d_X} \rightarrow \mathbf{R}^K$ is a vector of nonlinear functions, such as  basis functions in the series estimation. As we are not estimating a nonparametric function, the number $K$ of basis functions does not need to increase with the sample size $n$.}

\subsection{Specification of the Conditional Mean of $Y_{j,0}$ and Choice of $c_i$}
\label{subsec: Example of Conditional Mean Spec}

To construct a consistent estimator of $\mu_{j,0}$, we make the following assumption: for all $j \in N$,
\begin{align}
	\label{g_0 spec}
	\mathbf{E}[Y_{j,0}\mid G_{\mathsf{obs}},X] = F_0(X_j'\gamma_0),
\end{align}
where $F_0$ is a known distribution function. The covariate $X_j$ can include network characteristics of $G_{\mathsf{obs}}$, such as the average degree of agent $j$ or that of her neighbors. A part of the covariate vector $X_j$ can also include the average of certain characteristics of person $j$'s neighbors in $G_{\mathsf{obs}}$. Then we obtain the estimator of $\mu_{j,0}$ as follows:
\begin{align}
	\label{hat mu0}
   \hat \mu_{j,0} = F_0(X_j' \hat \gamma),
\end{align}
where $\hat \gamma$ is estimated using maximum likelihood estimation (MLE), that is,
\begin{align}
	\label{hat gamma}
   \hat \gamma = \arg \max_{\gamma} \sum_{j \in N} \left(Y_{j,0} \log F_0(X_j' \gamma) + (1 - Y_{j,0}) \log (1 - F_0(X_j' \gamma)) \right).
\end{align}
Under the usual conditions, we can show that $\hat \gamma$ is consistent and asymptotically normal. (See Section C in the Supplemental Note for details.)

For $c_i$, we consider the following choice:
\begin{align}
	\label{ci spec}
    c_i = F_1(X_i'\beta^*),
\end{align}
where $F_1$ is again a known distribution function (such as the distribution function of $N(0,1)$), and\footnote{It important to note that this choice of $c_i$ is \textit{not} based on any parametric specification of $\mathbf{E}[Y_{i,1} \mid \mathcal{F}]$. As mentioned before, any choice of $c_i$ that is $\mathcal{F}$-measurable can be used as long as we can use its estimator to construct a consistent and asymptotically normal estimator of $\mathsf{C}$. Our choice in (\ref{ci spec}) is motivated by its simplicity and good finite sample behavior (e.g., as compared to the choice of $c_i = 0$).}
\begin{align*}
   \beta^* = \arg \max_{\beta} \sum_{i \in N} \mathbf{E}\left[Y_{i,1} \log F_1(X_i' \beta) + (1 - Y_{i,1}) \log (1 - F_1(X_i' \beta))\mid \mathcal{F} \right].
\end{align*}
Then we take the following as an estimator of $c_i$:
\begin{align}
	\label{hat ci}
   \hat c_i = F_1(X_i'\hat \beta),
\end{align}
where $\hat \beta$ is a quasi-MLE defined as
\begin{align}
	\label{hat beta}
	\hat \beta = \arg \max_{\beta} \sum_{i \in N} \left(Y_{i,1} \log F_1(X_i' \beta) + (1 - Y_{i,1}) \log (1 - F_1(X_i' \beta))\right).
\end{align}

In this case, the quantity $\psi_i$ in (\ref{asymp lin hat ai}) takes the following form:
\begin{align}
	\label{psi_i}
	\psi_i = \Gamma_{n}  \times \pi_{i,0}  f_0(X_i'\gamma_0) X_i,
\end{align}
where, with $f_0$ denoting the density of $F_0$, $\pi_{i,0} = (Y_{i,0} - \mu_{i,0})/\sigma_{i,0}^2$,
\begin{align}
	\label{Gamma}
	\Gamma_{n} = \left(\frac{1}{n}\sum_{i \in N} \mathbf{E}\left[ (Y_{i,1} -c_i) \kappa_i \mid \mathcal{F}\right]\right) \times  \left(\frac{1}{n}\sum_{i \in N} \frac{f_0^2(X_i'\gamma_0)}{\sigma_{i,0}^2} X_i X_i'\right)^{-1},
\end{align}
and
\begin{align*}
	\kappa_i = \sum_{j \in N_{\mathsf{obs}}(i)} \frac{n w_j f_0(X_j'\gamma_0) X_j'}{\sigma_{j,0}^2} \left( \pi_{j,0}(2 \mu_{j,0} - 1) - 1 \right).
\end{align*}
We can obtain the estimator $\hat \psi_i$ of $\psi_i$ by taking
\begin{align}
	\label{hat psi_i}
	\hat \psi_i = \hat \Gamma_{n}  \times \hat \pi_{i,0} f_0(X_i'\hat \gamma) X_i,
\end{align}
where, with $\hat \sigma_{j,0}^2 = \hat \mu_{j,0}(1 - \hat \mu_{j,0})$, $\hat \pi_{i,0} = (Y_{i,0} - \hat \mu_{i,0})/\hat \sigma_{i,0}^2$,
\begin{align*}
	\hat \Gamma_{n} = \left(\frac{1}{n}\sum_{i \in N} (Y_{i,1} - \hat c_i) \hat \kappa_i \right) \times  \left(\frac{1}{n}\sum_{i \in N} \frac{f_0^2(X_i' \hat \gamma)}{\hat \sigma_{i,0}^2}  X_i X_i'\right)^{-1},
\end{align*}
and
\begin{align*}
	\hat \kappa_i = \sum_{j \in N_{\mathsf{obs}}(i)} \frac{n w_j f_0(X_j' \hat \gamma) X_j'}{\hat \sigma_{j,0}^2} \left( \hat \pi_{j,0}(2 \hat \mu_{j,0} - 1) - 1 \right).
\end{align*}
\medskip

\subsection{Asymptotic Validity}
Here we present the result that shows that the confidence intervals $\mathbb{C}_{1-\alpha}$ are asymptotically valid and introduce conditions for the estimators.
\begin{assumption}
	\label{assump: asymp approx}
	\begin{align*}
		\frac{1}{\sqrt{n}}\sum_{i \in N} (Y_{i,1} - c_i)(\hat a_i - a_i) &= \frac{1}{\sqrt{n}}\sum_{i \in N} (\psi_i - \mathbf{E}[\psi_i \mid \mathcal{F}])  + o_P(1), \text{ and }\\ \notag
		\frac{1}{\sqrt{n}}\sum_{i \in N} (\hat c_i - c_i) a_i &= o_P(1),
	\end{align*}
	where $\psi_i$ is a random variable such that for some $C >0$ that does not depend on $n$,
	\begin{align}
		\label{cond psi}
		\max_{i \in N} \mathbf{E}\left[\psi_i^4 \mid \mathcal{F} \right] \le C \max_{i \in N} \left(|N_{\mathsf{ctt}}(i)| + |N_{\mathsf{obs}}(i)| \right)^4,
	\end{align}
	and  $\psi_i$ is measurable with respect to $\sigma\left(Y_{j,0}: j \in \overline N_{\mathsf{cau}}(i) \cup \overline N_{\mathsf{obs}}(i)\right) \vee \mathcal{F}$.\footnote{Given two $\sigma$-fields, $\mathcal{F}_1$ and $\mathcal{F}_2$, we define $\mathcal{F}_1 \vee \mathcal{F}_2$ to be the smallest $\sigma$-field that contains both $\mathcal{F}_1$ and $\mathcal{F}_2$.}
\end{assumption}

\begin{assumption}
	\label{assump: rate est funs}
	(i) There exists $C>0$ such that $|\hat \mu_{j,0}| \le C$ for all $j \in N$ and all $n \ge 1$.
	
	(ii) There exists a constant $\kappa> 0$ such that as $n \rightarrow \infty$,
	\begin{align*}
		\max_{j \in N} |\hat \mu_{j,0} - \mu_{j,0}| + 	\max_{i \in N} |\hat c_i - c_i|  + \max_{i \in N} |\hat \psi_i - \psi_i| = O_P(n^{-\kappa}).
	\end{align*}

    (iii) There exists a constant $\overline w >0$ such that $w_j \le \overline w / n$ for all $j \in N$ and all $n \ge 1$.
\end{assumption}
Assumption \ref{assump: asymp approx} is a set of high level conditions for $\hat \mu_{j,0}$ and $\hat c_i$. These conditions are satisfied by the specifications of $\mu_{j,0}$ and $c_i$ which admit $\sqrt{n}$-consistent and asymptotically normal estimators. Asymptotic normality is obtained when the contact network is not too dense. (For the precise condition of the denseness of the contact network and the observed graph, see Assumption \ref{assump: network formation}, and the discussion that follows.)  Assumption \ref{assump: rate est funs}(i) requires that $\hat \mu_{j,0}$ is bounded uniformly over $j$ and $n \ge 1$. This is a mild condition that is mostly satisfied, when $\mu_{j,0}$ is parametrically specified as explained previously. For example, the choice $\hat \mu_{j,0}$ as in (\ref{hat mu0}) immediately satisfies this condition. Assumption \ref{assump: rate est funs}(ii) requires that the estimation errors of $\hat \mu_{j,0}$, $\hat c_i$ and $\hat \psi_i$ vanish in probability at a certain polynomial rate in $n$ uniformly over $i \in N$. We provide lower level conditions for Assumptions \ref{assump: asymp approx} and \ref{assump: rate est funs} in the Supplemental Note, using the setting in Section \ref{subsec: Example of Conditional Mean Spec}.

Here we introduce an assumption that ensures nondegeneracy of the asymptotic distribution. Recall the definition $\sigma^2$ in (\ref{sigma 2}).

\begin{assumption}
	\label{assump: non deg}
	(i) There exists small $c'>0$ such that for
	all $n\geq 1$,
	\begin{align*}
		\sigma^2 > c' \text{ and } \lambda_{\min}\left( \frac{1}{n}\sum_{i \in N} X_i X_i'\right) > c',
	\end{align*}
   where $\lambda_{\min}(A)$ for a symmetric matrix $A$ denotes the smallest eigenvalue of $A$.

   (ii) There exists $C>0$ such that for all $n \ge 1$ and $i \in N$, $\| X_i \|^2 \le C$.
\end{assumption}
\medskip
Assumption \ref{assump: non deg}(i) ensures the nondegeneracy of the distribution of the leading term in the asymptotic linear representation in (\ref{AL}). This condition is violated, if the randomness of $q_i$ (conditional on $\mathcal{F}$) disappears as $n \rightarrow \infty$. Since the conditional distribution of $q_i$ given $\mathcal{F}$ is highly unlikely to be degenerate in finite samples, it seems reasonable to rely on its asymptotic approximation using Assumption \ref{assump: non deg}(i). The bounded support condition in Assumption \ref{assump: non deg}(ii) has been used in the literature (see, for example,  \cite{Hirano/Imbens/Ridder:03:Eca}.) This condition can be relaxed to conditions involving moment conditions for the covariates, yet without adding insights.

The following assumption requires the contact network and the observed graph to be sparse.
\begin{assumption}
	\label{assump: network formation}
	(i) For some $C >0$,
	\begin{align*}
		\max_{i \in N} \left(|N_{\mathsf{ctt}}(i)| + |N_{\mathsf{obs}}(i)| \right) = O_P\left((\log n)^C\right).
	\end{align*}
    (ii) There exists $\tilde c>0$ such that
    \begin{align*}
    	\frac{1}{n}\sum_{i \in N}|N_{\mathsf{ctt}}(i)| \ge \tilde c,
    \end{align*}
    for all $n \ge 1$.
\end{assumption}
Assumption \ref{assump: network formation}(i) requires that the maximum degrees of the contact network and the observed graph are allowed to grow at most at a polynomial rate of the logarithm of the network size. While it is possible to relax this condition to a more complex, weaker condition, we nonetheless require this type of constraint on the denseness of the networks to build asymptotic inference for the ADM. Assumption \ref{assump: network formation}(ii) excludes the case where the contact network becomes dominated by isolated nodes asymptotically. This assumption simplifies the notation and is innocuous in our setting because we can simply redefine the contact network to consist of non-isolated nodes.

We study this assumption in terms of the primitives in a general network formation model. For simplicity, we focus on the contact network only. (The remarks below apply to the observed graph similarly.) Suppose that for each $t=1,2,...,t_1$, the contact neighborhoods $N_{\mathsf{ctt}}(i)$, $i \in N$, are formed by the following rule: for $i,j \in N$,
\begin{align}
	\label{network formation}
	 j \in N_{\mathsf{ctt}}(i)& \text{ if and only if } \varphi_{n,ij}(X,\eta) \ge v_{ij},
\end{align}
where $\eta = (\eta_i)_{i \in N}$ is a profile of individual-specific shocks, $\varphi_{n,ij}$ are nonstochastic maps, and $v_{ij}$ are edge-specific shocks that are conditionally independent across $i,j \in N$ given $(X,\eta)$. Suppose that for some $k >0$,
	\begin{align}
		\label{rate cond}
		\max_{i \in N} \sum_{j \in N \setminus \{i\}} P\left\{\varphi_{n,ij}(X,\eta) \ge v_{ij} \mid X,\eta \right\} = O_P\left((\log n)^k\right).
	\end{align}
This rate condition requires that the maximum expected number of neighbors in the contact network should not grow too fast, as $n \rightarrow \infty$. Then, we can use a result from \cite{Kojevnikov/Marmer/Song:JOE:2021} and show that Assumption \ref{assump: network formation} is satisfied for the contact network $G_{\mathsf{ctt}}$. (See Lemma \ref{lemm: dmx} in the Supplemental Note.) The network formation model (\ref{network formation}) is generic, encompassing various network formation models considered in the literature. One example is the following specification:
\begin{align*}
	\varphi_{n,ij}(X,\eta) = u_n(X_i,X_j) + \eta_{i} + \eta_{j},
\end{align*}
where $u_n(X_i,X_j)$ is a term that involves the homophily component in the payoff function of the agents.  (See Section 2.3.2 in \cite{Kojevnikov/Marmer/Song:JOE:2021} for further discussion and references on this model.)\footnote{This network formation model does not accommodate the model in \cite{Ridder/Sheng:20:WP}, where the payoff specification captures preference externalities from indirect friends, and the payoff parameters do not depend on the sample size.}

The following theorem establishes that the confidence interval $\mathbb{C}_{1- \alpha}$ is asymptotically valid.
\begin{theorem}
	\label{thm: asym val}
	Suppose that Assumptions \ref{assump: cond indep}-\ref{assump: network formation} hold, and $G_{\mathsf{obs}}$ satisfies the subgraph hypothesis. Then,
	\begin{align*}
		\liminf_{n \rightarrow \infty } P\left\{ \mathsf{ADM} \in\mathbb{C}_{1-\alpha}\right\} \ge 1-\alpha.
	\end{align*}
\end{theorem}
\medskip

\subsection{A Directional Test of the Subgraph Hypothesis}
\label{subsec: directional test}
\subsubsection{Directional Test}
A crucial condition for the observed graph $G_{\mathsf{obs}}$ that delivers the point-identification of the ADM as $\mathsf{C}$ is that the observed graph $G_{\mathsf{obs}}$ satisfies the subgraph hypothesis. In this subsection, we develop a test of the following hypothesis:\medskip

$H_0$: The observed graph $G_{\mathsf{obs}}$ satisfies the subgraph hypothesis.

$H_1$: The observed graph $G_{\mathsf{obs}}$ does not satisfy the subgraph hypothesis.\medskip

\noindent The main challenge for testing this hypothesis is that there are many ways in which the null hypothesis can be violated. This means that we may not be able to develop a test that ensures nontrivial power against all the ways in which the null hypothesis is violated.\footnote{It is well known in the literature of nonparametric tests that any omnibus test has nontrivial power against all but a finite number of directions of violations (see \cite{Janssen:00:AS}). Given such a result, we are not confident that it is possible to develop a test that shows a nontrivial power against all possible violations of the null hypothesis that $G_{\mathsf{obs}}$ satisfies the subgraph hypothesis.} Hence, we develop a simple directional test of $H_0$, which is designed to have nontrivial power against only certain violations of the subgraph hypothesis that lead to $\mathsf{ADM} \ne \mathsf{C}$.

The main idea is as follows. First, under Assumptions \ref{assump: cond indep} and \ref{assump: unconfounded}, we can write
\begin{align}
	\label{decom}
	\mathsf{ADM} = \mathsf{C} + \Delta\left(E_{\mathsf{cau}} \setminus E_{\mathsf{obs}} \right),
\end{align}
where, for a set $E$ of edges, we define
\begin{align*}
	\Delta\left(E \right) = \sum_{i,j \in N: ij \in E} \frac{\text{Cov}(Y_{j,0}, Y_{i,1} \mid \mathcal{F})w_j}{\text{Var}(Y_{j,0} \mid \mathcal{F})}.
\end{align*}
(Recall that $E_{\mathsf{cau}}$ denotes the set of edges in the causal graph $G_{\mathsf{cau}}$ and $E_{\mathsf{obs}}$ the set of edges in the observed graph $G_{\mathsf{obs}}$.) The decomposition in (\ref{decom}) shows that any discrepancy between $\mathsf{ADM}$ and $\mathsf{C}$ arises due to the presence of unobserved edges $ij$ in $G_{\mathsf{cau}}$. Under the subgraph hypothesis, we have $E_{\mathsf{cau}} \setminus E_{\mathsf{obs}}  = \varnothing$ so that
\begin{align*}
	\Delta\left(E_{\mathsf{cau}} \setminus E_{\mathsf{obs}} \right) = 0.
\end{align*}
Therefore, if $\Delta\left(E_{\mathsf{cau}}  \setminus E_{\mathsf{obs}} \right) \ne 0$ (that is, $\mathsf{ADM} \ne \mathsf{C}$), this indicates a violation of the subgraph hypothesis due to the presence of edges in $E_{\mathsf{cau}}  \setminus E_{\mathsf{obs}}$. An ideal directional test would be one that tests whether $\Delta(E_{\mathsf{cau}} \setminus E_{\mathsf{obs}}) = 0$ or not. However, this test is infeasible because we do not observe the causal graph $G_{\mathsf{cau}}$. Hence, we propose choosing a surrogate graph $G' = (N,E')$ that is disjoint with the observed graph $G_{\mathsf{obs}} = (N,E_{\mathsf{obs}})$ (that is, $E' \cap E_{\mathsf{obs}} = \varnothing$). (We will discuss how we construct such a graph $G'$ later.)

Let $\mathsf{C}'$ be the same as $\mathsf{C}$ in (\ref{CG}) except that $E_{\mathsf{obs}}$ is replaced by $E'$. Then, we can write
\begin{align*}
	\mathsf{C}' = \Delta\left(E_{\mathsf{cau}} \cap E' \right).
\end{align*}
Since $E_{\mathsf{cau}} \cap E'$ is contained in $E_{\mathsf{cau}} \setminus E_{\mathsf{obs}}$, under the null hypothesis, we have $E_{\mathsf{cau}} \cap E' = \varnothing$ and, hence, $\mathsf{C}' = 0$. If $\mathsf{C}' \ne 0$, it implies a violation of the subgraph hypothesis. Thus we propose testing the subgraph hypothesis by testing whether $\mathsf{C}' = 0$ or not. More specifically, our directional test proceeds as follows:
\begin{align*}
	\text{ Reject $H_0$, if and only if $\left| T(G') \right| > z_{1 - \alpha/2}$},
\end{align*}
where
\begin{align*}
	T(G') = \frac{\sqrt{n}\mathsf{\hat C}'}{\hat \sigma'},
\end{align*}
and $\mathsf{\hat C}'$ and $\hat \sigma'$ are the same as $\mathsf{\hat C}$ and $\hat \sigma$ except that $G'$ replaces $G_{\mathsf{obs}}$.

This test is a directional test intended to detect only such an alternative hypothesis that $G'$ contains edges in the causal graph $G_{\mathsf{cau}}$ that are not recorded in the observed graph $G_{\mathsf{obs}}$. Roughly speaking, the directional test detects a violation of the subgraph hypothesis by checking whether there are missing causal relationships between $Y_{j,0}$ and $Y_{i,1}$ with $ij \in E'$, which lead to nonzero covariances $\text{Cov}(Y_{j,0},Y_{i,1} \mid \mathcal{F})$.\footnote{See \cite{Stute:97:AS} and \cite{Escanciano/Song:10:JOE} for the approach of directional tests in nonparametric and semiparametric testing. Being a directional test, our test is not designed to detect all kinds of violations of the null hypothesis. As a consequence, the main merit of this test is that when one rejects the null hypothesis, one knows that it is because of the edges in $G'$ that belong to $G_{\mathsf{cau}}$ but are not captured by $G_{\mathsf{obs}}$.} From here on, we call the alternative graph $G'$ the \bi{direction} of the test.

\subsubsection{Choice of a Direction}

\begin{figure}[t]
	\begin{center}
		\includegraphics[scale=0.50]{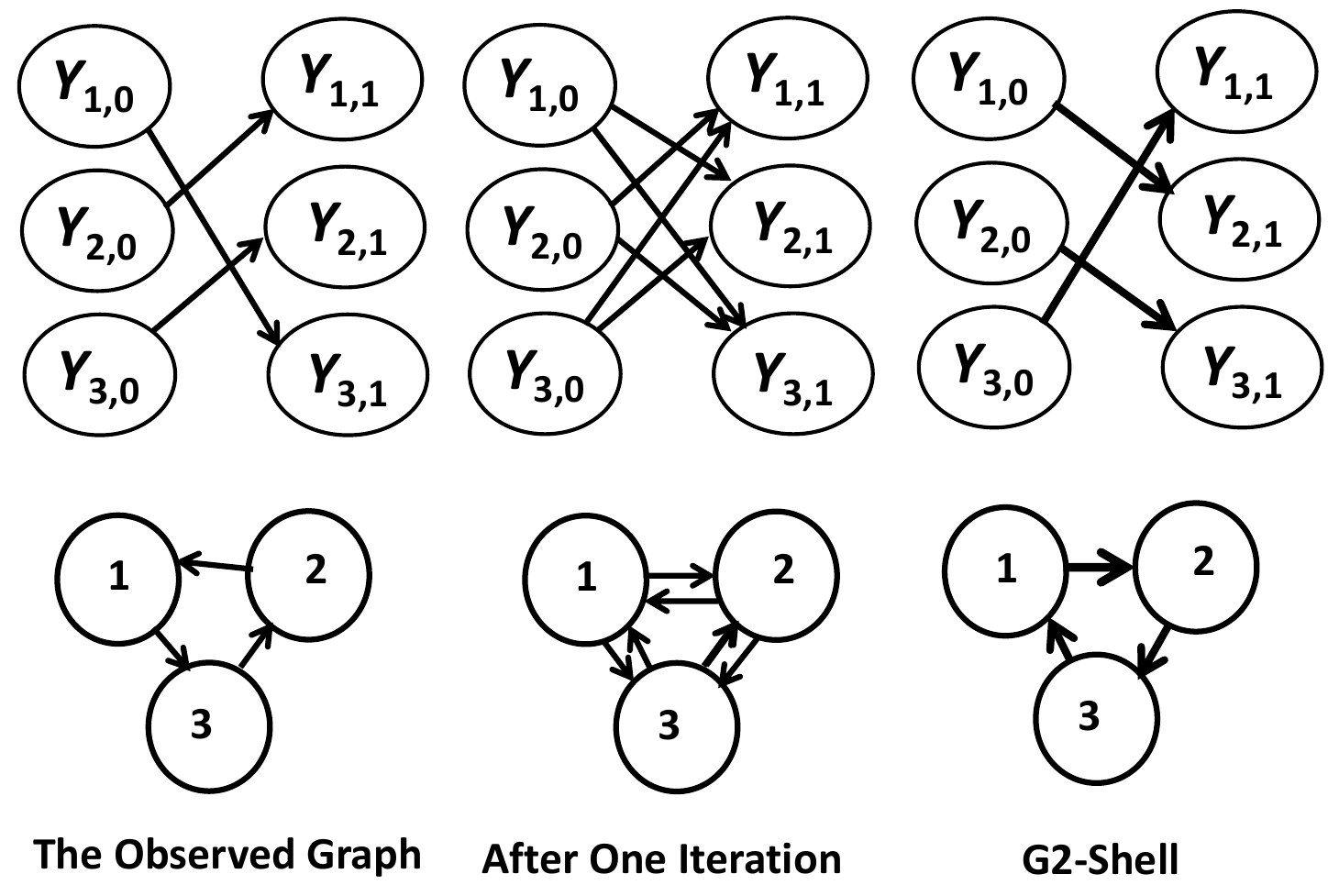}
		\caption{Constructing the G2-Shell from the Observed Graph: \footnotesize The figure illustrates the construction of the G2-shell from the observed graph, $G_{\mathsf{obs}}$, with three nodes, $N = \{1,2,3\}$ and the edge set $E_{\mathsf{obs}} = \{12, 23, 31\}$. First, we construct $G_{\mathsf{obs},2}$ by adding an edge between two nodes that are two-edges away from each other in $G_{\mathsf{obs}}$. For example, we add edge $21$ (that is, directed edge from $1$ to $2$) because, after one iteration of $G_{\mathsf{obs}}$, vertex $2$ is linked back to vertex $3$ which is linked back to vertex $1$. Then the G2-shell of the observed graph $G_{\mathsf{obs}}$ is taken to be the graph $G_{\mathsf{obs},2}$ after removing all the edges in the original observed graph $G_{\mathsf{obs}}$.}
		\label{fig:Directional Test}
	\end{center}
\end{figure}

Here, we consider the method of choosing $G'$ in practice. Intuitively, it is desirable to choose a direction $G'$ such that if the subgraph hypothesis fails, it is primarily due to the edges in $G'$. Even when the subgraph hypothesis fails, the observed graph may still contain information on the causal graph. Hence, it is likely that even if an edge $ij$ is not in the observed graph, the nodes $i$ and $j$ might still be causally related if they are indirectly connected in the observed graph. For example, suppose that there are two people $i$ and $j$ such that there exists a third person $k$, where $j$ influences $k$ and $k$ influences $i$. Suppose further that $ik, kj \in E_{\mathsf{obs}}$, but $ij \notin E_{\mathsf{obs}}$. In this case, by the transitivity of causality, $j$ influences $i$ indirectly, and hence $ij \in E_{\mathsf{cau}}$, despite that $ij \notin E_{\mathsf{obs}}$. We propose to construct direction $G'$ by collecting such edges $ij$. More specifically, we take the direction of the test to be the graph $G' = (N,E')$, where $ij \in E'$ if and only if $ij \notin E_{\mathsf{obs}}$ and for some $k \in N$, $ik, kj  \in E_{\mathsf{obs}}$. Thus, the edges $ij$ in $E'$ are such that $i$ and $j$ are connected through two edges but not adjacent in $G_{\mathsf{obs}}$. We call the graph $G'$ the \bi{G2-shell} of the observed graph $G_{\mathsf{obs}}$. See Figure \ref{fig:Directional Test} for an illustration of a G2-shell.

In practice, the edge set of the G2-shell of an observed graph can be large and it can weaken the power of the test through a large variance estimator $\hat \sigma^2$. We propose using subgraphs, $G_1,...,G_R$, that are randomly drawn from the G2-shell as follows. First, we set the cutoff degree $\overline d_n$ and let $G_{\textsf{2shell}} = (N, E_{\mathsf{2shell}})$ denote the G2 shell constructed from $G_{\mathsf{obs}}$ as previously. Then, we implement the following algorithm.\medskip

\noindent\rule{12cm}{1pt}

\textbf{Subgraph Generation Algorithm}

\small

\quad for $r=1 : R$,

\quad \quad Initialize graph $G_r = (N,E_r)$ to be an empty graph.

\quad \quad for $i = 1: n$,

\quad \quad \quad Find the in-degree $d_n(i)$ of $i$ in the G2-shell, that is, $d_n(i) = \left| \left\{j \in N: ij \in E_{\mathsf{2shell}} \right\} \right|.$

\quad \quad \quad If $d_n(i) \le \overline d_n $, place all its edges, $ij \in E_{\mathsf{2shell}}$, into $E_r$.

\quad \quad \quad If $d_n(i) > \overline d_n $, randomly select $\overline d_n$ number of edges from its edges, $ij \in E_{\mathsf{2shell}}$, and

\quad \quad \quad \quad place them in $E_r$.

\quad \quad end for

\quad \quad Store $G_r$

\quad end for

\noindent\rule{12cm}{1pt}
\normalsize

\medskip

In practice, we suggest setting the cutoff degree $\overline d_n = 5$ or choosing the cutoff degree $\overline d_n$ to be the smallest integer not greater than the average in-degree of $G_{\mathsf{obs}}$. For the number of subgraphs, $R$, we suggest choosing $R=100$. We find that these choices work well in our simulations.

Once we have generated subgraphs, say, $G_1,...,G_R$, using the subgraph generation algorithm, we proceed to perform a multiple testing procedure as follows. For each subgraph $G_r$, we construct the following hypothesis testing problem:
\begin{align*}
	H_{0,r}: \mathsf{C}_r = 0, \text{ vs. } H_{1,r}: \mathsf{C}_r \ne 0,
\end{align*}
where $\mathsf{C}_r$ is the same as $\mathsf{C}$ except that $G_r$ replaces $G_{\mathsf{obs}}$. For each $r=1,...,R$, we construct the individual test statistic $|T(G_r)|$. From the absolute value of the  standard normal distribution, we find individual $p$-values and perform a multiple testing procedure that controls the familywise error rate (FWER) at the level of the directional test. For example, the procedure proposed by \cite{Holm:79:SJS} suggests that we start with the smallest $p$-value and continue to reject the hypotheses $H_{0,m_r}$, $r=1,...,R$, as long as\footnote{See also Section 9.1 in \cite{Lehmann/Romano:05:TSH}. It is known that the Holm's step down procedure tends to be conservative because it does not take into account the dependence structure among the individual test statistics. \cite{Romano/Wolf:05:Eca} and \cite{Romano/Shaikh:10:Eca} propose an improved step-down procedure that considers the dependence structure. To apply their ideas, we should be able to derive the asymptotic distribution of the statistics of the form $\sqrt{n}\max_{m \in M'}\mathsf{\hat C}_m$ for all subsets $M' \subset M$. In our case, the index $m$ corresponds to each randomly selected subgraph of the G2-shell of the large observed graph. To the best of our knowledge, it is not obvious how one can derive such an asymptotic distribution without making further assumptions on the formation of the observed graph. We leave this question to future research.}
\begin{align}
	\label{Holm}
	\hat p_{(r)} < \frac{\alpha}{R - r + 1},
\end{align}
where $\hat p_{(1)} < \hat p_{(2)} < ... < \hat p_{(R)}$ are the p-values ordered from the smallest to the largest. Then we reject $H_0$ if any of the individual hypotheses are rejected. We investigate the finite sample performance of our directional testing procedure in the next section.

\section{Inference on Diffusion When the Subgraph Hypothesis Fails}
When the subgraph hypothesis, with the nonnegative spillover condition (\ref{MDC0}) holds, the spatio-temporal measure $\mathsf{C}$ serves as a lower bound for the ADM as shown in Lemma \ref{lemm: identification}. In this section, we consider this setting and propose a lower confidence bound for the ADM. We assume that $G_{\mathsf{obs}}$ does not satisfy the subgraph hypothesis, but the nonnegative spillover condition (\ref{MDC0}) holds. Then, as we saw before, $\mathsf{C} $ serves as a lower bound for the ADM. The lower confidence bound at level $1 - \alpha$ for the ADM that we propose is
\begin{align*}
	\hat L_{1-\alpha} = \mathsf{\hat C} - \frac{z_{1-\alpha}\hat \sigma}{\sqrt{n}},
\end{align*}
where $\mathsf{\hat C}$ and $\hat \sigma$ are as previously defined.

Here, we consider the conditions under which this lower confidence bound is asymptotically valid. The main challenge for inference when $G_{\mathsf{obs}}$ does not satisfy the subgraph hypothesis is that the cross-sectional dependence structure (shaped by the causal graph $G_{\mathsf{cau}}$) is unknown and the estimated long run variance $\hat \sigma^2$ is not consistent for the true long run variance based on causal graph $G_{\mathsf{cau}}$. We address this challenge by introducing a mild, technical condition on causal graph $G_{\mathsf{cau}}$ as follows.

\begin{assumption}
	\label{assump: non-DCG}
	There exist $\nu \in (0,1/2)$ and $c_1 >0$ such that for each $ij \in E_{\mathsf{ctt}} \setminus E_{\mathsf{obs}}$,
	\begin{align}
		\label{bound3}
		\text{Cov}\left(Y_{j,0},Y_{i,1} \mid \mathcal{F} \right) \ge c_1 n^{-1/2 + \nu},
	\end{align}
	for all $n \ge 1$.
\end{assumption}

Assumption \ref{assump: non-DCG} is satisfied if for all $ij \in E_{\mathsf{ctt}}$, $\text{Cov}(Y_{i,1},Y_{j,0}\mid \mathcal{F})$ is bounded away from zero by a sequence that decreases to zero not too fast. We can weaken this condition so that condition (\ref{bound3}) is satisfied by a large fraction of $ij$'s in $E_{\mathsf{ctt}} \setminus E_{\mathsf{obs}}$. (See the Supplemental Note for details.) This condition is violated when there are many pairs $ij$ in $E_{\mathsf{ctt}} \setminus E_{\mathsf{obs}}$ such that $\text{Cov}(Y_{i,1},Y_{j,0}\mid \mathcal{F})$ is positive and yet very close to zero. If the contact network is an empty graph (as in the case of no diffusion), or  $G_{\mathsf{obs}}$ contains the contact network as a subgraph, which is satisfied under the subgraph hypothesis, Assumption \ref{assump: non-DCG} is trivially satisfied.

To see the plausibility of Assumption \ref{assump: non-DCG}, consider the example where each state-switch represents an irreversible technology adoption by a farmer in a contact network, and the adoption decision is made based on linear threshold diffusion of the following form:
\begin{align*}
	\rho_{i,t}\left((A_{j,t-1})_{j \in N_{\mathsf{ctt}}(i)},U_{i,t};G_{\mathsf{ctt}}\right) = 1\left\{ \frac{\beta}{|N_{\mathsf{ctt}}(i)|} \sum_{j \in N_{\mathsf{ctt}}(i)} A_{j,t-1} \ge U_{i,t} \right\},
\end{align*}
where the $U_{i,t}$'s are independent and identically distributed (i.i.d.) across $t \ge 1$ and the $i$'s, follow the uniform distribution on $[0,1]$, and are conditionally independent of the $A_{i,0}$'s given $\mathcal{F}$. The parameter $\beta$ captures the influence from the previous actions by the farmers in the neighborhood in the contact network. Assume that $\beta$ does not depend on $n$. In this linear threshold diffusion model, Assumption \ref{assump: non-DCG} is satisfied, if $\beta>0$. (This is shown in Section D in the Supplemental Note.)\footnote{If $\beta = 0$, this model is the same as one that has an empty contact network. Once we replace the contact network by an empty network, Assumption \ref{assump: non-DCG} is satisfied trivially.} We are ready to present a theorem that establishes the asymptotic validity of the lower confidence bound for the ADM.
\begin{theorem}
	\label{thm: asym val2}
	Suppose that the condition of Lemma \ref{lemm: identification}(ii) and Assumptions \ref{assump: cond indep}-\ref{assump: network formation} and \ref{assump: non-DCG} hold. Then
	\begin{align*}
		\liminf_{n \rightarrow \infty } P\left\{ \mathsf{ADM} \ge \hat L_{1-\alpha}\right\} \ge 1-\alpha.
	\end{align*}
\end{theorem}

As we mentioned before, the main challenge for inference due to $G_{\mathsf{obs}}$ failing the subgraph hypothesis is that the cross-sectional dependence structure of the observations is not precisely known. Here, we provide the intuition for how we deal with this challenge to obtain Theorem \ref{thm: asym val2}. First, with $\sigma^2$ defined in (\ref{sigma 2}), we have
\begin{align}
	\label{t stat}
	\frac{\sqrt{n}(\mathsf{\hat C} - \mathsf{C} )}{\sigma} \rightarrow_d N(0,1).
\end{align}
Since we do not observe $G_{\mathsf{cau}}$, consistent estimation of $\sigma^2$ is not feasible. However, we can show that $\hat \sigma^2$ is consistent for $\tilde \sigma_\mathsf{obs}^2$, where
\begin{align*}
\tilde \sigma_\mathsf{obs}^2 &= \frac{1}{n}\sum_{i_1 i_2: \overline N_{\mathsf{obs}}(i_1) \cap \overline N_{\mathsf{obs}}(i_2) \ne \varnothing} \mathbf{E}[(q_{i_1} - h_{i_1})(q_{i_2} - h_{i_2})\mid \mathcal{F}].
\end{align*}
To resolve the gap between $\sigma^2$ and $\tilde \sigma_\mathsf{obs}^2$, we define $\overline N(i) = N(i) \cup \{i\}$ with $N(i) = N_{\mathsf{obs}}(i) \cup N_{\mathsf{cau}}(i)$, and
\begin{align*}
	\tilde \sigma^2 &= \frac{1}{n}\sum_{i_1 i_2: \overline N(i_1) \cap \overline N(i_2) \ne \varnothing}\mathbf{E}[(q_{i_1} - h_{i_1})(q_{i_2} - h_{i_2})\mid \mathcal{F}].
\end{align*}
Then, it follows that
\begin{align}
	\label{ineq}
	\sigma^2 \le \tilde \sigma^2.
\end{align}
Hence using $\tilde \sigma$ in place of $\sigma$ will make the limiting distribution of the statistic in (\ref{t stat}) less dispersed than $N(0,1)$, making the inference conservative yet asymptotically valid. On the other hand, $\hat \sigma^2$, which we use for the lower confidence bound is consistent for $\tilde \sigma_\mathsf{obs}^2$. Thus, it remains to deal with the gap between $\tilde \sigma^2$ and $\tilde \sigma_\mathsf{obs}^2$. This is where Assumption \ref{assump: non-DCG} plays a role. To see this, first, we write
\begin{align}
\label{lower bound}
\quad \quad P\left\{\mathsf{ADM} \ge \hat L_{1-\alpha}\right\}
=  P\left\{\frac{\sqrt{n}(\mathsf{C} - \mathsf{\hat C} )}{\tilde \sigma} \ge -z_{1-\alpha} +R_{1n} + R_{2n}\right\},
\end{align}
where
\begin{align*}
R_{1n} &= \frac{z_{1-\alpha} (\tilde \sigma - \tilde \sigma_\mathsf{obs}) - \sqrt{n}(\mathsf{ADM} - \mathsf{C} )}{\tilde \sigma},\text{ and }\\ \notag
R_{2n} &= - \frac{z_{1-\alpha} (\hat \sigma - \tilde \sigma_\mathsf{obs})}{\tilde \sigma}.
\end{align*}
We can show that $R_{2n} = o_P(1)$. For $R_{1n}$, we can show that under Assumption \ref{assump: non-DCG},
\begin{align*}
P\left\{z_{1-\alpha}(\tilde \sigma - \tilde \sigma_\mathsf{obs}) \le \sqrt{n}(\mathsf{ADM} - \mathsf{C} ) \right\} \rightarrow 1,
\end{align*}
as $n \rightarrow \infty$. Thus we have $P\left\{R_{1n} \le 0 \right\} \rightarrow 1$, as $n \rightarrow \infty$. This yields that the probability on the right-hand side of (\ref{lower bound}) is bounded from below by
\begin{align*}
P\left\{\frac{\sqrt{n}(\mathsf{C} - \mathsf{\hat C} )}{\tilde \sigma} \ge -z_{1-\alpha}\right\} + o(1).
\end{align*}
The leading probability is asymptotically bounded from below by $1 - \alpha$ by (\ref{t stat}) and (\ref{ineq}). Thus, we obtain the asymptotic validity of the lower confidence bound $\hat L_{1- \alpha}$.

\section{Monte Carlo Simulations}

\subsection{Simulation Design}\label{sim_sec}
In this section, we present the results from our Monte Carlo study on the finite sample behavior of our inference procedures. For the study, we first generate the contact network $G_{\mathsf{ctt}}$. Here, we mimic the network that we used in the empirical section. The adjacency matrix of the contact network is chosen as a block diagonal matrix and each block matrix is generated first by a Barabási-Albert model. Then we randomly rewire one link for each isolated node. We treat each block matrix as a village. We experiment on different numbers of villages to balance the sparsity and clustering of the network.

After generating the contact network, we generate binary actions $A_t = (A_{i,t})_{i \in N}$ that percolate through the contact network over time as follows. For each $j \in N$, we specify
\begin{align*}
	A_{j,0} = 1\left\{F_0(X_j'\gamma_0) \ge U_{j,0}\right\},
\end{align*}
where the $U_{j,0}$'s follow the uniform distribution on $[0,1]$ and are independent of the other random variables, and $F_0$ is the distribution function of $N(0,1)$. The covariates in the vector $X_j$ are drawn i.i.d. from the uniform distribution on $[0,1]$. We set $p=5$ and $\gamma_0 =  [-1.5, 0.6, -0.5, -0.4, 0.7, 0.5]'$, and the mean of $Y_{j,0}$ is around $0.16$.

For each $ i \in N$ and $t=1,...,t_1$, we specify
\begin{align*}
	A_{i,t} = \left\{ \begin{array}{ll}
		1\left\{\delta_0 \overline A_{i,t-1} + X_i'\beta_0 - U_{i,{t}} > 0 \right\},& \text{ if } A_{i,t-1} = A_{i,t-2} = ... = 0,\\
		0, & \text{ otherwise,}
	\end{array}
	\right.
\end{align*}
where the $U_{i,t}$'s are i.i.d. following the uniform distribution on $[0,1]$, and
\begin{align*}
	\overline A_{i,t-1} = \frac{1}{|N_{\mathsf{ctt}}(i)|} \sum_{j \in N_{\mathsf{ctt}}(i)} A_{j,t-1},
\end{align*}
and $N_{\mathsf{ctt}}(i)$ is the in-neighborhood of $i$ in the contact network $G_{\mathsf{ctt}}$.  In the simulation, we consider the choice of observation period $t_1 \in \{1,2,3\}$. We take the covariates to be time-invariant. We set $\beta_0 = [-1, 0.3, -0.4, -0.1, 0.4, 0.8]'$. The parameter $\delta_0$, when it takes a nonzero value, generates diffusion in our model over a given graph. We explore different values of $\delta_0 \in \{0,1.0,1.5\}$ which represent varied spillover effects.

To generate the observed graph $G_{\mathsf{obs}}$, we endow each pair $(i,j)$ of people an edge in the observed graph $G_{\mathsf{obs}}$ if and only if there exists a walk from $j$ to $i$ that has length less than or equal to $t_1$. The observed graph generated this way is the same as the causal graph $G_{\mathsf{cau}}$ and hence satisfies the subgraph hypothesis.

For the Monte Carlo simulations, we first draw the contact network, the observed graph, and the covariates, and store their values outside the Monte Carlo loop. In this simulation design, we do not allow for the unobserved common shock $\mathcal{C}$ in the definition of $\mathcal{F}$.

A summary of the network characteristics is given in Table \ref{table-1}. The observed graphs are denser than the contact networks when $t_1 = 2,3$. The observed graph becomes denser as $t_1$ increases. The true value of the ADM is reported in Table \ref{table-2}. The table shows that the value of the true ADM decreases over time. This is expected because one additional switcher in the initial period makes less difference at a later stage of diffusion.

\begin{table}[t]
	\caption{The Characteristics of the Networks }
	\begin{center}
		\small
		\begin{tabular}{cllccccccccc}
			\hline \hline
			& & & \multicolumn{3}{c}{$G_{\mathsf{ctt}}$}             & \multicolumn{3}{|c}{$G_{\mathsf{obs}}$ for $t_1 =2$} & \multicolumn{3}{|c}{$G_{\mathsf{obs}}$ for $t_1 =3$}\\\cline{4-12}
			
			&   &   & $n_V=20$ &  &  \multicolumn{1}{c|}{$n_V=30$} & $n_V = 20$  &  &\multicolumn{1}{c|}{ $n_V =30$} & $n_V = 20$  &  & $n_V =
			30$ \\
			\hline
			max. deg.	& & \multicolumn{1}{l|}{$n_H=50$}  & 11 & &\multicolumn{1}{c|}{12} &27& &\multicolumn{1}{c|}{30} &37&&39\\
			& & \multicolumn{1}{l|}{$n_H=200$}  & 29 & &\multicolumn{1}{c|}{28} &90 & &\multicolumn{1}{c|}{84} &145&&134 \\
			\hline
			ave. deg.	& & \multicolumn{1}{l|}{$n_H=50$} & 1.703 & &\multicolumn{1}{c|}{1.727} & 4.807 & &\multicolumn{1}{c|}{ 4.971} & 9.682 &&9.798 \\
			& & \multicolumn{1}{l|}{$n_H=200$} & 1.923 & &\multicolumn{1}{c|}{1.923} & 6.744 & & \multicolumn{1}{c|}{6.583}&17.42 &&17.06 \\
			\hline
			cluster 	& & \multicolumn{1}{l|}{$n_H=50$} & 0.043 & &\multicolumn{1}{c|}{0.048} & 0.283 & &\multicolumn{1}{c|}{ 0.268}&0.404 &&0.368\\
			& & \multicolumn{1}{l|}{$n_H=200$} & 0.024 & &\multicolumn{1}{c|}{0.017} & 0.140 & & \multicolumn{1}{c|}{0.146}&0.240 &&0.269 \\
			\hline
			
		\end{tabular}
	\end{center}
	\par
	\parbox{6.2in}{\footnotesize	\medskip
		Notes: The table presents the network characteristics of the contact network $G_{\mathsf{ctt}}$ and the observed graph $G_{\mathsf{obs}}$. For $t_1=1$, the observed graph $G_{\mathsf{obs}}$ is set to be the same as the contact network $G_{\mathsf{ctt}}$. We considered two different groups of villages, $n_V=20$ and $n_V=30$, and two different numbers of households in each village, $n_H = 50$ and $n_H = 200$.}
	\label{table-1}
\end{table}

\begin{table}[t]
	\caption{The True Value of the ADM over Time}
	\begin{center}
		\small
		\begin{tabular}{cllcccccc}
			\hline \hline
				& & & \multicolumn{3}{c}{$\delta_0 =1.0$}             & \multicolumn{3}{|c}{$\delta_0 =1.5$}\\\cline{4-9}
			
			&   &   & $t_1=1$ & $t_1=2$ &  \multicolumn{1}{c|}{$t_1=3$} & $t_1 = 1$  & $t_1 = 2$ & $t_1 =
			3$ \\
			\hline
			True ADM   & $n_V=20$ & \multicolumn{1}{l|}{$n_H=50$} & 0.3109  &0.1923 &\multicolumn{1}{l|}{0.0840} & 0.4443 &0.3060  & 0.1158 \\
			& & \multicolumn{1}{l|}{$n_H=200$}& 0.3088 & 0.1915&\multicolumn{1}{c|}{0.0736} &0.4426 &0.3007&0.1115
			\\
			
			\hline
			True ADM   & $n_V = 30$ & \multicolumn{1}{l|}{$n_H=50$} &  0.3087  &0.1912 &\multicolumn{1}{l|}{0.0824} &0.4441 &0.3063 & 0.1228\\
			& & \multicolumn{1}{l|}{$n_H=200$}& 0.3081 &0.1896 &\multicolumn{1}{c|}{0.0743} &0.4419 &0.3012 &0.1208
			\\
			
			\hline
		\end{tabular}
	\end{center}
	\par
	
	\parbox{6.2in}{\footnotesize \medskip
		Notes: This table shows the true values of the average diffusion at the margin (ADM) we used for the Monte Carlo study. The ADM decreases over time for each given network. We consider three periods with the number of villages $n_V=20$ and $n_V=30$ and cross-sectional dependency parameters $\delta_0 = 1.0$ and $\delta_0 =1.5$. The true ADMs were computed from $100,000$ simulations.
		\medskip}
	\label{table-2}
\end{table}

\subsection{Results}

\begin{table}[t]
	\caption{The Empirical Coverage Probability and the Mean Length of Confidence Intervals from the Normal
		Distribution at 95\% Nominal Level}
	
	\small
	The Empirical Coverage Probabilities
	\begin{center}
		\begin{tabular}{lllllllll}
			\hline \hline
			& & & \multicolumn{3}{c}{$n_V=20$}             & \multicolumn{3}{|c}{$n_V=30$}\\\cline{4-9}
			
			$\delta_0$   &   &   & $t_1=1$ & $t_1=2$ &  \multicolumn{1}{l|}{$t_1=3$} & $t_1 = 1$  & $t_1 = 2$ &
			$t_1 =3$ \\ \hline
			0   & & \multicolumn{1}{l|}{$n_H=50$}  & 0.9532  & 0.9333 &\multicolumn{1}{l|}{0.9243} &0.9562
			&0.9394 &0.9310 \\
			& & \multicolumn{1}{l|}{$n_H=200$} & 0.9523  & 0.9351 &\multicolumn{1}{l|}{0.9316 }&0.9527  & 0.9423 & 0.9359 \\
			\hline
			1.0 & & \multicolumn{1}{l|}{$n_H=50$}  & 0.9581  & 0.9365 &\multicolumn{1}{l|}{0.9271 }&0.9649  & 0.9390 & 0.9298 \\
			& & \multicolumn{1}{l|}{$n_H=200$} & 0.9590  & 0.9397 &\multicolumn{1}{l|}{0.9309 }&0.9656  & 0.9477 & 0.9349\\ \hline
			1.5   & & \multicolumn{1}{l|}{$n_H=50$}  & 0.9738  & 0.9382 &\multicolumn{1}{l|}{0.9239 }& 0.9711  & 0.9432 & 0.9297 \\
			& & \multicolumn{1}{l|}{$n_H=200$} & 0.9688  & 0.9404 &\multicolumn{1}{l|}{0.9280} &0.9744  & 0.9467 & 0.9322\\ \hline
		\end{tabular}%
	\end{center}
	\medskip
	The Mean-Length of Confidence Intervals
	\begin{center}
		\begin{tabular}{lllllllll}
			\hline \hline
			& & & \multicolumn{3}{c}{$n_V=20$}             & \multicolumn{3}{|c}{$n_V=30$}\\\cline{4-9}
			
			$\delta_0$   &   &   & $t_1 = 1$ & $t_1 = 2$ &  \multicolumn{1}{l|}{$t_1 =3$} & $t_1 = 1$  & $t_1 = 2$ &
			$t_1 =3$ \\
			\hline
			0   & & \multicolumn{1}{l|}{$n_H=50$}  & 0.2438  & 0.3867 &\multicolumn{1}{l|}{0.4963 }  &0.2017  & 0.3254 & 0.4115\\
			& & \multicolumn{1}{l|}{$n_H=200$} & 0.1276  & 0.2295 &\multicolumn{1}{l|}{0.3303 }  &0.1035  & 0.1858 & 0.2732 \\ \hline
			1.0 & & \multicolumn{1}{l|}{$n_H=50$}  & 0.2900  & 0.3996 &\multicolumn{1}{l|}{0.4798}
			&0.2409  & 0.3334 & 0.3979 \\
			& & \multicolumn{1}{l|}{$n_H=200$} & 0.1594  & 0.2447 &\multicolumn{1}{l|}{0.3276 }  &0.1288 & 0.1958 & 0.2670 \\ \hline
			1.5   & & \multicolumn{1}{l|}{$n_H=50$}  & 0.3262  & 0.3987 &\multicolumn{1}{l|}{0.4514 }  &0.2684  & 0.3334 & 0.3744 \\
			& & \multicolumn{1}{l|}{$n_H=200$} & 0.1841  & 0.2543 &\multicolumn{1}{l|}{0.3137}  &0.1482  & 0.2008 & 0.2558\\ \hline
		\end{tabular}
	\end{center}
	
	\parbox{6.2in}{\footnotesize
		\medskip
		Notes: The simulation study was based on a single generation of the random graphs. We performed $1,000$ simulations and compared the $20$ villages case with the $30$ villages case. The $30$ villages case shows better coverage probabilities and smaller mean lengths because these two networks share similar network characteristics and only differ in the number of total nodes.}
	
	\bigskip
	\label{table-3}
\end{table}

The finite sample performance of asymptotic inference on the ADM is shown in Table \ref{table-3}. The table reports the empirical coverage probability and mean length of confidence intervals based on the standard normal distribution.

First, the finite sample coverage probability of the asymptotic inference appears overall stable over a wide range of networks. For example, when $n_V = 30$ and $n_H=200$, the maximum degree of the contact network is $29$ and that of the observed graph is $145$. In this case, the empirical coverage probability at the nominal level of $95\%$ is around $0.932 \sim 0.974$, depending on $\delta_0$ and $t_1$, whereas the length of the confidence interval is around $0.104 \sim 0.273$.

Second, as the sample size increases, the length of the confidence interval decreases. This result holds across the range of graphs considered in the study. This reflects the law of large numbers, which arises due to the weak cross-sectional dependence among the outcomes. However, as the graph becomes denser, the confidence interval becomes longer. This is expected because, as the graph becomes denser, the cross-sectional dependence becomes extensive, increasing the uncertainty of the estimators.

We also perform a Monte Carlo study of the directional test for the subgraph hypothesis developed in Section \ref{subsec: directional test}. For this, we first generate a contact network $G_{\mathsf{ctt}}$ as before, using a block-diagonal adjacency matrix based on the Barab\'{a}si-Albert random graph. Let $G_{c,m}$ be a graph constructed by adding an edge between any nodes that are within the geodesic distance $m$ from each other in contact network $G_{\mathsf{ctt}}$. For the DGP for outcomes, we use $G_{\mathsf{ctt}}^0 = G_{\mathsf{ctt}}$ as the contact network under the null hypothesis $H_0$ and $G_{\mathsf{ctt}}^1 = G_{c,2}$ as the contact network under the alternative hypothesis. Under both the null hypothesis and the alternative hypothesis, we set the observed graph as follows: $G_{\mathsf{obs}} = G_{c,1}$ for $t_1 = 1$ and $G_{\mathsf{obs}} = G_{c,2}$ for $t_1 = 2$. Thus, by the design of the contact networks, $G_{\mathsf{obs}}$ satisfies the subgraph hypothesis under the null hypothesis but not under the alternative hypothesis. The graph generation for the directional tests is summarized in Table \ref{table_dir_graph}.

For the direction of the test, we set $n_V=30$ villages, each village with either $n_H = 50$ or $n_H = 200$ households. We apply Holm's multiple testing procedure by randomly drawing 100 subgraphs from the $G2$-shell of $G_{\mathsf{obs}}$ using the subgraph generation algorithm introduced earlier. We consider two ways of choosing the cutoff degree $\overline d_n$: one way is to set $\overline d_n$ to be 5 and the other way is to set $\overline d_n$ to be the smallest integer not greater than the average degree of $G_{\mathsf{obs}}$.

We compute the FWER using 1,000 Monte Carlo simulations. For brevity, we report only the results with the choice of $\overline d_n  = 5$. The results are similar for the case with $\overline d_n$ equal to the smallest integer not greater than the average degree of $G_{\mathsf{obs}}$. The finite sample FWER's are shown in Table \ref{table_dir_sim}. The FWERs are around the nominal level $\alpha$ under both the null hypothesis and the alternative hypothesis. The power of the test is reduced in the case of $t_1 = 2$. This is due to the fact that the observed graph is denser, resulting in a dense $G2$-shell from which we draw 100 subgraphs. When the $G2$-shell is dense, the 100 subgraphs may contain many edges that do not effectively capture the dependencies between nodes under the alternative hypothesis. This explains the weaker power of the directional test when the observed graph is denser.

\begin{table}[t]
	\caption{Generation of Graphs for Directional Tests}
	\small
	\begin{center}
		\begin{tabular}{c|ccc|ccc}
			\hline \hline
			& & $t_1 = 1$ & & & $t_1 = 2$ & \\
			\hline
			$H_0$ & $G_{\mathsf{ctt}} = G_{c,1}$ & $G_{\mathsf{cau}} = G_{c,1}$ & $G_{\mathsf{obs}} = G_{c,1}$ & $G_{\mathsf{ctt}} = G_{c,1}$ & $G_{\mathsf{cau}} = G_{c,2}$ & $G_{\mathsf{obs}} = G_{c,2}$\\
			\hline
			$H_1$ & $G_{\mathsf{ctt}} = G_{c,2}$ & $G_{\mathsf{cau}} = G_{c,2}$ & $G_{\mathsf{obs}} = G_{c,1}$ & $G_{\mathsf{ctt}} = G_{c,2}$ & $G_{\mathsf{cau}} = G_{c,4}$ & $G_{\mathsf{obs}} = G_{c,2}$\\
			\hline
		\end{tabular}
	\end{center}
	\parbox{6.2in}{\footnotesize \medskip
		Notes: This table reports the contact network and observed graph specification from $G_{c,m}$.}
	\label{table_dir_graph}
	\bigskip
	\bigskip
	\bigskip
\end{table}

\begin{table}	
	\caption{The Familywise Error Rate}
	\small
	\begin{center}
		\begin{tabular}{lllllllll}
			\hline \hline
			&& & \multicolumn{3}{c}{$H_0$}             & \multicolumn{3}{|c}{$H_1$}\\\cline{4-9}
			
			$\delta_0$&      &   & $\alpha=1\%$ & $\alpha=5\%$ &  \multicolumn{1}{l|}{$\alpha=10\%$} & $\alpha=1\%$  & $\alpha=5\%$ &
			$\alpha=10\%$ \\
			\hline
			1.0 & $n_H=50$ & \multicolumn{1}{l|}{$t_1=1$}& 0.006 & 0.039 &\multicolumn{1}{l|}{0.080}&1.000  & 1.000 & 1.000\\
			& & \multicolumn{1}{l|}{$t_1=2$} & 0.021  & 0.072 &\multicolumn{1}{l|}{0.126}&0.030  & 0.106 & 0.202\\
			\hline
			1.5 & $n_H=50$ & \multicolumn{1}{l|}{$t_1=1$} & 0.011  & 0.059 &\multicolumn{1}{l|}{0.092}&1.000  & 1.000 & 1.000\\
			& & \multicolumn{1}{l|}{$t_1=2$} & 0.010  & 0.059 &\multicolumn{1}{l|}{0.107}&0.115  & 0.320 & 0.473\\
			\hline
			1.0 & $n_H=200$ & \multicolumn{1}{l|}{$t_1=1$} & 0.009 & 0.033 &\multicolumn{1}{l|}{0.076}&1.000  & 1.000 & 1.000 \\
			&  & \multicolumn{1}{l|}{$t_1=2$} & 0.017  & 0.067 &\multicolumn{1}{l|}{0.116}&0.045  & 0.135 & 0.239\\
			\hline
			1.5	& $n_H=200$  & \multicolumn{1}{l|}{$t_1=1$} & 0.005  & 0.038 &\multicolumn{1}{l|}{0.076} &1.000  & 1.000 & 1.000\\
			&   & \multicolumn{1}{l|}{$t_1=2$} & 0.013  & 0.045 &\multicolumn{1}{l|}{0.095} &0.172  & 0.491 & 0.703\\
			\hline
		\end{tabular}%
	\end{center}
	\parbox{6.2in}{\footnotesize \medskip
		Notes: This table reports the results of the directional tests, with the cutoff degree $\overline d_n = 5$. Each sample consists of $30$ villages with $50$ or $200$ households in each village. The FWERs were computed according to Holm's multiple testing procedure over $100$ hypotheses, using $1,000$ simulations.}
	\medskip
	\label{table_dir_sim}
	\bigskip
	\bigskip
	\bigskip
\end{table}

\section{Empirical Application: Diffusion of Microfinancing Decisions over Social Networks}
\label{sec: emp app}
\subsection{Data}
We apply our procedure to estimate the ADM of microfinancing decisions in 43 rural villages in southern India. The data set used here originated from the data used by \cite{Banerjee/Chandrasekhar/Duflo/Jackson:13:Science}. In 2006, social network data were collected for 75 rural villages in Karnataka, a state in southern India. The network was measured along the twelve dimensions, such as: visiting each other; kinship; borrowing or lending money, rice, or kerosene; giving or exchanging advice; or going to the place of prayer together. The data contain information on microfinancing decisions observed over the trimesters after the program began operations as well as other individual characteristics such as caste, village leader indicator, savings indicator, and education. In this empirical study, the cross-sectional units are households.

We distinguish between two definitions of initial state-switchers. First, we define the \bi{leaders} as those who first learned about the micro-financing program. In 2007, a microfinancing institution, Bharatha Swamukti Samsthe (BSS), began operations in 43 of the villages, first by inviting a pre-determined set of people (such as teachers, shopkeepers, and saving group leaders) to a private meeting with credit officers, and asking the leaders to spread the information on the microfinancing program. The leaders were pre-determined based on their expected well-connectedness within the village. Hence, they fit well with the definition of the leaders in this data. Second, we define the \bi{leader-adopters} as those who are leaders and participated in the micro-financing program in the first trimester.

First, we focus on the ADM based on the initial switchers as households, one of whose members is a leader-adopter, that is, we define $Y_{j,0} = 1$ if and only if household $j$ has a leader-adopter. Then, we take $Y_{i,1} = 1$ if and only if household $i$ adopted microfinancing in the first trimester or later. This ADM measures the diffusion of microfinancing decisions triggered by the village leaders participating in the microfinancing program. Second, we define the initial switchers to be those households that have a leader, that is, we define $Y_{j,0} = 1$ if and only if the household $j$ has a village leader, and take $Y_{i,1}$ as before. The ADM in this case measures the diffusion triggered by the village leaders' obtaining the information on the microfinancing program.

For the study, we construct directed networks at the household level. Similar to \cite{Leung:15:JOE}, we obtain directed links based on the individual survey questions that collected social network data along the twelve dimensions.\footnote{The networks are constructed as follows. For each graph, the household-level adjacency matrix is constructed from the individual data, where a node is a household. One household is linked to another if any of its members is linked through the relationships indicated by the social network.} We define graph $G_{\mathsf{ee}}$ to be the social network where two households are linked if and only if material exchanges (borrowing/lending rice, kerosene or money) occurred, and graph $G_{\mathsf{sc}} $ to be the social network, where two households are linked if and only if some social activities (such as seeking advice, or going to the same temple or church) occurred. We also consider graph $G_{\mathsf{all}}$, which is the union of $G_{\mathsf{ee}}$ and $G_{\mathsf{sc}}$ so that two households are linked if and only if any of the 12 dimensions in the social network data (as mentioned before) occurred between the households. The summary statistics for the three networks are listed in Table \ref{table-em1}.

\begin{table}[t]
	\caption{Social Network Characteristics from Survey Data on Indian Villages}
	\small
	\begin{center}
		\begin{tabular}{c|cccc}
			\hline\hline
			Networks & Size & Maximum Degree & Average Degree & Median Degree \\ \hline
			$G_{\mathsf{ee}}$ & 4413 & 42 & 3.7088 & 3 \\
			$G_{\mathsf{sc}}$ & 4413 & 76 & 5.6646 & 5 \\
			$G_{\mathsf{all}}$ & 4413 & 78 & 6.1854 & 5 \\ \hline
		\end{tabular}
	\end{center}
	\par
	\parbox{6.2in}{\footnotesize \medskip
		
		Notes: The network $G_{\mathsf{ee}}$ is defined based on material exchanges between households
		(such as borrowing rice, kerosene or money). The network $G_{\mathsf{sc}}$ is defined based on
		social activities (such as seeking advice or going to the same temple or church).
		The network $G_{\mathsf{all}}$ is the union of the two networks $G_{\mathsf{ee}}$ and $G_{\mathsf{sc}}$.		
		\medskip}
	\label{table-em1}
\end{table}

We include similar covariates from the survey data included in \cite{Banerjee/Chandrasekhar/Duflo/Jackson:13:Science}. The demographic controls in \cite{Banerjee/Chandrasekhar/Duflo/Jackson:13:Science} include the number of households, self-help group participation in the village, savings participation in the village, caste composition, and village households that are designated as leaders. Except for the number of the households which is the same across households in the same village, all the other controls are dummy variables. We construct the set of covariates from the household averages. The household average covariates are constructed by averaging the individual covariates from the same household. For example, if any member of a household participates in a self-help group(SHG)/has savings/is a leader in the village, then the dummy for SHG/savings/leaders for the household average equals 1. The caste composition dummy is an indicator of minority group, which includes the scheduled caste and scheduled tribes, the relatively disadvantaged groups. In the following experiment, we use the leaders dummy to construct the initial outcome and therefore exclude it from the covariates.

The data appears to fit our framework well. The leader status was pre-determined primarily based on individuals' positions in the social networks. They were the first people who were exposed to the information on the microfinancing program, and learned about the program through private meetings. Therefore, it is plausible that their leadership status and initial adoptions are conditionally independent across different households, and independent from the potential outcomes of their neighbors, given the networks and observed characteristics. Furthermore, the researcher does not observe the actual process of diffusion. The administrative data provided by BSS contained information on microfinancing decisions over several trimesters. It is plausible that within each trimester, the diffusion progressed without the researcher observing it. Our framework accommodates this feature of misalignment between the actual timing of the diffusion and the researcher's observation.

\subsection{Results}
\begin{figure}[t]
	\begin{center}
		\includegraphics[scale=0.75]{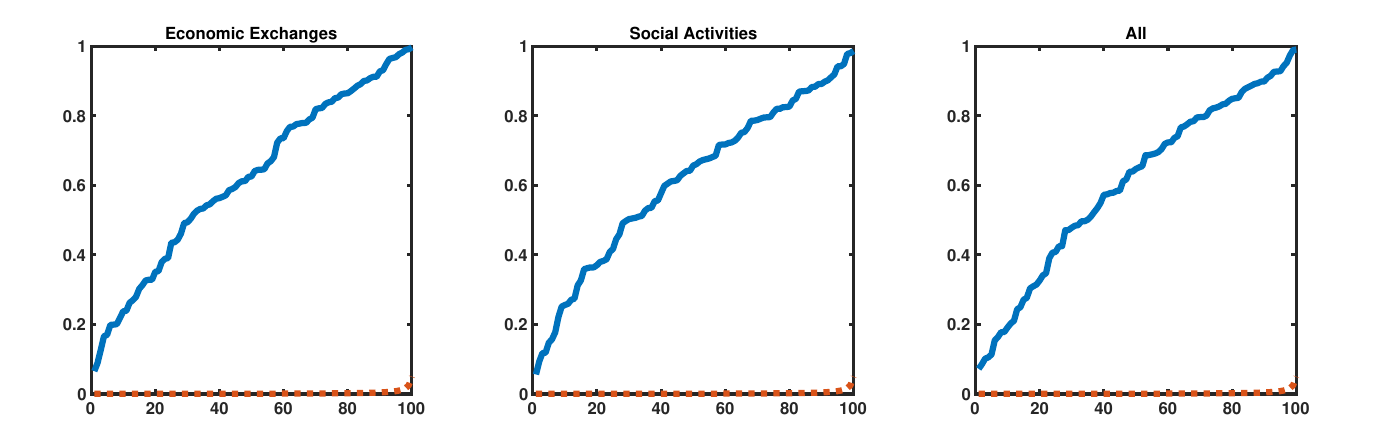}
		\caption{Directional Tests of the Subgraph Hypothesis Using the ADM Based on Leader-Adopters and Multiple Subgraphs of the G2-Shell with the Cutoff Degree Equal to 5: \footnotesize The three panels show the directional testing results for three types of networks, based on economic exchanges, social activities, and the union of all relations, using the original graph in the data. The tests are based on the ADM with leader-adopters as initial state-switchers.  The cutoff degree is set to be 5 for all the three types of networks. The solid line is the sorted $p$-values and the dotted line is the Holm thresholds at FWER control at 5\%. We reject the null hypothesis of the subgraph hypothesis, if the minimum value of the dotted line lies above the blue line at some point. The null hypothesis is not rejected for all three networks with the original graph. Thus, the results show no evidence against the subgraph hypothesis.}
		\label{dir_g1}
	\end{center}

\end{figure}

\subsubsection{Directional Tests of the Subgraph Hypothesis}

Our proposal offers two routes to the researcher to make inference on the ADM. First, she can construct the confidence interval for the ADM under the subgraph hypothesis for the observed graph used. Second, she can replace the subgraph hypothesis by the assumption of nonnegative spillover effect in (\ref{MDC0}), and construct the lower confidence bound for the ADM.

In this empirical application, we first investigate whether there is any evidence against the subgraph hypothesis for the three observed networks, $G_{\mathsf{ee}}$, $G_{\mathsf{sc}}$, and $G_{\mathsf{all}}$, using the directional tests we introduced earlier. We performed this test in two different ways: first, using the multiple testing approach (the Holm procedure) based on 100 random subgraphs from the G2-shell of the observed graphs, and second, using a single test based on the G2-shell of the observed graph.

The results from the multiple testing approach are shown in Figure \ref{dir_g1}. Here, the tests are based on the ADM with leader-adopters as initial state-switchers. The cutoff degree is set as $\overline d_n = 5$ for all the three kinds of networks. The solid lines indicate the sorted $p$-values and the dotted lines represent the Holm's thresholds (the term on the right-hand side of (\ref{Holm})). The $p$-values for the different networks are all substantially and uniformly greater than the Holm's thresholds at 5\% of FWER. Hence we do not reject any of the individual hypotheses at 5\% of FWER. This result shows no evidence against the subgraph hypothesis for the observed graphs, $G_{\mathsf{ee}}$, $G_{\mathsf{sc}}$, and $G_{\mathsf{all}}$.\footnote{We also performed multiple testing, setting the cut-off degree to be the ceiling of the average degree of the observed graph, and with leaders as initial state-switchers. The results are similar, showing no evidence against the subgraph hypothesis at the 5\% significance level. These results are omitted for brevity.}

We also perform a single test of the subgraph hypothesis by checking if the ADM is significantly different from zero when the G2-shell is used as the observed graph. Under the subgraph hypothesis, the ADM is zero. Hence, if there are links not captured by the observed graph but captured by the G2-shell, the ADM constructed using the G2-shell may be different from zero. The results of this directional test are reported in Table \ref{table-em2}. The estimated ADM using the G2-shell is not significantly different from zero at $99\%$, $95\%$, and $90\%$ across the different definitions of social networks. This result suggests no evidence against the subgraph hypothesis.

In summary we investigated whether the subgraph hypothesis is plausible or not in our data, using both multiple-testing and single-testing approaches. We have not found any evidence against the subgraph hypothesis for the observed graphs.

\begin{table}[t]
	\caption{Directional Tests of the Subgraph Hypothesis Using the G2-Shell}
	\small
	\begin{center}
		
		\begin{tabular}{c|cccc}
			\hline\hline
			&&&  CI& \\ \cline{3-5}
			& $\mathsf{ADM}$  & 99\% & 95\%& 90\%  \\ \hline
			$G_{\mathsf{ee}}$ &  0.113 &[-0.400,0.625]&[-0.277,0.503]&[-0.215,0.440]
			\\
			$G_{\mathsf{sc}}$ & -0.170 &[-1.087,0.747]&[-0.868,0.527]&[-0.756,0.415]
			\\
			$G_{\mathsf{all}}$ & -0.088 &[-0.926,0.750]&[-0.726,0.550]&[-0.623,0.447]
			\\ \hline
		\end{tabular}
	\end{center}

	\par
	\parbox{6.2in}{\footnotesize \medskip
		Notes: The subgraph hypothesis implies that the ADM measured with the G2-shell of the observed graph is zero. The confidence intervals for the ADM contain zero at 99\%, 95\% and 90\%. This suggests no evidence against the subgraph hypothesis in the directional test using the ADM measured with the G2-shell.}
	\label{table-em2}
	\medskip
\end{table}

\subsubsection{Estimated ADM}
We measure the ADM in two different ways. First, we measure the ADM, taking the initial state-switchers as households having a leader-adopter. Second, we measure the ADM, taking the initial state-switchers as households having a leader. The results are shown in Table \ref{table-em3}.

The estimated ADMs with leader-adopters as initial switchers are significantly different from zero at 95\% except for $G_{\mathsf{ee}}$ when we use the subgraph hypothesis. When we drop the subgraph hypothesis but employ the assumption of nonnegative spillover effects, the estimated ADM is significantly different from zero at 95\% for all three graphs. This shows the role of social networks in channeling the diffusion of microfinancing decisions that were triggered by the leader's microfinancing decisions. The expected increase in the number of switchers by the end of the observation period due to one additional leader-adopter is estimated to be around 0.22-0.43, depending on the definition of the graphs.

When we considered the ADM using the leaders as initial switchers, the estimated ADMs were insignificant across all the graphs, in terms of the lower confidence bound or the confidence intervals. This shows that even if there was a diffusion of information on the microfinancing programs, their actual impacts may be significantly different across the leaders, depending on whether the leader actually participated in the program or not. For example, it is conceivable that the influence of the leaders who participated in the microfinancing program was stronger than the influence of the leaders who learned about the program but did not participate in it. Our results demonstrate that the measured diffusion can depend crucially on how we define the initial triggers.

\begin{table}[t]
	\caption{The Estimated ADM of Microfinancing over Social Networks}
	\small
	\begin{center}
		\textbf{The Case with the Initial Switchers Being Leader-Adopters.}
		\begin{tabular}{c|cccc|ccc}
			\hline\hline
			&&&  LB &&& CI \\ \cline{3-5} \cline{6-8}
			& $\mathsf{ADM}$  & 99\% & 95\%& 90\%& 99\% & 95\% & 90\%\\ \hline
			$G_{\mathsf{ee}}$ & 0.215 &-0.075 &0.010& 0.055 &[-0.106,0.536] &[-0.029,0.460]&[0.010,0.420]
			\\
			$G_{\mathsf{sc}}$ & 0.434 & 0.035 &0.152& 0.214 &[-0.008,0.875]&[0.098,0.770]&[0.152,0.716]
			\\
			$G_{\mathsf{all}}$ & 0.435&0.020 & 0.141& 0.206 &[-0.025,0.895]&[0.085,0.785]&[0.141,0.728]
			\\ \hline
		\end{tabular}
		\medskip
		\medskip
		
		\textbf{The Case with the Initial Switchers Being Leaders.}
		\begin{tabular}{c|cccc|ccc}
			\hline\hline
			&&&  LB &&& CI \\ \cline{3-5} \cline{6-8}
			& $\mathsf{ADM}$  & 99\% & 95\%& 90\%& 99\% & 95\% & 90\%\\
			\hline
			$G_{\mathsf{ee}}$ & -0.052 & -0.137 & -0.112& -0.099 & [-0.146,0.041]& [-0.123,0.019] &[-0.112,0.007]
			\\
			$G_{\mathsf{sc}}$ & -0.049 & -0.165 &-0.131 & -0.113  & [-0.178,0.080]& [-0.146,0.049] &[-0.131,0.033]
			\\
			$G_{\mathsf{all}}$ & -0.051 & -0.167 &-0.133& -0.115  & [-0.180,0.079]&[-0.149,0.048] &[-0.133, 0.032]
			\\ \hline
		\end{tabular}
	\end{center}
	
	\par
	\parbox{6.2in}{\footnotesize \medskip
		Notes: The estimated ADM of microfinancing decisions depends on how we define initial switchers. When we take the initial switchers to be leader-adopters, the results are significantly positive overall. When we take them to be just leaders, the significance disappears. The household covariates include four basic controls: number of households in the village, dummies for self-help group participation, savings participation, and caste composition.\medskip}
	\label{table-em3}
\end{table}

\section{Conclusion}
In this paper, we proposed a measure of diffusion of binary outcomes over a large network using the potential outcome approach in causal inference. We introduced a spatio-temporal dependence measure of outcomes and showed that it is identified as the measure of diffusion under a certain condition for the observed network. We showed that when the condition fails but the spillover effect is nonnegative, the dependence measure is a lower bound for the measure of diffusion. Then, we developed an inference method and established its asymptotic validity, despite the fact that the observations exhibit heterogeneity and a complex form of cross-sectional dependence. Our empirical application demonstrated the usefulness of our framework.

There are different directions for extensions that appear potentially promising. First, we may consider a setting where the unconfoundedness condition fails, because the initial switches are observed a while after the diffusion already progressed. It would be interesting to see if we can develop an instrumental variable approach in this situation. Second, it would be useful to explore methods that can be used to assess the role of the covariates in the causal interpretation of the spatio-temporal dependence measure. While \cite{Song:22:ET} provides a method of decomposition to capture the spurious diffusion, the method assumes the subgraph hypothesis. It would be interesting to characterize the behavior of the spatio-temporal dependence measure as a lower bound when the subgraph hypothesis fails.

\bigskip
\bigskip
\bigskip

\putbib[measuring_diffusion]
\end{bibunit}

\begin{bibunit}[econometrica]
\newpage
\appendix

\vspace*{5ex minus 1ex}
\begin{center}
	\Large \textsc{Supplemental Note to ``Measuring Diffusion over a Large Network''}
	\bigskip
\end{center}

\date{%
	\today%
}

\vspace*{7ex minus 1ex}
\begin{center}
	Xiaoqi He and Kyungchul Song\\
	\textit{Central University of Finance and Economics and University of British Columbia}
	\bigskip
\end{center}

This note is the Supplemental Note to \cite{He/Song:22:arXiv}. It is divided into four parts. Part A explains the general notion of dependency causal graph which captures the stochastic dependence between variables that arises due to their causal connections. This notion is used to produce a condition that is weaker than the subgraph hypothesis in the paper. We prove all the results in the paper under this general condition. Part B presents the proofs of the results in the paper along with auxiliary results. Part C provides low level conditions for Assumptions \ref{assump: asymp approx} - \ref{assump: rate est funs}. Part D provides some further details on Assumption \ref{assump: non-DCG} using a linear threshold diffusion model.

\section*{A. Dependency Causal Graphs }
\setcounter{section}{1}
\setcounter{subsection}{0}
\setcounter{equation}{0}
\setcounter{lemma}{0}

The main proposal of \cite{He/Song:22:arXiv} for measuring diffusion is built on several crucial assumptions. In particular, we focus on two assumptions used in the paper. The first assumption is that the initial switches and the unobserved heterogeneities are cross-sectionally independent conditional on the contact network, the observed graph, the entire profile of covariates, and the unobserved common shock. Second, for the point-identification of the ADM, the paper uses the subgraph hypothesis. In this section, we show how these two conditions can be relaxed into the requirement that both the causal graph and the observed graph be what we call Dependency Causal Graphs.

For any set $A \subset N$, we write $Y_{A,1} = (Y_{i,1})_{i \in A}, \text{ and } Y_{A,0} = (Y_{j,0})_{j \in A}$, and for any graph $G = (N,E_G)$, let
\begin{align*}
	N_G(A) = \bigcup_{i \in A} N_G(i) \text{ and } \overline N_G(A) = N_G(A) \cup A,
\end{align*}
where $N_G(i) = \{j \in N: ij \in E_G\}$. The set $\overline N_G(A)$ represents the set of people in $A$ and their in-neighbors in $G$. Recall our notation $Y_1 = (Y_{i,1})_{i \in N}$ and $Y_0 = (Y_{j,0})_{j \in N}$.

\begin{definition}
	Given a graph $G$ that is $\mathcal{F}$-measurable, we say that $G$ is a \textit{\textbf{Dependency Causal Graph (DCG) for $(Y_1,Y_0)$ given $\mathcal{F}$}}, if for any subsets $A,B,A',B' \subset N$ such that
	\begin{align*}
		(\overline N_G(A) \cup B) \cap (\overline N_G(A') \cup B') = \varnothing,
	\end{align*}
	we have
	\begin{align*}
		(Y_{A,1} , Y_{\overline N_G(A)\cup B,0}) \CI (Y_{A',1}, Y_{\overline N_G(A')\cup B',0}) \mid \mathcal{F},
	\end{align*}
	where $Z_1 \CI Z_2 \mid \mathcal{F}$ denotes the conditionally independence of $Z_1$ and $Z_2$ given $\mathcal{F}$.
\end{definition}

When $(Y_1,Y_0)$ has graph $G$ as a DCG given $\mathcal{F}$, this implies that $Y_{j,0}$'s are conditionally independent given $\mathcal{F}$, that is, any source of cross-sectional dependence among the initial period outcomes is already included in $\mathcal{F}$. Furthermore, conditional on the \textit{same} $\mathcal{F}$, any state-switches by next observation time $t_1$, $Y_{i_1,1}$ and $Y_{i_2,1}$, are allowed to be correlated only when they share certain people in their in-neighborhoods in $G$. The definition of DCG also involves sets $B$ and $B'$, so that $Y_{j,0}$ is conditionally independent of $Y_{i,1}$ given $\mathcal{F}$, whenever $j$ is not in the in-neighborhood of $i$, that is, the initial state-switch of person $j$ does not influence that of person $i$ by period $t_1$, and $Y_{j_1,0}$ and $Y_{j_2,0}$ are conditionally independent given $\mathcal{F}$ whenever $j_1 \ne j_2$.

We introduce a condition that is weaker than Assumption \ref{assump: cond indep}.
\begin{assumption}
	\label{assump: DCG}
	The causal graph $G_{\mathsf{cau}}$ is a DCG for $(Y_1,Y_0)$ given $\mathcal{F}$.     	
\end{assumption}

This assumption means that if there is any dependence between $Y_{j,0}$ and $Y_{i,1}$, this must be because person $j$ is in the in-neighborhood of $i$ in the causal graph $G_{\mathsf{cau}}$. Recall the notation $U_i = (U_{i,1},...,U_{i,t_1})$, a vector of the unobserved attributes for person $i$ which affects her actions from time $t=1$ until time $t=t_1$. We obtain the following result.

\begin{lemma}
	\label{lemm: DCG2}
	Suppose that Assumption \ref{assump: cond indep} holds, namely that $(U_i,Y_{i,0})$'s are conditionally independent across $i$'s given $\mathcal{F}$. Then, $G_{\mathsf{cau}}$ is a DCG for $(Y_1,Y_0)$ given $\mathcal{F}$.
\end{lemma}

\noindent \textbf{Proof: } Note that $Y_{i,1}$ in  (\ref{actual outcome2}) is determined by $(U_j,Y_{j,0})_{j \in \overline N_{\mathsf{cau}}(i)}$ and $G_{\mathsf{ctt}}$, where $\overline N_{\mathsf{cau}}(i) = N_{\mathsf{cau}}(i) \cup \{i\}$ and $N_{\mathsf{cau}}(i)$ is the in-neighborhood of $i$ in $G_{\mathsf{cau}}$. Take subsets $A$, $B$, $A'$, $B'$ of $N$ such that $\overline N_{\mathsf{cau}}(A) \cup B$ and $\overline N_{\mathsf{cau}}(A') \cup B'$ are disjoint. We observe from (\ref{actual outcome2}) that $(Y_{A,1},Y_{\overline N_{\mathsf{cau}}(A) \cup B,0})$ is a function of $((U_j,Y_{j,0})_{j \in \overline N_{\mathsf{cau}}(A) \cup B}, G_{\mathsf{ctt}})$, and similarly, $(Y_{A',1},Y_{\overline N_{\mathsf{cau}}(A') \cup B',0})$ is a function of $((U_j,Y_{j,0})_{j \in \overline N_{\mathsf{cau}}(A') \cup B'}, G_{\mathsf{ctt}})$. Since $\overline N_{\mathsf{cau}}(A) \cup B$ and $\overline N_{\mathsf{cau}}(A') \cup B'$ are disjoint, $(Y_{A,1},Y_{\overline N_{\mathsf{cau}}(A) \cup B,0})$ and $(Y_{A',1},Y_{\overline N_{\mathsf{cau}}(A') \cup B',0})$ are conditionally independent given $\mathcal{F}$, by the assumption of the lemma. Thus, $G_{\mathsf{cau}}$ is a DCG for $(Y_1,Y_0)$ given $\mathcal{F}$. $\blacksquare$\medskip

The lemma says that the causal graph $G_{\mathsf{cau}}$ is a DCG, if the unobserved heterogeneities and initial state-switches are cross-sectionally conditionally independent given $\mathcal{F}$. It is also worth noting that a DCG is not uniquely determined by a conditional joint distribution of $(Y_1,Y_0)$ given $\mathcal{F}$. If $G_{\mathsf{cau}}$ is a DCG, then any graph,  say, $G$ that contains $G_{\mathsf{cau}}$ as a subgraph is also a DCG. This is because any two sets $A$ and $B$ with non-overlapping neighborhoods in $G$ are also non-overlapping in $G_{\mathsf{cau}}$ when $G_{\mathsf{cau}}$ is a subgraph of $G$. This observation yields the following lemma saying that the subgraph hypothesis is stronger than the hypothesis that $G_\mathsf{obs}$ is a DCG for $(Y_1,Y_0)$ given $\mathcal{F}$.
\begin{lemma}
	\label{assump: DCG2}
	Suppose that $G_{\mathsf{cau}}$ is a DCG for $(Y_1,Y_0)$ given $\mathcal{F}$, and that $G_\mathsf{obs}$ satisfies the subgraph hypothesis. Then,  $G_\mathsf{obs}$ is a DCG for $(Y_1,Y_0)$ given $\mathcal{F}$.
\end{lemma}

The subgraph hypothesis is simpler but stronger than the DCG assumption. For example, some people in the neighborhood in the contact network $G_{\mathsf{ctt}}$ may stop exerting causal influence after a certain period (such as in the case where the function $\rho_{i,t}$ in (\ref{actual outcome}) no longer depends on such people after some time). In this case, the observed graph $G_{\mathsf{obs}}$ can fail the subgraph hypothesis and yet still be a DCG. This is because the presence of an edge between two nodes in the causal graph $G_{\mathsf{cau}}$ \textit{does not necessarily mean that there is causal influence from one to the other}. It only indicates that such causal influence is allowed in the model.

The literature of probabilistic graphical models associates a causal graph with conditional independence restrictions. (See \cite{Koller/Friedman:09:ProbGraph} and references therein.) The conditional independence restrictions in this paper are different from those used in this literature. Suppose that the intersection of the in-neighborhoods of vertices $i_1$ and $i_2$ is equal to $\{j\}$. Then a probabilistic graphical model assumes that $Y_{i_1,1}$ and $Y_{i_2,1}$ are conditionally independent given $Y_{j,0}$. While such conditional independence can be accommodated in our framework, it is not used in the definition of a DCG here.

\section*{B. Mathematical Proofs}
\setcounter{section}{2}
\setcounter{subsection}{0}
\setcounter{equation}{0}
\setcounter{lemma}{0}

For the mathematical proofs of all the results in the paper, we use Assumption \ref{assump: DCG} in place of Assumption \ref{assump: cond indep}, and the assumption that $G_{\mathsf{obs}}$ is a DCG for $(Y_1,Y_0)$ given $\mathcal{F}$ in place of the subgraph hypothesis for $G_{\mathsf{obs}}$.

\subsection{Proofs of Lemma \ref{lemm: identification} and Proposition \ref{prop: spurious diffusion}}
\begin{lemma}
	\label{lemm: DCG}
	Suppose that $G_1 = (N,E_1)$ and $G_2 = (N,E_2)$ are DCGs for $(Y_1,Y_0)$ given $\mathcal{F}$. Then, for each $j \in N$,
	\begin{align*}
		\sum_{i \in N: ij \in E_1} \text{Cov}(Y_{i,1}, Y_{j,0}\mid \mathcal{F}) = \sum_{i \in N: ij \in E_2} \text{Cov}(Y_{i,1}, Y_{j,0}\mid \mathcal{F})
		= \sum_{i \in N: ij \in E_1 \cap E_2} \text{Cov}(Y_{i,1}, Y_{j,0}\mid \mathcal{F}).
	\end{align*}
\end{lemma}

\noindent \textbf{Proof:} The result follows because whenever $ij \notin E_1$ or $ij \notin E_2$, $\text{Cov}(Y_{i,1}, Y_{j,0}\mid \mathcal{F}) = 0$, by the condition that $G_1$ and $G_2$ are DCGs. $\blacksquare$\medskip

\begin{lemma}
	\label{lemm: Cov}
	Suppose that Assumption \ref{assump: unconfounded} holds. Then, for each pair $i,j \in N$,
	\begin{align*}
		\mathbf{E}[Y_{ij}^*(1) - Y_{ij}^*(0)\mid \mathcal{F}] = \frac{\text{Cov}(Y_{i,1},Y_{j,0}\mid \mathcal{F})}{\mu_{j,0}(1 - \mu_{j,0})},
	\end{align*}
where we recall the definition: $\mu_{j,0} = \mathbf{E}[Y_{j,0}\mid \mathcal{F}]$.
\end{lemma}
\noindent \textbf{Proof: } Since $Y_{j,0} \in \{0,1\}$, we write
\begin{align*}
	Y_{i,1} Y_{j,0}  = Y_{ij}^*(1) Y_{j,0} \text{ and }
	Y_{i,1} (1-Y_{j,0}) = Y_{ij}^*(0) (1-Y_{j,0}).
\end{align*}
Hence, taking conditional expectations given $\mathcal{F}$, and, using Assumption \ref{assump: unconfounded},
\begin{align*}
	\frac{\mathbf{E}[Y_{i,1} Y_{j,0} \mid \mathcal{F}]}{\mu_{j,0}} - \frac{\mathbf{E}[Y_{i,1} (1-Y_{j,0}) \mid \mathcal{F}]}{1-\mu_{j,0}}
	= \mathbf{E}[Y_{ij}^*(1) - Y_{ij}^*(0)\mid \mathcal{F}].
\end{align*}
We write the difference on the left-hand side as
\begin{align*}
	& \frac{(1 - \mu_{j,0})\mathbf{E}[Y_{i,1} Y_{j,0} \mid \mathcal{F}] - \mu_{j,0}\mathbf{E}[Y_{i,1} (1-Y_{j,0}) \mid \mathcal{F}]}{\mu_{j,0}(1 - \mu_{j,0})} = \frac{\text{Cov}(Y_{i,1},Y_{j,0}\mid \mathcal{F})}{\mu_{j,0}(1 - \mu_{j,0})}.
\end{align*}
The equality follows because $\mu_{j,0} = \mathbf{E}[Y_{j,0}\mid \mathcal{F}]$. $\blacksquare$\medskip

\noindent \textbf{Proof of Lemma \ref{lemm: identification}:} (i) By Lemma \ref{lemm: Cov}, we obtain that
\begin{align*}
	\mathsf{ADM} &= \sum_{j \in N} \frac{w_j}{\sigma_{j,0}^2} \sum_{i \in N: ij \in E_{\mathsf{cau}}} \text{Cov}(Y_{i,1}, Y_{j,0}\mid \mathcal{F})\\
	&= \sum_{j \in N} \frac{w_j}{\sigma_{j,0}^2} \sum_{i \in N: ij \in E_{\mathsf{obs}}} \text{Cov}(Y_{i,1}, Y_{j,0}\mid \mathcal{F}) = \mathsf{C}.
\end{align*}
The second equality follows from Lemma \ref{lemm: DCG} because both $G_{\mathsf{cau}}$ and $G_{\mathsf{obs}}$ are DCGs.\medskip

(ii) Using Lemma \ref{lemm: Cov}, we write
\begin{align*}
	\mathsf{ADM} &= \sum_{j \in N} \frac{w_j}{\sigma_{j,0}^2}  \sum_{i \in N: ij \in E_{\mathsf{cau}} \setminus E_{\mathsf{obs}}} \text{Cov}(Y_{i,1}, Y_{j,0}\mid \mathcal{F}) + \sum_{j \in N} \frac{w_j}{\sigma_{j,0}^2} \sum_{i \in N: ij \in E_{\mathsf{cau}} \cap E_{\mathsf{obs}}} \text{Cov}(Y_{i,1}, Y_{j,0}\mid \mathcal{F})\\
	&= \sum_{j \in N} \frac{w_j}{\sigma_{j,0}^2} \sum_{i \in N: ij \in E_{\mathsf{cau}} \setminus E_{\mathsf{obs}}} \text{Cov}(Y_{i,1}, Y_{j,0}\mid \mathcal{F}) + \mathsf{C},
\end{align*}
because for each $ij \in E_{\mathsf{obs}} \setminus E_{\mathsf{cau}}$, $\text{Cov}(Y_{i,1},Y_{j,0}\mid \mathcal{F})  = 0$ by the assumption that $G_{\mathsf{cau}}$ is a DCG. The desired result follows by (\ref{MDC0}) and Lemma \ref{lemm: Cov}. $\blacksquare$\medskip

We let $\varepsilon_{i,1} = Y_{i,1} - \mathbf{E}[Y_{i,1}\mid \mathcal{F}]$ and $\varepsilon_{j,0} = Y_{j,0} - \mathbf{E}[Y_{j,0}\mid \mathcal{F}]$.\medskip

\noindent \textbf{Proof of Proposition \ref{prop: spurious diffusion}: } Let $\mathcal{F}_{-S} = \sigma(G_{\mathsf{ctt}}, G_{\mathsf{obs}}, X_{-S},\mathcal{C})$. Define
\begin{align*}
	\varepsilon_{j,0,-S} = Y_{j,0} - \mathbf{E}[Y_{j,0}\mid \mathcal{F}_{-S}] \text{ and }
	\varepsilon_{i,1,-S} = Y_{i,1} - \mathbf{E}[Y_{i,1} \mid \mathcal{F}_{-S}].
\end{align*}
Let us write
\begin{align*}
	\varepsilon_{j,0,-S} = \varepsilon_{j,0} + \mu_{j,0,-S}^\Delta \text{ and }
	\varepsilon_{i,1,-S} = \varepsilon_{i,1} + \mu_{i,1,-S}^\Delta,
\end{align*}
where
\begin{align*}
	\mu_{j,0,-S}^\Delta = \mathbf{E}[Y_{j,0} \mid \mathcal{F}] - \mathbf{E}[Y_{j,0}\mid \mathcal{F}_{-S}] \text{ and } \mu_{i,1,-S}^\Delta = \mathbf{E}[Y_{i,1}\mid \mathcal{F}] - \mathbf{E}[Y_{i,1}\mid \mathcal{F}_{-S}].
\end{align*}
Then we can write $\mathsf{C}_{-S}$ as
\begin{align}
	\label{develop}
	&\sum_{j \in N} \sum_{i\in N: ij \in E_{\mathsf{cau}}} \frac{\text{Cov}(Y_{i,1}, Y_{j,0}\mid \mathcal{F}_{-S}) w_j}{\text{Var}(Y_{j,0}\mid \mathcal{F}_{-S})}\\ \notag
	&= \sum_{j \in N} \sum_{i\in N: ij \in E_{\mathsf{cau}}} \frac{\text{Cov}(\varepsilon_{i,1}, \varepsilon_{j,0}\mid \mathcal{F}_{-S}) w_j}{\text{Var}(Y_{j,0}\mid \mathcal{F}_{-S})} + \sum_{j \in N} \sum_{i\in N: ij \in E_{\mathsf{cau}}} \frac{\text{Cov}\left(\mu_{i,1,-S}^\Delta,\mu_{j,0,-S}^\Delta\mid \mathcal{F}_{-S}\right) w_j}{\text{Var}(Y_{j,0}\mid \mathcal{F}_{-S})}.
\end{align}
The equality above follows because $\mathcal{F}_{-S} \subset \mathcal{F}$, $\mathbf{E}[\varepsilon_{j,0}\mid \mathcal{F}] = 0$, and  $\mathbf{E}[\varepsilon_{i,1}\mid \mathcal{F}] = 0$. The leading term on the right-hand side is zero because whenever $i \ne j$, $Y_{i,1}$ and $Y_{j,0}$ are conditionally independent given $\mathcal{F}$ due to the no-diffusion assumption. (Note that the conditional independence assumption in this proposition implies Assumption \ref{assump: cond indep}.) For the last covariance, since $Y_{j,0}$ is conditionally independent of $X_{-j}$ given $(G_{\mathsf{obs}},G_{\mathsf{ctt}},\mathcal{C})$, we find that $\mathbf{E}[Y_{j,0}\mid \mathcal{F}]$ is a function of $(X_j,G_{\mathsf{obs}}, G_{\mathsf{ctt}},\mathcal{C})$ only, and so is $\mu_{j,0,-S}^\Delta$.

Similarly, since $(U_i,Y_{i,0})$ is conditionally independent of $X_{-i}$ given $(G_{\mathsf{obs}},G_{\mathsf{ctt}},\mathcal{C})$, using (\ref{obs outcome2}) and (\ref{actual outcome22}), we find that $\mathbf{E}[Y_{i,1}\mid \mathcal{F}]$ is also a function of $(X_i,G_{\mathsf{obs}}, G_{\mathsf{ctt}},\mathcal{C})$ only, and so is $\mu_{i,1,-S}^\Delta$. Since $X_i$'s are conditionally independent given $(G_{\mathsf{obs}}, G_{\mathsf{ctt}},\mathcal{C})$, the conditional covariance in the last term in (\ref{develop}) is zero. $\blacksquare$

\subsection{Preliminary Results on Moments}

For the rest of the proofs, notation $C$ represents a positive constant that does not depend on $n$. Constant $C$ may take different values in different places. For each $i \in N$, let
\begin{align*}
	N(i) = \{j \in N: ij \in E_{\mathsf{cau}} \text{ or } ij \in E_{\mathsf{obs}}\}.
\end{align*}
Hence $N(i)$ is the union of the in-neighborhoods of $i$ in the causal graph $G_{\mathsf{cau}}$ and the observed graph $G_{\mathsf{obs}}$. Let $G = (N,E)$ be the graph defined by the neighborhoods $N(i)$, that is, $ij \in E$ if and only if $j \in N(i)$. Define
\begin{align}
	\label{max deg and ave deg}
	d_{mx} = \max_{i \in N} |N(i)| \text{ and } d_{av} = \frac{1}{n}\sum_{i \in N} |N(i)|.
\end{align}
Hence $d_{mx}$ and $d_{av}$ are the maximum and average degrees of the union of $G_{\mathsf{cau}}$ and $G_{\mathsf{obs}}$. Let $\overline N(i) = N(i) \cup \{i\}$. From here on, we assume that Assumption \ref{assump: DCG} (in place of Assumption \ref{assump: cond indep}) and Assumptions \ref{assump: unconfounded}-\ref{assump: network formation} hold.

For each $i \in N$, define
\begin{align*}
	e_{i,1} = Y_{i,1} - c_i.
\end{align*}
Recall that we have chosen $c_i$ to be an $\mathcal{F}$-measurable random variable.

\begin{lemma}\label{lemm: Var ea}
	(i)
	\begin{align*}
	\text{Var}\left( \frac{1}{\sqrt{n}}\sum_{i \in N} a_i \mid \mathcal{F}\right)
	=O_P\left(d_{av} d_{mx} \right).
	\end{align*}
	
	(ii)
	\begin{align*}
	\text{Var}\left( \frac{1}{\sqrt{n}}\sum_{i \in N}e_{i,1}a_i \mid \mathcal{F}\right)
	=O_P\left(d_{av} d_{mx}^3\right).
	\end{align*}
\end{lemma}

\noindent \textbf{Proof: } (i) We write
\begin{align}
	\label{der32}
	\text{Var}\left( \frac{1}{\sqrt{n}}\sum_{i \in N}a_i \mid \mathcal{F}\right) = \frac{1}{n}\sum_{i_1, i_2 \in N}\sum_{j \in N_{\mathsf{obs}}(i_1) \cap N_{\mathsf{obs}}(i_2)} \frac{n^2 w_j^2 \mathbf{E}\left[\varepsilon_{j,0}^2\mid \mathcal{F}\right]}{\sigma_{j,0}^4}.
\end{align}
The last equality follows because $\mathbf{E}[Y_{j,0}\mid \mathcal{F}] = \mathbf{E}[Y_{j,0}\mid G_{\mathsf{obs}},X]$, and $Y_{j,0}$'s are conditionally independent across $j$'s given $\mathcal{F}$ by Assumption \ref{assump: cond indep}. By Assumptions \ref{assump: unconfounded}(ii) and \ref{assump: rate est funs}(iii), we have
\begin{align*}
	\max_{j \in N} \frac{n^2 w_j^2\mathbf{E}\left[\varepsilon_{j,0}^2\mid \mathcal{F}\right]}{\sigma_{j,0}^4} \le \max_{j \in N} \frac{n^2 w_j^2}{\sigma_{j,0}^4} \le \frac{\overline w^2}{c^4}.
\end{align*}
Hence the last term in (\ref{der32}) is bounded by
\begin{align*}
	\frac{\overline w^2}{c^4} \times \frac{1}{n}\sum_{i_1, i_2 \in N}|N_{\mathsf{obs}}(i_1) \cap N_{\mathsf{obs}}(i_2)|.
\end{align*}
The last sum over $i_1, i_2 \in N$ is bounded by the number of the pairs $(i_1,i_2)$ such that $N_{\mathsf{obs}}(i_1) \cap N_{\mathsf{obs}}(i_2) \ne \varnothing$. This number is bounded by
\begin{align}
	\label{bound}
	\sum_{i_1 \in N} \sum_{k \in N_{\mathsf{obs}}(i_1)}|N_{\mathsf{obs}}(k)| \le n d_{av}d_{mx}.
\end{align}
This gives the desired rate.\medskip

(ii) Whenever $i_1, i_2$ are such that $\overline N_{\mathsf{cau}}(i_1) \cap \overline N_{\mathsf{cau}}(i_2) = \varnothing$, for any $j_1 \in N_\mathsf{obs}(i_1)$ and $j_2 \in N_\mathsf{obs}(i_2)$, we have $\text{Cov}\left(\varepsilon_{j_1,0} e_{i_1,1},\varepsilon_{j_2,0} e_{i_2,1} \mid \mathcal{F}\right) = 0$, because $G_{\mathsf{cau}}$ is a DCG. We write
\begin{align*}
	& \text{Var}\left( \frac{1}{\sqrt{n}}\sum_{i \in N} e_{i,1}a_i \mid \mathcal{F}\right) \\
	&= \frac{1}{n}\sum_{i_1, i_2 \in N: \overline N_{\mathsf{cau}}(i_1) \cap \overline N_{\mathsf{cau}}(i_2) \ne \varnothing}\sum_{j_1 \in N_{\mathsf{obs}}(i_1),j_2 \in N_{\mathsf{obs}}(i_2)}\frac{n^2 w_{j_1} w_{j_2} \text{Cov}\left(\varepsilon_{j_1,0} e_{i_1,1},\varepsilon_{j_2,0} e_{i_2,1}\mid \mathcal{F}\right)}{\sigma_{j_1,0}^2\sigma_{j_2,0}^2}\\
	&\le \frac{\overline w^2}{c^4} \frac{1}{n}\sum_{i_1, i_2 \in N: \overline N_{\mathsf{cau}}(i_1) \cap \overline N_{\mathsf{cau}}(i_2) \ne \varnothing} |N_{\mathsf{obs}}(i_1)| |N_{\mathsf{obs}}(i_2)| \le \frac{\overline w^2 d_{av} d_{mx}^3}{c^4}.
\end{align*}
The last inequality comes from the same arguments involving the bound (\ref{bound}). $\blacksquare$

\begin{lemma}\label{P4}
	\begin{align*}
		\text{Var}\left( \frac{1}{\sqrt{n}}\sum_{i \in N}q_i\mid \mathcal{F} \right)
		=O_P\left(d_{av}d_{mx}^3\right).
	\end{align*}
\end{lemma}

\noindent \textbf{Proof: } Note that
\begin{align}
	\label{decomp}
	\text{Var}\left( \frac{1}{\sqrt{n}}\sum_{i \in N}q_i \mid \mathcal{F} \right) &\le 2\text{Var}\left(
	\frac{1}{\sqrt{n}}\sum_{i \in N}e_{i,1}a_i \mid \mathcal{F}\right) + 2 \text{Var}\left( \frac{1}{\sqrt{n}}\sum_{i \in N}\psi_i \mid \mathcal{F}\right).
\end{align}
The leading term on the right-hand side is $O_P(d_{av} d_{mx}^3)$ by Lemma \ref{lemm: Var ea}(ii). Since $G_{\mathsf{cau}}$ is a DCG for $(Y_0,Y_1)$ given $\mathcal{F}$, $Y_{j,0}$'s are conditionally independent across $j$'s given $\mathcal{F}$. Therefore, by Assumption \ref{assump: asymp approx}, $\psi_{i_1}$ and $\psi_{i_2}$ are conditionally independent given $\mathcal{F}$, whenever
\begin{align*}
	\left(\overline N_{\mathsf{cau}}(i_1) \cup \overline N_{\mathsf{obs}}(i_1)\right) \cap \left(\overline N_{\mathsf{cau}}(i_2) \cup \overline N_{\mathsf{obs}}(i_2) \right) = \varnothing.
\end{align*}
Hence, following the same arguments in the proof of Lemma \ref{lemm: Var ea}(i), we find that the last term on the right-hand side of (\ref{decomp}) is $O_P(d_{av}d_{mx}^3)$. $\blacksquare$\medskip

\begin{lemma}
	\label{lemm: dmx}
	Suppose that the contact network is formed according to (\ref{network formation}) and the condition (\ref{rate cond}) is satisfied. Then
	\begin{align*}
		d_{mx} = O_P\left((\log n)^{k t_1 + 1}\right),
	\end{align*}
    where $k$ is the constant that appears in (\ref{rate cond}).
\end{lemma}

\noindent \textbf{Proof: } Define
\begin{align*}
	\pi_{n,c} = \max_{i \in N} \sum_{j \in N \setminus \{i\}} P\left\{\varphi_{n,ij}(X,\eta) \ge v_{ij} \mid X,\eta \right\}.
\end{align*}
We let $d_c$ be the geodesic distance on $N$ in the contact network. Define
\begin{align*}
	N_{\mathsf{ctt}}^\partial(i;s) = \left\{j \in N: d_c(i,j) = s\right\}.
\end{align*}
First, by (B.6) in the Supplemental Note of \cite{Kojevnikov/Marmer/Song:JOE:2021}, for each $s = 1,...,(t_1 \wedge n)$,
\begin{align}
	\label{bound322}
	n^{-3.3} &\ge P\left\{ |N_{\mathsf{ctt}}^\partial (i;s)| \ge 5.7 s^2(\pi_{n,c} \vee 1)^s \log n \right\}\\ \notag
	&\ge P\left\{ |N_{\mathsf{ctt}}^\partial (i;s)| \ge 5.7 t_1^2(\pi_{n,c} \vee 1)^{t_1} \log n \right\},
\end{align}
where $a \vee b$ denotes the maximum between $a$ and $b$. For each $i \in N$, we have
\begin{align*}
	N_{\mathsf{cau}}(i) \subset \bigcup_{s=1}^{t_1} N_{\mathsf{ctt}}^\partial (i;s).
\end{align*}
(Note that for any two nodes $j$ and $i$ that are adjacent in the causal graph $G_{\mathsf{cau}}$, the shortest path between the two nodes in the contact network $G_{\mathsf{ctt}}$ has a length less than or equal to $t_1$.) Furthermore, $N_{\mathsf{ctt}}^\partial (i;s)$'s are djsjoint across $s$'s. Hence
\begin{align*}
    & P\left\{ |N_{\mathsf{cau}}(i)| \ge 5.7 t_1^2(\pi_{n,c} \vee 1)^{t_1} \log n \right\} \\ \notag
	&\le P\left\{|N_{\mathsf{ctt}}^\partial (i;s)| \ge 5.7 t_1^2(\pi_{n,c} \vee 1)^{t_1} \log n, \text{ for some } s = 1,...,t_1 \right\} \le t_1 n^{-3.3},
\end{align*}
by (\ref{bound322}). Therefore,
\begin{align*}
	& P\left\{ \max_{i \in N} |N_{\mathsf{cau}}(i)| \ge 5.7 t_1^2(\pi_{n,c} \vee 1)^{t_1} \log n \right\}\\ \notag
	&\le \sum_{i \in N} P\left\{ |N_{\mathsf{cau}}(i)| \ge 5.7 t_1^2(\pi_{n,c} \vee 1)^{t_1} \log n \right\} \le t_1 n^{-2.3}.
\end{align*}
Using (\ref{network formation}), we find that $\max_{i \in N} |N_{\mathsf{cau}}(i)| = O_P\left( (\log n)^{k t_1 + 1}\right)$. Using a similar argument, we find that $\max_{i \in N} |N_{\mathsf{obs}}(i)| = O_P\left( (\log n)^{k t_1 + 1}\right)$.
Since $d_{mx} \le \max_{i \in N} |N_{\mathsf{cau}}(i)| + \max_{i \in N} |N_{\mathsf{obs}}(i)|$, we obtain the desired result. $\blacksquare$

\subsection{Asymptotic Normality}

Define
\begin{align*}
	\hat e_{i,1} = Y_{i,1}-\hat c_i \text{ and }
	\hat{\varepsilon}_{j,0} = Y_{j,0}-\hat \mu_{j,0}.
\end{align*}
Then we can write
\begin{align*}
	\mathsf{\hat C} = \frac{1}{n} \sum_{i \in N} \hat e_{i,1}\hat{a}_i,
\end{align*}
where we recall $\hat{a}_i = \sum_{j\in N: ij \in E_{\mathsf{obs}}} n w_j \hat \varepsilon_{j,0} / \hat \sigma_{j,0}^2$ 
and $\hat \sigma_{j,0}^2 = \hat \mu_{j,0}(1 - \hat \mu_{j,0})$.

\begin{lemma}
	\label{lemm: asymp lin}
	\begin{align*}
		\sqrt{n}(\mathsf{\hat C} - \mathsf{C}) =  \frac{1}{\sqrt{n}}\sum_{i \in N} (q_i - \mathbf{E}[q_i\mid \mathcal{F}]) + o_{P}(1).
	\end{align*}
\end{lemma}

\noindent \textbf{Proof: } First, we show that
\begin{align}
	\label{uniform rate hat a_i}
	\max_{i \in N} | \hat a_i - a_i | = O_P\left(n^{-\kappa} (\log n)^C \right),
\end{align}
where $\kappa$ is the constant in Assumption \ref{assump: rate est funs}(ii) and $C$ is the constant in Assumption \ref{assump: network formation}. For $i \in N$, we write
\begin{align*}
	\hat a_i - a_i &= \sum_{j \in N: ij \in E_{\mathsf{obs}}} nw_j\left(A_{1j} + A_{2j} + A_{3j}\right),
\end{align*}
where
\begin{align*}
	A_{1j} = \frac{\mu_{j,0} - \hat \mu_{j,0}}{\sigma_{j,0}^2}, 	\quad A_{2j} = (\mu_{j,0} - \hat \mu_{j,0})\left(\frac{1}{\hat \sigma_{j,0}^2} - \frac{1}{\sigma_{j,0}^2}\right), \text{ and } A_{3j} = \varepsilon_{j,0}\left(\frac{1}{\hat \sigma_{j,0}^2} - \frac{1}{\sigma_{j,0}^2}\right).
\end{align*}
By Assumptions \ref{assump: unconfounded}(ii) and \ref{assump: rate est funs}(ii), we have
\begin{align*}
	\max_{j \in N} |A_{1j}| = O_P(n^{-\kappa}).
\end{align*}
As for $A_{2j}$, again, using Assumptions \ref{assump: unconfounded}(ii) and \ref{assump: rate est funs}(ii), we bound
\begin{align*}
	\max_{j \in N} |A_{2j}| &\le  \max_{j \in N} \left|(\mu_{j,0} - \hat \mu_{j,0}) \left( \frac{\hat \mu_{j,0}(1 - \hat \mu_{j,0}) - \mu_{j,0}(1 - \mu_{j,0})}{ c^4 + o_P(1)} \right)\right|\\
	&\le \max_{j \in N} \frac{C(\mu_{j,0} - \hat \mu_{j,0})^2}{ c^4 + o_P(1)} = O_P\left(n^{-2\kappa}\right).
\end{align*}
Using the same argument, we also obtain that $\max_{j \in N} |A_{3j}| = O_P\left(n^{-\kappa}\right)$. Hence,
\begin{align*}
	\max_{i \in N} | \hat a_i - a_i | = O_P\left(n^{-\kappa} d_{mx} \right) = O_P\left(n^{-\kappa} (\log n)^C \right),
\end{align*}
by Assumption \ref{assump: network formation}, obtaining (\ref{uniform rate hat a_i}).

By Assumption \ref{assump: rate est funs}(ii),
\begin{align*}
	\frac{1}{\sqrt{n}}\sum_{i \in N} (\hat c_i - c_i) (\hat a_i - a_i) &\le \sqrt{n} \max_{i \in N} |\hat c_i - c_i| \max_{i \in N}|\hat a_i - a_i| 
	=O_P\left(n^{-2\kappa} (\log n)^C\right) = o_P(1).
\end{align*}
Using this result and Assumption \ref{assump: asymp approx}, we can write
	\begin{align*}
		\sqrt{n}\{\mathsf{\hat C} - \mathsf{C} \}
		&=  \frac{1}{\sqrt{n}}\sum_{i \in N} \left(\hat e_{i,1} \hat a_i -\mathbf{E}[e_{i,1} a_i \mid \mathcal{F}] \right)= \frac{1}{\sqrt{n}}\sum_{i \in N} \left((Y_{i,1} - \hat c_i) \hat a_i -\mathbf{E}[e_{i,1} a_i \mid \mathcal{F}] \right)\\ \notag
		&=\frac{1}{\sqrt{n}}\sum_{i \in N} (Y_{i,1} - c_i) (\hat a_i - a_i) - \frac{1}{\sqrt{n}}\sum_{i \in N} (\hat c_i - c_i) a_i \\
		&\qquad + \frac{1}{\sqrt{n}}\sum_{i \in N}  \left(e_{i,1} a_i -\mathbf{E}[e_{i,1} a_i \mid \mathcal{F}] \right) + o_P(1)\\
		&= \frac{1}{\sqrt{n}}\sum_{i \in N}\left(q_i - \mathbf{E}[q_i\mid \mathcal{F}]\right) + o_P(1).
	\end{align*}
$\blacksquare$\medskip

\begin{lemma}
	\label{lemm: max q}
	$\max_{i \in N} \mathbf{E}\left[q_i^4\mid \mathcal{F}\right] = O_P\left(d_{mx}^4 \right) \text{ and } \max_{i \in N} h_i^4  = O_P\left(d_{mx}^4 \right)$.
\end{lemma}	

\noindent \textbf{Proof: } For the first statement, note that
\begin{align}
	\label{bd78b}
	\quad \max_{i \in N} \mathbf{E}\left[q_i^4\mid G_{\mathsf{obs}},X\right] &\le 8 \max_{i \in N} \mathbf{E}\left[e_{i,1}^4a_i^4\mid G_{\mathsf{obs}},X\right] \\ \notag
	& \quad + 8 \max_{i \in N} \mathbf{E}\left[\psi_i^4\mid G_{\mathsf{obs}},X\right] = O_P\left(d_{mx}^4\right),
\end{align}
because $a_i = \sum_{j \in N_{\mathsf{obs}}(i)} n w_j \varepsilon_{j,0}/\sigma_{j,0}^2$, and due to (\ref{cond psi}).

For the second statement, let
\begin{align*}
	M = \frac{1}{n}\sum_{i \in N} X_i X_i' \text{ and } b = \frac{1}{n}\sum_{i \in N} X_i \mathbf{E}\left[q_i \mid \mathcal{F} \right].
\end{align*}
Then, by Cauchy-Schwarz inequality,
\begin{align*}
	h_i^2 &= ((M^{-1} b)' X_i)^2 \le (M^{-1} b)'(M^{-1} b) X_i' X_i \\
	&= \text{tr}\left( M^{-2} b b' \right) \|X_i\|^2 \le \frac{1}{c'^2} \max_{i \in N} \|X_i\|^4 \max_{i \in N} \mathbf{E}\left[q_i^2\mid \mathcal{F}\right] = O(d_{mx}^2),
\end{align*}
by (\ref{bd78b}) and Assumption \ref{assump: non deg}(ii). $\blacksquare$\medskip

Define a graph $G^* = (N,E^*)$, where for $i_1 \ne i_2$, $i_1i_2 \in E^*$ if and only if $\overline N(i_1) \cap \overline N(i_2) \ne \varnothing$, and let $\overline E^* = E^* \cup \{ii: i \in N\}$. Then $G^* = (N,E^*)$ contains $G = (N,E)$ as a subgraph. Let $d_{mx}^*$ and $d_{av}^*$ be the maximum degree and the average degree of $G^*$, that is,
\begin{align*}
	d_{mx}^* = \max_{i \in N} |N^*(i)| \text{ and } d_{av}^* = \frac{1}{n} \sum_{i \in N} |N^*(i)|.
\end{align*}
By definition, we have
\begin{align}
	\label{dmx star}
	d_{mx}^* = O(d_{mx}^2), \text{ and } d_{av}^* = O(d_{mx}d_{av}),
\end{align}
where $d_{mx}$ and $d_{av}$ are defined in (\ref{max deg and ave deg}).
We also let $E_{\mathsf{obs}}^*$ be the collection of $i_1i_2$ such that $\overline N_{\mathsf{obs}}(i_1) \cap \overline N_{\mathsf{obs}}(i_2) \ne \varnothing$, and similarly, let $E_{\mathsf{cau}}^*$ be the collection of $i_1i_2$ such that $\overline N_{\mathsf{cau}}(i_1) \cap \overline N_{\mathsf{cau}}(i_2) \ne \varnothing$. We let $\overline E_{\mathsf{obs}}^* = E_{\mathsf{obs}}^* \cup \{ii: i \in N\}$ and $\overline E_{\mathsf{cau}}^* = E_{\mathsf{cau}}^* \cup \{ii: i \in N\}$. Then we define the graphs $G_{\mathsf{obs}}^*$ and $G_{\mathsf{cau}}^*$ as follows:
\begin{align}
	\label{graphs G^*}
	G_{\mathsf{obs}}^* = (N,E_{\mathsf{obs}}^*) \text{ and } G_{\mathsf{cau}}^* = (N,E_{\mathsf{cau}}^*).
\end{align}

Recall the definition:
\begin{align*}
\sigma^2 = \mathbf{E}\left[ \left(\frac{1}{\sqrt{n}}\sum_{i \in N} (q_i - \mathbf{E}[q_i\mid \mathcal{F}]) \right)^2 \mid \mathcal{F} \right].
\end{align*}

\begin{lemma}\label{lemm: asym norm}
	\begin{align*}
	\sup_{u \in \mathbf{R}}\left\vert P\left\{ \frac{\sqrt{n}(\mathsf{\hat C} - \mathsf{C})}{\sigma} \le u \mid \mathcal{F}\right\} -\Phi (u)\right\vert \rightarrow 0\text{\textit{, as }} n \rightarrow \infty,
	\end{align*}
	where $\Phi$ is the distribution function of $N(0,1).$
\end{lemma}

\noindent \textbf{Proof: } For each $i \in N$, $q_i$ is a function of $(Y_{i,1},c_i,a_i,\psi_i)$ which is measurable with respect to the $\sigma$-field generated by $Y_{i,1}$, $(Y_{j,0})_{j \in \overline N_{\mathsf{cau}}(i) \cup \overline N_{\mathsf{obs}}(i)}$, and $\mathcal{F}$, by (\ref{actual outcome2}), and Assumption \ref{assump: asymp approx}. By Assumption \ref{assump: DCG}, $\{q_i\}_{i \in N}$ has $G^*$ as a conditional dependency graph given $\mathcal{F}$, which is a special case of conditional neighborhood dependency introduced in \cite{Lee/Song:19:BJ}. We apply their Corollary 3.1 to deduce that
\begin{align*}
	& \sup_{u \in \mathbf{R}}\left\vert P\left\{ \frac{1}{\sigma \sqrt{n}}\sum_{i \in N} (q_i - \mathbf{E}[q_i\mid \mathcal{F}]) \leq u \mid \mathcal{F}\right\} -\Phi (u)\right\vert \\
	&\le
	C \left( \frac{\sqrt{d_{mx}^*d_{av}^*\mu_3^3}}{n^{1/4}} - \log\left(\frac{d_{mx}^* d_{av}^* \mu_3^3}{\sqrt{n}}\right) \frac{\sqrt{\left(d_{mx}^* \right)^2 d_{av}^* \mu_4^4}}{\sqrt{n}} \right),
\end{align*}
for some constant $C>0$ that does not depend on $n$, where
\begin{align*}
	\mu_p^p = \max_{i \in N} \mathbf{E}\left[\left|\frac{q_i - \mathbf{E}[q_i\mid \mathcal{F}]}{\sigma} \right|^p \mid \mathcal{F} \right].
\end{align*}
 Thus, the desired result follows from this and Assumption \ref{assump: network formation} and Lemmas \ref{lemm: asymp lin} and \ref{lemm: max q}. $\blacksquare$\medskip

 Recall the definitions of $\tilde \sigma^2$ and $\tilde \sigma_\mathsf{obs}^2$ which we write as
 \begin{align*}
 	\tilde \sigma^2 &= \frac{1}{n}\sum_{i_1 i_2 \in \overline E^*} \mathbf{E}\left[ (q_{i_1} - h_{i_1})(q_{i_2} - h_{i_2})\mid \mathcal{F} \right] \text{ and }\\
 	\tilde \sigma_\mathsf{obs}^2 &= \frac{1}{n}\sum_{i_1 i_2 \in \overline E_{\mathsf{obs}}^*} \mathbf{E}\left[ (q_{i_1} - h_{i_1})(q_{i_2} - h_{i_2})\mid \mathcal{F} \right],
 \end{align*}
 where $h_i$ is as defined in (\ref{h i tau}). Note that when $G_{\mathsf{obs}}$ is a DCG, we have $\tilde \sigma^2 = \tilde \sigma_{\mathsf{obs}}^2$.

\begin{lemma}
	\label{lemm: positive semidefinte}
	$\tilde \sigma^2 \ge \sigma^2$.
\end{lemma}

\noindent \textbf{Proof: } Let $A$ be the $n \times n$ matrix whose $(i,j)$-th entry is $1/n$ if $ij \in \overline E^*$ and zero otherwise. Furthermore, $A$ is positive semidefinite.\footnote{As shown in \cite{Kojevnikov:21:WP}, this follows because $c' A c  = \frac{1}{n}\sum_{j \in N} b_j^2$ with $b_j = \sum_{i \in \overline N(j)} c_i$ for any vector $c = [c_1,...,c_n]'$.} Define
\begin{align*}
	q =[q_1,...,q_n]' \text{ and } h = [h_1,...,h_n]',
\end{align*}
where $h_i$'s are as defined in (\ref{h i tau}). Then, note that
\begin{align*}
	\sigma^2 &= \mathbf{E}\left[(q - \mathbf{E}[q\mid \mathcal{F}])' A (q - \mathbf{E}[q\mid \mathcal{F}]) \mid \mathcal{F} \right]\\
	&= \mathbf{E}\left[(q - h)' A (q - h) \mid \mathcal{F} \right]
	- (\mathbf{E}[q\mid \mathcal{F}] - h)' A (\mathbf{E}[q\mid \mathcal{F}] - h)\\
	& = \tilde \sigma^2 - (\mathbf{E}[q\mid \mathcal{F}] - h)' A (\mathbf{E}[q\mid \mathcal{F}] - h),
\end{align*}
because $A$ and $h$ are measurable with respect to $\mathcal{F}$. The last term is nonnegative because $A$ is positive semidefinite. $\blacksquare$

\begin{lemma}
	\label{lemm: asym bound}
	For each $c >0$,
	\begin{align*}
		P\left\{ \left|\frac{\sqrt{n}(\mathsf{\hat C} - \mathsf{C})}{\tilde \sigma} \right| \le c\right\}
		\ge P\left\{ \left|\mathbb{Z} \right| \le c \right\} + o(1),
	\end{align*}
	where $\mathbb{Z} \in \mathbf{R}$ is a random variable distributed as $N(0,1)$.
\end{lemma}

\noindent \textbf{Proof: } First, by Lemma \ref{lemm: asym norm}, we find that for any Borel set $B$,
\begin{align*}
	P\left\{ \frac{\sqrt{n}(\mathsf{\hat C} - \mathsf{C})}{\tilde \sigma} \in B \right\} = P\left\{ \frac{\sigma \mathbb{Z}}{\tilde \sigma} \in B \right\} + o(1).
\end{align*}
Note that
\begin{align*}
	P\left\{ \left|\frac{\sqrt{n}(\mathsf{\hat C} - \mathsf{C})}{\tilde \sigma} \right| \le c \right\} &= P\left\{ \left|\frac{\sqrt{n} (\mathsf{\hat C} - \mathsf{C})}{\sigma} \right| \le \frac{\tilde \sigma c}{\sigma} \right\} \\
	&= P\left\{ \left| \mathbb{Z} \right| \le \frac{\tilde \sigma c}{\sigma} \right\} + o(1) \ge P\left\{ \left| \mathbb{Z} \right| \le c \right\} + o(1),
\end{align*}
by Lemma \ref{lemm: positive semidefinte}. $\blacksquare$

\subsection{Consistency of Variance Estimators}

\begin{lemma}
	\label{lemm: conv rate D hat}
	$\mathsf{\hat C} = \mathsf{C} + o_P(1).$
\end{lemma}

\noindent \textbf{Proof: } The statement follows from Lemma \ref{P4}, Assumption \ref{assump: network formation}, and Lemma \ref{lemm: asymp lin}. $\blacksquare$\medskip

\begin{lemma}
	\label{lemm: unif rate for hat qi}
	$\max_{i \in N} |\hat q_i - q_i | = O_P\left(n^{-\kappa}(\log n)^C\right)$.
\end{lemma}

\noindent \textbf{Proof: } By Assumption \ref{assump: rate est funs}, it suffices to show that
\begin{align*}
	\max_{i \in N}  |\hat e_{i,1} \hat a_i - e_{i,1} a_i | = O_P\left(n^{-\kappa} (\log n)^C \right).
\end{align*}
Since $c_i$ takes values from $[0,1]$,
\begin{align*}
   \max_{i \in N} 	|\hat e_{i,1} \hat a_i - e_{i,1} a_i| &\le \max_{i \in N}  |\hat e_{i,1} (\hat a_i - a_i)| + \max_{i \in N}  |(\hat e_{i,1} - e_{i,1}) a_i |\\
   &\le \max_{i \in N}  |\hat c_i - c_i| |\hat a_i - a_i| + 2 \max_{i \in N}  |\hat a_i - a_i| + \max_{i \in N}  |\hat c_i - c_i||a_i| = O_P\left(n^{-\kappa} (\log n)^C\right),
\end{align*}
from Assumptions \ref{assump: rate est funs} and \ref{assump: network formation}, and (\ref{uniform rate hat a_i}). $\blacksquare$

\begin{lemma}
	\label{lemm: Lemma B14}
	\begin{align*}
		\frac{d_{mx}^6}{n} \sum_{i \in N}(\hat q_i - q_i)^2 = o_P(1) \text{ and }
		\frac{d_{mx}^6}{n} \sum_{i \in N}(\hat h_i - h_i)^2 = o_P(1).
	\end{align*}
\end{lemma}

\noindent \textbf{Proof: } The first result comes from Lemma \ref{lemm: unif rate for hat qi} and Assumption \ref{assump: network formation}. As for the second result, $\hat h = [\hat h_1,...,\hat h_n]'$ is the projection of $\hat q = [\hat q_1,...,\hat q_n]'$ onto the range space of $X = [X_1,....,X_n]'$. Let $P_X$ be the projection matrix projecting from $\mathbf{R}^n$ onto the range space of $X$. Define $\tilde h = [\tilde h_1,...,\tilde h_n]'$, where $\tilde h_i = \tilde \lambda ' X_i$, with
\begin{align*}
	\tilde \lambda = \left( \frac{1}{n}\sum_{i \in N} X_i X_i' \right)^{-1} \frac{1}{n}\sum_{i \in N} X_i q_i.
\end{align*}
Then
\begin{align*}
	\|\hat h - \tilde h \|^2 \le \text{tr}\left( P_X (\hat q - q) (\hat q - q)' P_X \right) = \text{tr}\left( P_X (\hat q - q) (\hat q - q)' \right) \le \|\hat q - q\|^2,
\end{align*}
because the maximum eigenvalue of $P_X$ is at most 1. Hence
\begin{align}
	\label{state1}
	\frac{d_{mx}^6}{n}\sum_{i \in N} (\hat h_i - \tilde h_i)^2 \le \frac{d_{mx}^6}{n}\sum_{i \in N} (\hat q_i - q_i)^2 = o_P(1).
\end{align}
Now, consider
\begin{align*}
		\frac{1}{n} \|\tilde h - h \|^2 &= \left(\frac{1}{n} \sum_{i \in N} (q_i - \mathbf{E}[q_i \mid \mathcal{F}]) X_i'\right) \left(\frac{1}{n} \sum_{i \in N} X_i X_i' \right)^{-1} \left(\frac{1}{n} \sum_{i \in N} (q_i - \mathbf{E}[q_i \mid \mathcal{F}]) X_i\right).
\end{align*}
 Let $X_{ik}$ be the $k$-th entry of $X_i$. Then,
\begin{align*}
	\mathbf{E}\left[\left(\frac{1}{n} \sum_{i \in N} (q_i - \mathbf{E}[q_i \mid \mathcal{F}]) X_{ik} \right)^2 \mid \mathcal{F} \right] &= \frac{1}{n^2}\sum_{i_1 i_2 \in \overline E^*} \text{Cov}\left(q_{i_1} X_{i_1 k}, q_{i_2} X_{i_2 k} \mid \mathcal{F} \right)\\
	&= O_P\left( \frac{d_{av}^*}{n}\right) = O_P\left( \frac{d_{av} d_{mx}}{n}\right),
\end{align*}
where the second to the last equality uses the assumption that $\max_{i \in N} \| X_i \|^2 <C$ in Assumption \ref{assump: non deg} (ii) and the last equality uses (\ref{dmx star}). Hence we have
\begin{align*}
	\frac{d_{mx}^6}{n} \sum_{i \in N}(\tilde h_i - h_i)^2 = O_P\left( \frac{d_{mx}^7d_{av}}{n}\right) =o_P(1),
\end{align*}
by Assumption \ref{assump: network formation}. Combining this with (\ref{state1}), we obtain the second result of this lemma. $\blacksquare$\medskip

Define $\eta_i = q_i - h_i$, $\hat \eta_i = \hat q_i - \hat h_i$, and
\begin{align*}
	V = \frac{1}{n} \sum_{i_1i_2 \in \overline E_{\mathsf{obs}}^*} \left(\hat{\eta}_{i_1}\hat{\eta}_{i_2} - \eta_{i_1} \eta_{i_2}\right),
\end{align*}
where we recall $\overline E_{\mathsf{obs}}^* = E_{\mathsf{obs}}^* \cup \{ii: i \in N\}$.

\begin{lemma}\label{lemm: V}
	$V =  o_P(1).$
\end{lemma}
\noindent \textbf{Proof: } We write
\begin{align*}
	V = \frac{1}{n} \sum_{i_1i_2 \in \overline E_{\mathsf{obs}}^*} \left(\hat{\eta}_{i_1} - \eta_{i_1}\right)\hat \eta_{i_2} +\frac{1}{n} \sum_{i_1i_2 \in \overline E_{\mathsf{obs}}^*}  \left(\hat{\eta}_{i_2} - \eta_{i_2}\right)\eta_{i_1} = A_{1n} + A_{2n}, \text{ say}.
\end{align*}
We consider first $A_{1n}$. We write it as $A_{1n}' + A_{1n}''$, where
\begin{align*}
	A_{1n}' &= \frac{1}{n} \sum_{i_1i_2 \in \overline E_{\mathsf{obs}}^*} \left(\hat{\eta}_{i_1} - \eta_{i_1}\right) \eta_{i_2}, \text{ and }\\
	A_{1n}'' &= \frac{1}{n} \sum_{i_1i_2 \in \overline E_{\mathsf{obs}}^*} \left(\hat{\eta}_{i_1} - \eta_{i_1}\right)(\hat \eta_{i_2} - \eta_{i_2}).
\end{align*}
Write $A_{1n}'$ as
\begin{align*}
	\frac{1}{n} \sum_{i_1i_2 \in \overline E_{\mathsf{obs}}^*} (\hat{\eta}_{i_1} - \eta_{i_1})\eta_{i_2}
	&=\frac{1}{n} \sum_{i_1i_2 \in \overline E_{\mathsf{obs}}^*} (\hat q_{i_1} - q_{i_1})(q_{i_2} - h_{i_2})
	\\
	&\quad - \frac{1}{n} \sum_{i_1i_2 \in \overline E_{\mathsf{obs}}^*} (\hat h_{i_1} - h_{i_1})(q_{i_2} - h_{i_2}) = A_{11n}' - A_{12n}', \text{ say.}
\end{align*}
We bound $A_{11n}'$ by
\begin{align*}
	\sqrt{\frac{1}{n} \sum_{i_1 \in N} |\overline N_\mathsf{obs}^*(i_1)| \left(\hat q_{i_1} - q_{i_1}\right)^2}\sqrt{\frac{1}{n} \sum_{i_1 \in N} \sum_{i_2 \in \overline N_\mathsf{obs}^*(i_1)}(q_{i_2} - h_{i_2})^2}.
\end{align*}
Defining $P_X$, $q$ and $h$ as in the proof of Lemma \ref{lemm: Lemma B14},
\begin{align*}
  \sum_{i \in N} \mathbf{E}\left[ h_i^2\mid G_{\mathsf{obs}},X \right] &= \mathbf{E}\left[\|h \|^2 \mid G_{\mathsf{obs}},X \right] \le \mathbf{E}\left[\| P_X \mathbf{E}\left[q \mid \mathcal{F}\right] \|^2 \mid G_{\mathsf{obs}},X \right] \\ \notag
  &\le \mathbf{E}\left[\mathbf{E}\left[ \| P_X q  \|^2 \mid \mathcal{F} \right] \mid G_{\mathsf{obs}},X \right] \le \sum_{i \in N} \mathbf{E}\left[ q_i^2\mid G_{\mathsf{obs}},X \right].
\end{align*}
Hence we have
\begin{align*}
\frac{1}{n} \sum_{i_1 \in N} \sum_{i_2 \in \overline N_\mathsf{obs}^*(i_1)}(q_{i_2} - h_{i_2})^2 &\le \frac{2}{n} \sum_{i_1 \in N} \sum_{i_2 \in \overline N_\mathsf{obs}^*(i_1)}q_{i_2}^2 + \frac{2}{n} \sum_{i_1 \in N}  \sum_{i_2 \in \overline N_\mathsf{obs}^*(i_1)} h_{i_2}^2 \\
&\le \frac{2 d_{mx}^*}{n} \left(\| q\|^2 + \| h \|^2 \right) \le \frac{4 d_{mx}^* \| q\|^2 }{n}  = O_P(d_{mx}^4),
\end{align*}
because $\|h \|^2 \le \|q\|^2$, and by (\ref{bd78b}). By Lemma \ref{lemm: Lemma B14}, $A_{11n}' = o_P(1)$. Using the same argument and Lemma \ref{lemm: Lemma B14}, we also find that $A_{12n}' = o_P(1)$.

Let us turn to $A_{1n}''$ which is bounded by
\begin{align*}
	& \sqrt{\frac{1}{n} \sum_{i_1 \in N}|\overline N_\mathsf{obs}^*(i_1)| \left(\hat{\eta}_{i_1} - \eta_{i_1}\right)^2}\sqrt{\frac{1}{n} \sum_{i_1 \in N}\sum_{i_2 \in \overline N_\mathsf{obs}^*(i_1)} \left(\hat \eta_{i_2} -  \eta_{i_2}\right)^2}\\
	&\le \frac{d_{mx}^*+1}{n} \sum_{i \in N} \left(\hat{\eta}_i - \eta_i \right)^2.
\end{align*}
The last term is bounded by
\begin{align*}
	\frac{2 (d_{mx}^*+1)}{n} \sum_{i \in N} \left(\hat q_i - q_i\right)^2
	+   \frac{2 (d_{mx}^*+1)}{n} \sum_{i \in N} \left(\hat h_i - h_i\right)^2 =  o_P(1),
\end{align*}
by Lemma \ref{lemm: Lemma B14}. Finally, note that $A_{2n}$ is the same as $A_{1n}'$. Thus the proof is complete. $\blacksquare$\medskip

Let
\begin{align*}
W= \frac{1}{n}\sum_{i_1i_2 \in \overline E_{\mathsf{obs}}^*}\left(\eta_{i_1}\eta_{i_2} - \mathbf{E}[\eta_{i_1}\eta_{i_2}\mid \mathcal{F}]\right).
\end{align*}

\begin{lemma}\label{lemm: W}
	$W= o_P(1).$
\end{lemma}

\noindent \textbf{Proof: } First, observe that by Lemma \ref{lemm: max q},
\begin{align}
\label{bd31}
\max_{i_1,i_2 \in N} \mathbf{E}\left[\eta_{i_1}^2\eta_{i_2}^2\mid \mathcal{F} \right] = O_P(d_{mx}^4).
\end{align}
Write
\begin{align*}
	\mathbf{E}\left[W^2\mid \mathcal{F} \right] = \frac{1}{n^2} \sum_{i_1i_2 \in \overline E_{\mathsf{obs}}^*}\sum_{i_1'i_2' \in \overline E_{\mathsf{obs}}^*} \text{Cov}\left(\eta_{i_1}\eta_{i_2},\eta_{i_1'} \eta_{i_2'} \mid \mathcal{F} \right).
\end{align*}
Since $G_{\mathsf{cau}}$ is a DCG, for any $i_1i_2, i_1'i_2' \in \overline E_{\mathsf{obs}}^*$, $\text{Cov}\left(\eta_{i_1}\eta_{i_2},\eta_{i_1'} \eta_{i_2'} \mid \mathcal{F} \right)$ is not zero only if $\{i_1,i_2\}$ and $\{i_1',i_2'\}$ are adjacent in $G^*$. Thus the number of the terms in the double sum above is of the same order as the number of ways one chooses paths $i_1 k_1 i_2 k_2 i_1' k_3 i_2'$ in $G$, which is $O(nd_{av} d_{mx}^5)$. With (\ref{bd31}), this implies that by Assumption \ref{assump: network formation},
\begin{align*}
	\mathbf{E}\left[W^2\mid \mathcal{F} \right] = O\left(\frac{d_{av} d_{mx}^9}{n}\right) = o_P(1).
\end{align*}
$\blacksquare$
\medskip

\begin{lemma}\label{B18}
	$\hat \sigma^2 = \tilde \sigma_\mathsf{obs}^2+ o_P(1).$
\end{lemma}

\noindent \textbf{Proof: } Note that $\hat \sigma^2 - \tilde \sigma_\mathsf{obs}^2 = V + W= o_P(1)$,
by Lemmas \ref{lemm: V} and \ref{lemm: W}. $\blacksquare$\medskip

\subsection{Proofs of Theorems \ref{thm: asym val} and \ref{thm: asym val2}}

\noindent \textbf{Proof of Theorem \ref{thm: asym val}: } Since $G_{\mathsf{obs}}$ is a DCG, we have $\mathsf{ADM} = \mathsf{C}$ and $\tilde \sigma^2 = \tilde \sigma_\mathsf{obs}^2$. Hence
\begin{align*}
  \hat \sigma^2 = \tilde \sigma_\mathsf{obs}^2 + o_P(1) \ge \sigma^2 + o_P(1),
\end{align*}
by Lemmas \ref{lemm: positive semidefinte} and \ref{B18}. We find from Lemma \ref{lemm: asym bound} that
\begin{align*}
		P\left\{ \mathsf{C} \in\mathbb{C}_{1-\alpha}\right\} \ge P\left\{z_{\alpha/2} \le \mathbb{Z} \le z_{1-\alpha/2}\right\} + o(1) = 1 - \alpha + o(1).
\end{align*}
$\blacksquare$\medskip

We define for any $\delta > 0$,
\begin{align*}
	\overline N_{\mathsf{ctt}}(i;\delta) = \left\{j \in N: \text{Cov}(Y_{i,1},Y_{j,0}\mid \mathcal{F}) \ge \delta \right\} \cap \overline N_{\mathsf{ctt}}(i),
\end{align*}
and
\begin{align*}
	d_{\mathsf{ctt}}^{\Delta}(\delta) &= \frac{1}{n}\sum_{i \in N} |\overline N_{\mathsf{ctt}}(i;\delta) \setminus \overline N_{\mathsf{obs}}(i)|, \text{ and }\\
	d_{\mathsf{cau}}^{\Delta} &= \frac{1}{n}\sum_{i \in N} |\overline N_{\mathsf{cau}}(i) \setminus \overline N_{\mathsf{obs}}(i)|.
\end{align*}
We let
\begin{align*}
	e_n(\delta) = \left\{ \begin{array}{ll}
		d_{\mathsf{ctt}}^{\Delta}(\delta) / d_{\mathsf{cau}}^{\Delta}, &\text{ if } d_{\mathsf{cau}}^{\Delta}>0,\\
		\infty, & \text{ if } d_{\mathsf{cau}}^{\Delta} = 0.
	\end{array}
	\right.
\end{align*}
The quantity $e_n(\delta)$ decreases in $\delta$. If the observed graph $G_{\mathsf{obs}}$ contains the causal graph $G_{\mathsf{cau}}$ as a subgraph (so that $G_{\mathsf{obs}}$ is a DCG), then $e_n(\delta) = \infty$. The following assumption requires that $e_n(\delta_n)$ does not shrink to zero too quickly as $\delta_n$ decreases to zero.

\begin{assumption}[Nontrivially Effective Dependency Causal Graph]
	\label{assump: bound}
	There exists a positive sequence $\delta_n$ such that for any constant $C>0$,
	\begin{align}
		\label{bound4}
		P\left\{\left(d_{mx}^\mathsf{cau} d_{mx}\right)^2 \le C \sqrt{n} e_n(\delta_n) \delta_n \right\} \rightarrow 1,\text{ as } n \rightarrow \infty,
	\end{align}
	where $d_{mx}^\mathsf{cau}$ denotes the maximum degree of $G_{\mathsf{cau}}$.
\end{assumption}
Suppose that Assumption \ref{assump: non-DCG} holds, and we take $\delta_n$ to be such that $\delta_n \le c_1 n^{-1/2 + \nu}$, where $c_1>0$ is the constant that appears in Assumption \ref{assump: non-DCG}. Then,
\begin{align*}
	e_n(\delta_n) \ge d_{mx}^{-(t_1-1)}.
\end{align*}
because, for all $i \in N$,
\begin{align*}
    |\overline N_{\mathsf{cau}}(i) \setminus \overline N_{\mathsf{obs}}(i)| \le |\overline N_{\mathsf{ctt}}(i) \setminus \overline N_{\mathsf{obs}}(i)| \left(d_{mx}^{\mathsf{ctt}} \right)^{t_1-1}.
\end{align*}
Hence, we have (\ref{bound4}) by Assumption \ref{assump: network formation}. Therefore, Assumption \ref{assump: bound} is weaker than Assumption \ref{assump: non-DCG}.

\begin{lemma}
	\label{lemm: bound}
	Suppose that the conditions of Lemma \ref{lemm: identification}(ii) and Assumption \ref{assump: bound} hold. Then, 
	\begin{align}
		\label{prob bound}
		P\left\{z_{1-\alpha}(\tilde \sigma - \tilde \sigma_\mathsf{obs}) \le \sqrt{n}(\mathsf{ADM} - \mathsf{C} )\right\} \rightarrow 1,
	\end{align}
    as $n \rightarrow \infty$.
\end{lemma}
\medskip

\noindent \textbf{Proof: } Let
\begin{align*}
	\overline N_{\mathsf{cau}}^*(i) &= \left\{ i' \in N: \overline N_{\mathsf{cau}}(i) \cap \overline N_{\mathsf{cau}}(i') \ne \varnothing \right\}, \text{ and }\\
	 \overline N_{\mathsf{obs}}^*(i) &= \left\{ i' \in N: \overline N_{\mathsf{obs}}(i) \cap \overline N_{\mathsf{obs}}(i') \ne \varnothing \right\}.
\end{align*}
For any sets $A, A',B,B'$, we have
\begin{align*}
	(A \cap A') \setminus (B \cap B') &= ((A \cap A') \setminus B) \cup ((A \cap A') \setminus B')\\
	&= (A' \cap (A \setminus B)) \cup (A \cap (A' \setminus B')).
\end{align*}
Hence
\begin{align}
	\label{bound2}
	\frac{1}{n}\sum_{i \in N} |\overline N_{\mathsf{cau}}^*(i) \setminus \overline N_{\mathsf{obs}}^*(i)| 
	&\le \frac{1}{n}\sum_{i \in N} \sum_{i' \in N} \left| (\overline N_{\mathsf{cau}}(i) \cap \overline N_{\mathsf{cau}}(i')) \setminus (\overline N_{\mathsf{obs}}(i) \cap \overline N_{\mathsf{obs}}(i'))\right|\\ \notag
	& \le \frac{1}{n}\sum_{i \in N} \sum_{i' \in N} \left|\overline N_{\mathsf{cau}}(i') \cap (\overline N_{\mathsf{cau}}(i) \setminus \overline N_{\mathsf{obs}}(i)) \right|\\ \notag
	& \qquad + \frac{1}{n}\sum_{i \in N} \sum_{i' \in N} \left|\overline N_{\mathsf{cau}}(i) \cap (\overline N_{\mathsf{cau}}(i') \setminus \overline N_{\mathsf{obs}}(i')) \right|.
\end{align}
The last two double summations are identical, and hence
\begin{align*}
	 \frac{1}{n}\sum_{i \in N} |\overline N_{\mathsf{cau}}^*(i) \setminus \overline N_{\mathsf{obs}}^*(i)|  \le \frac{2}{n}\sum_{i \in N} \sum_{i' \in N} \left|\overline N_{\mathsf{cau}}(i') \cap (\overline N_{\mathsf{cau}}(i) \setminus \overline N_{\mathsf{obs}}(i)) \right|.
\end{align*}
As for the last term, we write it as
\begin{align*}
	&\frac{2}{n}\sum_{i \in N} \sum_{i': \overline N_{\mathsf{cau}}(i') \cap (\overline N_{\mathsf{cau}}(i) \setminus \overline N_{\mathsf{obs}}(i)) \ne \varnothing} \left|\overline N_{\mathsf{cau}}(i') \cap (\overline N_{\mathsf{cau}}(i) \setminus \overline N_{\mathsf{obs}}(i)) \right|\\
	&\le \frac{2}{n}\sum_{i \in N} \sum_{i': \overline N_{\mathsf{cau}}(i') \cap (\overline N_{\mathsf{cau}}(i) \setminus \overline N_{\mathsf{obs}}(i)) \ne \varnothing} \left| \overline N_{\mathsf{cau}}(i) \setminus \overline N_{\mathsf{obs}}(i) \right|\\
	&\le \frac{2\left(d_{mx}^{\mathsf{cau}}\right)^2}{n}\sum_{i \in N} \left| \overline N_{\mathsf{cau}}(i) \setminus \overline N_{\mathsf{obs}}(i) \right| = 2 \left(d_{mx}^{\mathsf{cau}}\right)^2 d_{mx}^\Delta.
\end{align*}
Let us first find an upper bound for $\tilde \sigma^2 - \tilde \sigma_\mathsf{obs}^2$. Now, observe that
\begin{align}
	\label{bound343}
	\tilde \sigma^2 - \tilde \sigma_\mathsf{obs}^2 &= \frac{1}{n}\sum_{i_1 i_2 \in \overline{E}^* \setminus \overline{E}_\mathsf{obs}^*} \mathbf{E}[\eta_{i_1} \eta_{i_2}\mid \mathcal{F}] \\ \notag
	&\le O_P(d_{mx}^2) \times \frac{1}{n}\sum_{i_1 \in N} \sum_{i_2 \in N} 1\left\{ i_2 \in \overline N_{\mathsf{cau}}^*(i_1), i_2 \notin \overline N_{\mathsf{obs}}^*(i_1) \right\}\\ \notag &= O_P(d_{mx}^2) \times \frac{1}{n}\sum_{i \in N} \left|\overline N_{\mathsf{cau}}^*(i) \setminus \overline N_{\mathsf{obs}}^*(i) \right| \le O_P(d_{mx}^2) \times 2 \left(d_{mx}^\mathsf{cau} \right)^2 \cdot d_{\mathsf{cau}}^{\Delta},
\end{align}
by (\ref{bound2}), and because $\max_{i \in N} \mathbf{E}[\eta_i^4 \mid \mathcal{F}] = O_P\left(d_{mx}^4\right)$, which is due to Lemma \ref{lemm: max q}. Let $\delta_n$ be the sequence in Assumption \ref{assump: bound}. Recalling $\varepsilon_{i,1} = Y_{i,1} - \mathbf{E}[Y_{i,1}\mid \mathcal{F}]$, we note that
\begin{align}
	\label{der4}
	\mathsf{ADM} - \mathsf{C} &= \frac{1}{n} \sum_{i \in N}  \sum_{j \in N_{\mathsf{cau}}(i) \setminus N_{\mathsf{obs}}(i)} \frac{n w_j\mathbf{E}\left[\varepsilon_{i,1}\varepsilon_{j,0} \mid \mathcal{F} \right]}{\sigma_{j,0}^2} \\ \notag
	&\ge  \frac{1}{n} \sum_{i \in N}  \sum_{j \in N_{\mathsf{ctt}}(i;\delta_n) \setminus N_{\mathsf{obs}}(i)} \frac{n w_j\mathbf{E}\left[\varepsilon_{i,1}\varepsilon_{j,0} \mid \mathcal{F} \right]}{\sigma_{j,0}^2} \ge \frac{\overline w d_{\mathsf{ctt}}^{\Delta}(\delta_n) \delta_n}{c^2},
\end{align}
for the constant $c>0$ in Assumption \ref{assump: unconfounded}(ii). The first inequality follows by the condition (\ref{MDC0}) assumed for $(Y_{ij}^*(1),Y_{ij}^*(0))$ and by Lemma \ref{lemm: Cov}.

The probability on the left-hand side of (\ref{prob bound}) is equal to
\begin{align*}
	P\left\{\frac{z_{1-\alpha}(\tilde \sigma^2 - \tilde \sigma_\mathsf{obs}^2)}{\tilde \sigma + \tilde \sigma_\mathsf{obs}} \le \sqrt{n}(\mathsf{ADM} - \mathsf{C} ) \right\} \ge P\left\{\frac{z_{1-\alpha}(\tilde \sigma^2 - \tilde \sigma_\mathsf{obs}^2)}{\sqrt{c'}} \le \sqrt{n}(\mathsf{ADM} - \mathsf{C} ) \right\},
\end{align*}
by Assumption \ref{assump: non deg} and Lemma \ref{lemm: positive semidefinte}. By (\ref{bound343}) and (\ref{der4}), for any $\epsilon>0$, there exist $n_0 \ge 1$ and $C_\epsilon>0$ such that for all $n \ge n_0$, the last probability is bounded from below by
\begin{align*}
	& P\left\{z_{1-\alpha} d_{\mathsf{cau}}^{\Delta} \left(d_{mx}^\mathsf{cau} d_{mx}\right)^2 \le C_\epsilon \sqrt{n} d_{\mathsf{ctt}}^{\Delta}(\delta_n)\delta_n \right\} - \epsilon \\
	&= P\left\{z_{1-\alpha} \left(d_{mx}^\mathsf{cau} d_{mx} \right)^2 \le C_\epsilon \sqrt{n} e_n(\delta_n)\delta_n \right\} - \epsilon \rightarrow 1 - \epsilon,
\end{align*}
as $n \rightarrow \infty$, by Assumption \ref{assump: bound}. Since the choice of $\epsilon>0$ was arbitrary, we obtain the desired result. $\blacksquare$\medskip

\noindent \textbf{Proof of Theorem \ref{thm: asym val2}: } By Lemma \ref{lemm: asymp lin}, we have
\begin{align*}
	\sqrt{n}\{\mathsf{\hat C} - \mathsf{C} \}
	=  \frac{1}{\sqrt{n}}\sum_{i \in N}\left(q_i - \mathbf{E}[q_i\mid \mathcal{F}]\right)
	+ o_P(1).
\end{align*}
We write
\begin{align}
	\label{lower bound2}
	\quad \quad \quad
	P\{\mathsf{ADM}  \ge \hat L_{1-\alpha}\}
	=  P\left\{\frac{\sqrt{n}(\mathsf{C} - \mathsf{\hat C} )}{\tilde \sigma} \ge -z_{1-\alpha} +R_{1n} + R_{2n}\right\},
\end{align}
where
\begin{align}
	R_{1n} &= \frac{z_{1-\alpha} (\tilde \sigma - \tilde \sigma_{\mathsf{obs}}) - \sqrt{n}(\mathsf{ADM}  - \mathsf{C} )}{\tilde \sigma},\text{ and }\\ \notag
	R_{2n} &= - \frac{z_{1-\alpha} (\hat \sigma - \tilde \sigma_{\mathsf{obs}})}{\tilde \sigma}.
\end{align}
By Lemmas \ref{lemm: positive semidefinte} and \ref{B18}, and Assumption \ref{assump: non deg}, we have $R_{2n} = o_P(1)$. On the other hand, by Lemma \ref{lemm: bound}, $P\{R_{1n} \le 0 \} \rightarrow 1$ as $n \rightarrow \infty$. The probability on the right-hand side of (\ref{lower bound2}) is bounded from below by
\begin{align}
	P\left\{\frac{\sqrt{n}(\mathsf{C} - \mathsf{\hat C} )}{\tilde \sigma}  \ge -z_{1-\alpha}\right\} + o(1)
	&= P\left\{\frac{\sqrt{n}(\mathsf{C} - \mathsf{\hat C} )}{\sigma}  \ge -\frac{\tilde \sigma}{\sigma} z_{1-\alpha}\right\} + o(1)\\ \notag
	&\ge P\left\{\frac{\sqrt{n}(\mathsf{C} - \mathsf{\hat C} )}{\sigma} \ge - z_{1-\alpha}\right\} + o(1),
\end{align}
since $\tilde \sigma^2 \ge \sigma^2$ by Lemma \ref{lemm: positive semidefinte}. The last probability converges to $1 - \alpha$ by Lemma \ref{lemm: asym norm}. $\blacksquare$\medskip

\section*{C. Low Level Conditions for Assumptions \ref{assump: asymp approx} and \ref{assump: rate est funs}}
\setcounter{section}{3}
\setcounter{subsection}{0}
\setcounter{equation}{0}
\setcounter{lemma}{0}

In this section, we focus on the specifications in (\ref{g_0 spec}) and (\ref{ci spec}), and their estimators constructed using (\ref{hat gamma}) and (\ref{hat beta}), as explained in Section \ref{subsec: Example of Conditional Mean Spec}. We provide low level conditions for Assumptions \ref{assump: asymp approx} and \ref{assump: rate est funs}. For the results below, we assume that Assumptions \ref{assump: DCG}, \ref{assump: unconfounded}, \ref{assump: non deg} and \ref{assump: network formation} hold.

While we follow the standard arguments (e.g., \cite{White:82:Eca}, \cite{Amemiya:85:AdvEcon}, and \cite{Newey/McFadden:94:Handbook}), some modifications are necessary, as the setting here is different from the standard situation for two reasons. First, the ``population'' objective functions, $Q_0(\gamma)$ and $Q_1(\beta)$, depend on the sample size $n$ and are random, due to the conditioning on $\mathcal{F}$. Second, the second period outcomes, $Y_{i,1}$, are cross-sectionally dependent along a large network, conditional on $\mathcal{F}$. In dealing with this dependence, we rely on the results on empirical processes with conditional dependency graphs in \cite{Lee/Song:19:BJ}.

Define
\begin{align*}
  F_{j,0}(\gamma) &= F_0(X_j'\gamma), \quad f_{j,0}(\gamma) = f_0(X_j'\gamma),  \quad f_{j,0}'(\gamma) = f_0'(X_j'\gamma), \text{ and }\\
  F_{i,1}(\beta) &= F_1(X_i'\beta), \quad f_{i,1}(\beta) = f_1(X_i'\beta),  \quad f_{i,1}'(\beta) = f_1'(X_i'\beta),
\end{align*}
where $f_0$ and $f_1$ are the first order derivatives of $F_0$ and $F_1$, and $f_0'$ and $f_1'$ are the first order derivatives of $f_0$ and $f_1$. We define the sample objective functions:
\begin{align*}
  \hat Q_0(\gamma) &= \frac{1}{n}\sum_{j \in N}  \left( Y_{j,0}\log F_{j,0}(\gamma) + (1 -Y_{j,0})\log(1 - F_{j,0}(\gamma)) \right) \text{ and }\\
  \hat Q_1(\beta) &= \frac{1}{n}\sum_{i \in N}  \left( Y_{i,1}\log F_{i,1}(\beta) + (1 -Y_{i,1})\log(1 - F_{i,1}(\beta)) \right).
\end{align*}
We take the following as their ``population'' objective functions:
\begin{align*}
   Q_0(\gamma) = \mathbf{E}\left[ \hat Q_0(\gamma)\mid \mathcal{F} \right] \text{ and }  Q_1(\beta) = \mathbf{E}\left[ \hat Q_1(\beta)\mid \mathcal{F} \right].
\end{align*}
We let
\begin{align*}
H_0(\gamma) = \frac{\partial^2 Q_0(\gamma)}{\partial \gamma \partial \gamma'} \text{ and } H_1(\beta) = \frac{\partial^2 Q_1(\beta)}{\partial \beta \partial \beta'},
\end{align*}
and
\begin{align*}
\hat H_0(\gamma) = \frac{\partial^2 \hat Q_0(\gamma)}{\partial \gamma \partial \gamma'} \text{ and } \hat H_1(\beta) = \frac{\partial^2 \hat Q_1(\beta)}{\partial \beta \partial \beta'}.
\end{align*}
Let $\beta^*$ be a maximizer of $Q_1(\beta)$ over $\beta \in B$. We introduce a set of conditions, and later show that under these conditions, Assumptions \ref{assump: asymp approx} and \ref{assump: rate est funs} hold.

\begin{assumption}
	\label{assump: ident}
	\noindent (i) The parameter spaces $\Gamma$ and $B$ for $\gamma_0$ and $\beta^*$ are compact, and $\gamma_0$ and $\beta^*$ are in the interior of the parameter spaces $\Gamma$ and $B$ respectively.\medskip
	
	(ii) The densities $f_0$ and $f_1$ of $F_0$ and $F_1$ are log-concave.\medskip
	
	(iii) $F_0$ and $F_1$ are three times continuously differentiable with bounded derivatives, and for any compact set $K \subset \mathbf{R}$, there exists a constant $c_K>0$ that depends only on $K$ such that
	\begin{align*}
		\inf_{z \in K} F_0(z)(1- F_0(z)) > c_K \text{ and } \inf_{z \in K} F_1(z)(1- F_1(z)) > c_K.
	\end{align*}

    (iv) There exists a constant $\bar c>0$ such that for all $n \ge 1$,
	 \begin{align}
		\label{min eigen}
	\lambda_{\min}(-H_0(\gamma_0)) \ge \bar c \text{ and }  \lambda_{\min}(-H_1(\beta^*)) \ge \bar c,
	\end{align}
    with probability one.
\end{assumption}

Assumption \ref{assump: ident} ensures that the unique maximizers, $\gamma_0$ and $\beta^*$, are well separated from other points in the parameter spaces in a way that is uniform over the sample size $n \ge 1$. The negative definiteness of the hessian matrix $H_1(\beta^*)$ is known to be a necessary condition for the identifiability of $\beta^*$. (See Theorem 3.1 of \cite{White:82:Eca}.) Assumption \ref{assump: ident}(iv) strengthens this negative definiteness to be uniform over $n \ge 1$. This condition prevents the hessian matrix from being nearly singular as $n \rightarrow \infty$. The log-concavity condition is made to ensure that the log likelihood function is globally concave in $\gamma$ and $\beta$. (See \cite{Pratt:81:JASA}.) This condition is satisfied, for example, when $F_0$ and $F_1$ are distribution functions of a normal distribution or a logistic distribution. In Assumption \ref{assump: ident}(iii), we require the functions $F_0$ and $F_1$ to be three times continuously differentiable, instead of twice continuously differentiable. This ensures that the hessian matrices $H_0(\gamma)$ and $H_1(\beta)$ behave smoothly so that the first order derivatives of the matrices with respect to  $\gamma$ and $\beta$ around $\gamma_0$ and $\beta^*$ are bounded uniformly over $n \ge 1$.

\subsection{Asymptotic Linear Representation of $\hat \gamma$ and $\hat \beta$}

Let us first establish the asymptotic linear representation of estimators $\hat \gamma$ and $\hat \beta$. From here on, we assume that Assumption \ref{assump: ident} holds. Our first focus is on the consistency of the estimators $\hat \gamma$ and $\hat \beta$. For this, we establish the following uniform convergence result.
\begin{lemma}
	\label{lemm: unif conv}
	(i) 
    \begin{align*}
        \sup_{\gamma \in \Gamma} \| \hat Q_0(\gamma) - Q_0(\gamma)\| = O_P(n^{-1/2}) \text{ and } \sup_{\beta \in B} \| \hat Q_1(\beta) - Q_1(\beta)\| = O_P(n^{-1/2} d_{mx}^2).
    \end{align*}

    (ii)
    \begin{align*}
        \sup_{\gamma \in \Gamma} \| \hat H_0(\gamma) - H_0(\gamma)\| = O_P(n^{-1/2}) \text{ and } \sup_{\beta \in B} \| \hat H_1(\beta) - H_1(\beta)\| = O_P(n^{-1/2} d_{mx}^2).
    \end{align*}
\end{lemma}

\noindent \textbf{Proof: } We focus on proving (i). The statement (ii) is proved similarly. We also prove only the second statement of (i). The first statement of (i) is simpler to prove because $Y_{j,0}$'s are conditionally independent given $\mathcal{F}$. We write
\begin{align*}
	\hat Q_1(\beta) - Q_1(\beta) = \frac{1}{n}\sum_{i \in N} (Y_{i,1} - \mathbf{E}[Y_{i,1} \mid \mathcal{F}])\left(\log(F_{i,1}(\beta)) - \log (1 - F_{i,1}(\beta))\right).
\end{align*}
Define $\mathcal{H} = \{h(\cdot,\cdot; \beta): \beta \in B\}$, where $h(x,y;\beta) = y\left(\log\left(F_1(x'\beta)\right) - \log\left(1 - F_1(x'\beta)\right)\right)$, $x \in \mathbf{R}^p, y \in \{0,1\}$. Define the norm for any measurable function $h: \mathbf{R}^{p} \times \{0,1\} \rightarrow \mathbf{R}$,
\begin{align*}
	\overline \rho_n(h) = \sqrt{ \frac{1}{n}\sum_{i \in N} \mathbf{E}\left[ h^2(X_i,Y_{i,1}) \mid \mathcal{F} \right]}.
\end{align*}
For any $\beta,\tilde \beta \in B$, we have $|h(x,y;\beta) - h(x,y;\tilde \beta)| \le \Psi(x) \| \tilde \beta - \beta \|$, where
\begin{align*}
	\Psi(x) = \sup_{\beta \in B}  \left\| \left(\frac{f_1(x'\beta)}{F_1(x'\beta)} + \frac{f_1(x'\beta)}{1 - F_1(x'\beta)}\right) x \right\|.
\end{align*}
By Theorem 2.7.11 of \cite{vanderVaart/Wellner:96:WeakConvg}, p.164, for any $\epsilon>0$,
\begin{align}
	\label{bd32}
	N_{[]}(2 \epsilon \overline \rho_n(\Psi), \mathcal{H}, \overline \rho_n) \le N(\epsilon, B, \| \cdot \|) \le C \epsilon^{-p},
\end{align}
for some constant $C$ that does not depend on $n$, where $N_{[]}(\epsilon, \mathcal{H}, \overline \rho_n)$ denotes the $\epsilon$-bracketing number of $\mathcal{H}$ with respect to the norm $\overline \rho_n$ and $N(\epsilon, B, \| \cdot \|)$ denotes the $\epsilon$-covering number of $B$ with respect to the Euclidean norm $\| \cdot \|$. (See Sections 2.6 and 2.7 of \cite{vanderVaart/Wellner:96:WeakConvg}.) The last inequality follows because $B$ is a compact set in $\mathbf{R}^p$. Note that $(Y_{i,1},X_i)$'s have $G_{\mathsf{cau}}^*$ (defined in (\ref{graphs G^*})) as a conditional dependency graph given $\mathcal{F}$, because $(Y_1,Y_0)$ has $G_{\mathsf{cau}}$ as a DCG given $\mathcal{F}$. By Assumption \ref{assump: non deg}(ii) and Assumption \ref{assump: ident}(iii), there exists a constant $C>0$ that does not depend on $n$ such that $\overline \rho_n(\Psi) \le C$. By Lemma 3.4 of \cite{Lee/Song:19:BJ}, there exist $C,C'>0$ such that for all $n \ge 1$,
\begin{align*}
	\mathbf{E}\left[\sqrt{n} \sup_{\beta \in B} |\hat Q_1(\beta) - Q_1(\beta)| \mid \mathcal{F} \right] &\le C (d_{mx}^* + 1) \int_0^{\overline \rho_n(\Psi)} \sqrt{1 + \log N_{[]}\left( \epsilon, \mathcal{H},\overline \rho_n\right)} d\epsilon\\ \notag
	&= C (d_{mx}^* + 1) \int_0^{C'} \sqrt{1 + \log N_{[]}\left( \epsilon \overline \rho_n(\Psi), \mathcal{H},\overline \rho_n\right)} d\epsilon \overline \rho_n(\Psi).
\end{align*}
By (\ref{bd32}), we obtain the desired result. $\blacksquare$\medskip

The lemma below establishes the consistency of the estimators, $\hat \gamma$ and $\hat \beta$.
\begin{lemma}
	\label{lemm: consistency gamma}
	As $n \rightarrow \infty$, $\|\hat \gamma - \gamma_0\| \rightarrow_P 0$, and $\|\hat \beta - \beta^*\| \rightarrow_P 0$.
\end{lemma}

\noindent \textbf{Proof: } We prove the second statement only. The proof of the first statement is similar. The proof follows the proof of Theorem 2.1 in \cite{Newey/McFadden:94:Handbook} with minor modifications. We fix $\epsilon>0$. Assumption \ref{assump: ident}(ii) implies that $Q_1(\beta)$ is globally concave, due to the result from \cite{Pratt:81:JASA}.  By Assumption \ref{assump: ident}(iv), there exists $c_\epsilon>0$ such that
\begin{align}
	\label{unique id}
	Q_1(\beta^*) \ge \sup_{\beta \in B: \|\beta - \beta^*\| > \epsilon} Q_1(\beta) + c_\epsilon,
\end{align}
with probability one for all $n \ge 1$. Since $\hat \beta$ is a maximizer of $\hat Q_1(\beta)$ over $\beta \in B$, by Lemma \ref{lemm: unif conv}, we have with probability approaching one,
\begin{align*}
	Q_1(\hat \beta) > \hat Q_1(\hat \beta) - \frac{c_\epsilon}{2} \ge \hat Q_1(\beta^*) - \frac{c_\epsilon}{2} > Q_1(\beta^*) - c_\epsilon
	\ge  \sup_{\beta \in B: \|\beta - \beta^*\| > \epsilon} Q_1(\beta).
\end{align*}
This implies that $\|\hat \beta - \beta^*\|\le \epsilon$ with probability approaching one. $\blacksquare$

\begin{lemma}
	\label{lemm: ident}
	There exist constants $C>0$ and $\bar \epsilon>0$ that do not depend on $n$ such that, for each $\epsilon \in (0,\bar \epsilon)$, and for all $n \ge 1$,
	\begin{align*}
		Q_0 (\gamma_0) &\ge \sup_{\beta \in B: \epsilon < \|\gamma - \gamma_0\| \le \bar \epsilon} Q_0(\gamma) + C \epsilon^2 , \text{ and }\\
	Q_1 (\beta^*)  &\ge \sup_{\beta \in B: \epsilon < \|\beta - \beta^*\| \le \bar \epsilon} Q_1(\beta) + C\epsilon^2.
	\end{align*}
\end{lemma}

\noindent \textbf{Proof: } We prove the second statement only. Using the same arguments, we can arrive at the first statement as well. First, we write
\begin{align*}
	H_1(\beta^*) = \frac{1}{n}\sum_{i \in N} \xi_i(\beta^*)X_i X_i',
\end{align*}
where
\begin{align*}
	\xi_i(\beta^*) &= \frac{\mathbf{E}[Y_{i,1} \mid \mathcal{F}]\left(f_{i,1}'(\beta^*) F_{i,1}(\beta^*) - f_{i,1}^2(\beta^*)\right)}{F_{i,1}^2(\beta^*)} \\
	&\quad -  \frac{(1-\mathbf{E}[Y_{i,1} \mid \mathcal{F}])\left(f_{i,1}'(\beta^*) (1- F_{i,1}(\beta^*)) + f_{i,1}^2(\beta^*)\right)}{\left(1-F_{i,1}(\beta^*)\right)^2}.
\end{align*}
Therefore, by Assumption \ref{assump: non deg}(ii), we find that
\begin{align}
	\label{Lip}
	\left| \lambda_{\min}(-H_1(\beta)) - \lambda_{\min}(-H_1(\beta^*))\right|
	\le \|H_1(\beta) - H_1(\beta^*)\| \le \tilde C \|\beta - \beta^*\|,
\end{align}
where the first inequality follows by the result in \cite{Hoffman/Wielandt:53:DMJ} and the constant $\tilde C$ depends only on the supremum of the derivatives of $F_1$ and $c_K$ in Assumption \ref{assump: ident}(iii), with $K = \{x'\beta: \beta \in B, x \in \mathbf{R}^p, \|x\| \le C\}$ for $C>0$ in Assumption \ref{assump: non deg}.

Now, take a small enough $\bar \epsilon>0$ such that $\bar c - \tilde C \bar \epsilon > \bar c/2$, where $\bar c$ is the constant in Assumption \ref{assump: ident}(iv). Take any $\epsilon \in (0,\bar \epsilon)$ and $\beta \in B$ such that $\epsilon < \| \beta - \beta^*\| \le \bar \epsilon$. For any $\tau \in [0,1]$ and $\tilde \beta = \tau \beta + (1-\tau) \beta^*$, we have
\begin{align*}
	\lambda_{\min}(-H_1(\tilde \beta)) &= \lambda_{\min}(-H_1(\beta^*)) + (\lambda_{\min}(-H_1(\tilde \beta)) - \lambda_{\min}(-H_1(\beta^*)))\\
	&\ge  \lambda_{\min}(-H_1(\beta^*)) - \tilde C \| \tilde \beta - \beta^*\|\\
	&\ge  \lambda_{\min}(-H_1(\beta^*)) - \tilde C \|\beta - \beta^*\| \ge \bar c - \tilde C \bar \epsilon,
\end{align*}
by (\ref{Lip}) and Assumption \ref{assump: ident}(iv). Since $\beta^*$ is a maximizer of $Q_1(\beta)$,
\begin{align*}
	Q_1(\beta^*) - Q_1(\beta) &= -\frac{1}{2}(\beta - \beta^*)' H_1(\tilde \beta) (\beta - \beta^*)\\
	&\ge \frac{1}{2}\lambda_{\min}(-H_1(\tilde \beta))\|\beta - \beta^*\|^2 \ge \frac{\bar c - \tilde C \bar \epsilon}{2} \|\beta - \beta^*\|^2 > \frac{\bar c \epsilon^2}{4},
\end{align*}
where $\tilde \beta$ is a point on the line segment between $\beta$ and $\beta^*$. $\blacksquare$\medskip

\begin{lemma}
	\label{lemm: consistency}
	For any vanishing sequences $\epsilon_{n,0}$ and $\epsilon_{n,1}$ such that $n^{1/4} \epsilon_{n,0} \rightarrow \infty$ and $n^{1/4} \epsilon_{n,1} / d_{mx} \rightarrow \infty$ as $n \rightarrow \infty$,
	\begin{align*}
		\hat \gamma - \gamma_0 = O_P(\epsilon_{n,0}), \text{ and }  \hat \beta - \beta^* = O_P(\epsilon_{n,1}).
	\end{align*}
\end{lemma}

\noindent \textbf{Proof: } We prove the second statement only. Let $\mathcal{N}(\epsilon_{n,1}) = \{\beta \in B: \|\beta - \beta^*\| \le \epsilon_{n,1}\}$ for the sequence $\epsilon_{n,1}$ chosen to satisfy the condition of the lemma. Let $C>0$ and $\bar \epsilon>0$ be the constants in Lemma \ref{lemm: ident}. Let $n_0$ be such that, for all $n \ge n_0$, $\epsilon_{n,1} < \bar\epsilon$. Then, for all $n \ge n_0$,
\begin{align*}
	P\left\{\hat \beta \in B \setminus \mathcal{N}(\epsilon_{n,1})\right\} = P\left\{\hat \beta \in B \setminus \mathcal{N}(\bar \epsilon)\right\}  + P\left\{\hat \beta \in \mathcal{N}(\bar \epsilon) \setminus \mathcal{N}(\epsilon_{n,1})\right\}.
\end{align*}
The first probability on the right-hand side vanishes by Lemma \ref{lemm: consistency gamma}. The last probability is bounded by
\begin{align*}
	P\left\{Q_1(\hat \beta) \le \sup_{\beta \in \mathcal{N}(\bar \epsilon) \setminus \mathcal{N}(\epsilon_{n,1})} Q_1(\beta) \right\} \le  P\left\{Q_1(\hat \beta) \le Q_1(\beta^*) - C \epsilon_{n,1}^2 \right\},
\end{align*}
by Lemma \ref{lemm: ident}. Since $\hat Q_1(\hat \beta) \ge \hat Q_1(\beta^*)$, by Lemma \ref{lemm: unif conv}, the last probability is bounded by
\begin{align*}
	P\left\{Q_1(\beta^*) \le Q_1(\beta^*) - C \epsilon_{n,1}^2 + O_P\left(n^{-1/2} d_{mx}^2 \right) \right\} \rightarrow 0,
\end{align*}
as $n \rightarrow \infty$, because $n^{1/4} \epsilon_{n,1}/d_{mx} \rightarrow \infty$. Hence, $P\left\{\hat \beta \in B \setminus \mathcal{N}(\epsilon_{n,1})\right\}  \rightarrow 0$ as $n \rightarrow \infty$. $\blacksquare$\medskip

\begin{lemma}
	\label{lemm: Al gamma beta}
	For any sequences $\epsilon_{n,0}$ and $\epsilon_{n,1}$ that satisfy the conditions of Lemma \ref{lemm: consistency},
	\begin{align*}
	  \sqrt{n}(\hat \gamma - \gamma_0) &= -H_0^{-1}(\gamma_0) \frac{1}{\sqrt{n}}\sum_{j \in N} \pi_{j,0} f_{j,0}(\gamma_0) X_j + O_P(\epsilon_{n,0}), \text{ and }\\
	  \sqrt{n}(\hat \beta - \beta^*) &=  -H_1^{-1}(\beta^*) \frac{1}{\sqrt{n}}\sum_{i \in N} (\pi_{i,1} - \mathbf{E}[\pi_{i,1}\mid \mathcal{F}]) f_{i,1}(\beta^*) X_i + O_P\left(\sqrt{d_{av}d_{mx}} \epsilon_{n,1}\right),
	\end{align*}
	where
	\begin{align*}
	  \pi_{j,0} = \frac{Y_{j,0} - \mu_{j,0}}{\mu_{j,0}( 1 - \mu_{j,0})} \text{ and } \pi_{i,1} = \frac{Y_{i,1} - F_{i,1}(\beta^*)}{F_{i,1}(\beta^*)(1 - F_{i,1}(\beta^*))}.
	\end{align*}
\end{lemma}
\medskip

\noindent \textbf{Proof: } By the definition of the estimators, we can write
\begin{align}
	\label{expression}
   \sqrt{n}(\hat \gamma - \gamma_0) = \hat H_0^{-1}(\overline \gamma) \left( - \sqrt{n} \frac{\partial \hat Q_0(\gamma_0)}{\partial \gamma} \right),
\end{align}
where $\overline \gamma$ is a point in the line segment between $\hat \gamma$ and $\gamma_0$. By Lemma \ref{lemm: consistency},
\begin{align}
\label{consistency}
   \hat \gamma = \gamma_0 + O_P(\epsilon_{n,0}).
\end{align}
Since
\begin{align*}
    \sqrt{n}\frac{\partial \hat Q_0(\gamma_0)}{\partial \gamma}  = \frac{1}{\sqrt{n}}\sum_{j \in N} \pi_{j,0} f_{j,0}(\gamma_0) X_j,
\end{align*}
and $Y_{j,0}$'s are independent across $j$'s conditional on $\mathcal{F}$, we find that
\begin{align}
\label{bound32}
  \mathbf{E}\left[ \left\|\sqrt{n} \frac{\partial \hat Q_0(\gamma_0)}{\partial \gamma} \right\|^2 \mid \mathcal{F} \right]
  = \frac{1}{n}\sum_{j \in N} \mathbf{E}[\pi_{j,0}^2\mid \mathcal{F}] f_{j,0}^2(\gamma_0) \|X_j\|^2 = O_P(1),
\end{align}
by Assumptions \ref{assump: ident}(iii) and \ref{assump: non deg}(ii). On the other hand,
\begin{align*}
	\left\| \hat H_0^{-1}(\overline \gamma) - H_0^{-1}(\gamma_0) \right\| &\le \left\| \hat H_0^{-1}(\overline \gamma)  \right\| \left\| H_0^{-1}(\gamma_0) \right\|  \left\| \hat H_0(\overline \gamma) - H_0(\gamma_0) \right\| \\
	&\le \left(\left\| H_0^{-1}(\gamma_0) \right\| ^2 + o_P(1) \right) \left\| \hat H_0(\overline \gamma) - H_0(\gamma_0) \right\|,
\end{align*}
by Lemmas \ref{lemm: unif conv} and \ref{lemm: consistency gamma}. As for the last term,
\begin{align*}
	\left\| \hat H_0(\overline \gamma) - H_0(\gamma_0) \right\| &\le \left\| \hat H_0(\overline \gamma) - H_0(\overline \gamma) \right\| + \left\| H_0(\overline \gamma) - H_0(\gamma_0) \right\|\\
	&\le O_P(n^{-1/2}) + \tilde C \| \overline \gamma - \gamma_0\| = O_P(\epsilon_{n,0}),
\end{align*}
by Lemmas \ref{lemm: unif conv} and \ref{lemm: consistency} and (\ref{Lip}). Therefore, by Assumption \ref{assump: ident}(iv), we find from (\ref{expression}) that
\begin{align*}
	\sqrt{n}(\hat \gamma - \gamma_0) = H_0^{-1}(\gamma_0) \left( - \sqrt{n} \frac{\partial \hat Q_0(\gamma_0)}{\partial \gamma} \right) + O_P(\epsilon_{n,0}).
\end{align*}

Let us turn to the second statement. We write
\begin{align*}
   \sqrt{n} \frac{\partial \hat Q_1(\beta^*)}{\partial \beta} &= \sqrt{n} \left(\frac{\partial \hat Q_1(\beta^*)}{\partial \beta} - \mathbf{E}\left[ \frac{\partial \hat Q_1(\beta^*)}{\partial \beta}\mid \mathcal{F} \right]\right)\\
   &= \frac{1}{\sqrt{n}}\sum_{i \in N} \frac{Y_{i,1} - \mathbf{E}[Y_{i,1} \mid \mathcal{F}]}{F_{i,1}(\beta^*)(1 -F_{i,1}(\beta^*))} f_{i,1}(\beta^*) X_i 
   =  \frac{1}{\sqrt{n}}\sum_{i \in N} (\pi_{i,1} - \mathbf{E}[\pi_{i,1}\mid \mathcal{F}]) f_{i,1}(\beta^*) X_i.
\end{align*}
The first equality follows because $\beta^*$ is in the interior of $B$ and the unique maximizer of $Q_1(\beta) = \mathbf{E}[\hat Q_1(\beta) \mid \mathcal{F} ]$ over $\beta \in B$. We can show that
\begin{align}
	\label{eq43}
	\mathbf{E}\left[ \left\|\sqrt{n} \frac{\partial \hat Q_1(\beta^*)}{\partial \beta} \right\|^2 \mid \mathcal{F} \right]
	&=\mathbf{E}\left[ \left\|  \frac{1}{\sqrt{n}}\sum_{i \in N} (\pi_{i,1} - \mathbf{E}[\pi_{i,1}\mid \mathcal{F}])f_{i,1}(\beta^*) X_i\right\|^2\mid \mathcal{F} \right]\\ \notag
	&\le \frac{C}{n}\sum_{i_1, i_2 \in N} \left|\text{Cov}\left(\pi_{i_1,1}, \pi_{i_2,1} \mid \mathcal{F} \right)\right| = O_P(d_{av} d_{mx}),
\end{align}
for some constant $C$ that does not depend on $n$. The last equality follows because $\pi_{i,1}$'s have $G^*$ as a conditional dependency graph given $\mathcal{F}$. The rest of the proof is the same as in the first statement and is omitted. $\blacksquare$

\subsection{Verifying Assumptions \ref{assump: asymp approx} and \ref{assump: rate est funs}}

Let $\hat \mu_{j,0}$, $\hat c_i$, and $\hat \psi_j$ be constructed as in (\ref{hat mu0}), (\ref{hat ci}), and (\ref{hat psi_i}).
\begin{lemma}
	\label{lemm: unif consist}
	\begin{align*}
		\max_{j \in N} |\hat \mu_{j,0} - \mu_{j,0}| &= O_P\left(n^{-1/2}\right),  \\
		\max_{i \in N} |\hat c_{i} - c_{i}| &= O_P\left(n^{-1/2} \sqrt{d_{av} d_{mx}}\right), \text{ and }\\
		\max_{j \in N} | \hat \psi_j - \psi_j| &= O_P\left(n^{-1/2} \sqrt{d_{av} d_{mx}}\right).
	\end{align*}
\end{lemma}

\noindent \textbf{Proof: } By Lemma \ref{lemm: Al gamma beta}, (\ref{min eigen}), and (\ref{bound32}), $\|\hat \gamma - \gamma_0\| = O_P(n^{-1/2})$. Hence, the first result of Lemma \ref{lemm: unif consist} follows from Assumption \ref{assump: ident}(iii). As for the second result, the result follows similarly from $\| \hat \beta - \beta^* \| = O_P\left(n^{-1/2} \sqrt{d_{av} d_{mx}}\right)$, which is due to (\ref{eq43}) and Assumption \ref{assump: ident}(iv). The third statement comes from the first two statements and expanding summands in $\hat \psi_j$. Since the arguments are standard, details are omitted. $\blacksquare$

\begin{lemma}
	Assumptions \ref{assump: asymp approx} and \ref{assump: rate est funs} hold.
\end{lemma}

\noindent \textbf{Proof: } Let us first show Assumption \ref{assump: asymp approx}. Recall the notation $e_{i,1} = Y_{i,1} - c_i$, and write
\begin{align*}
	\frac{1}{\sqrt{n}}\sum_{i \in N} e_{i,1}(\hat a_i - a_i) = A_{n,1} + A_{n,2},
\end{align*}
where
\begin{align*}
	A_{n,1} &=\frac{1}{\sqrt{n}}\sum_{i \in N} e_{i,1}\sum_{j \in N_{\mathsf{obs}}(i)} n w_j\varepsilon_{j,0} \left(\frac{1}{\hat \sigma_{j,0}^2} - \frac{1}{\sigma_{j,0}^2} \right), \text{ and }\\
	A_{n,2} &=\frac{1}{\sqrt{n}}\sum_{i \in N} e_{i,1}\sum_{j \in N_{\mathsf{obs}}(i)} \frac{n w_j(\hat \varepsilon_{j,0} - \varepsilon_{j,0})}{\hat \sigma_{j,0}^2}.
\end{align*}
Let us focus on $A_{n,1}$. We write it as
\begin{align*}
	&\frac{1}{\sqrt{n}}\sum_{i \in N} e_{i,1}\sum_{j \in N_{\mathsf{obs}}(i)} n w_j\varepsilon_{j,0} \frac{\sigma_{j,0}^2 - \hat \sigma_{j,0}^2}{\sigma_{j,0}^4(1 + o_P(1))}\\
	&=\frac{1}{\sqrt{n}}\sum_{i \in N} e_{i,1}\sum_{j \in N_{\mathsf{obs}}(i)} n w_j\varepsilon_{j,0} \frac{(\hat \mu_{j,0} - \mu_{j,0})(2 \mu_{j,0} -1)}{\sigma_{j,0}^4(1 + o_P(1))} + R_n,
\end{align*}
where
\begin{align*}
    R_n = \frac{1}{\sqrt{n}}\sum_{i \in N} e_{i,1}\sum_{j \in N_{\mathsf{obs}}(i)} n w_j\varepsilon_{j,0} \frac{(\hat \mu_{j,0} - \mu_{j,0})^2}{\sigma_{j,0}^4(1 + o_P(1))}.
\end{align*}
By Lemma \ref{lemm: unif consist} and Assumption \ref{assump: unconfounded},
\begin{align*}
	R_n \le O_P(n^{-1}) \times \frac{1}{\sqrt{n}}\sum_{i \in N} \sum_{j \in N_{\mathsf{obs}}(i)} n w_j|\varepsilon_{j,0}| = O_P(n^{-1/2} d_{av}) = o_P(1).
\end{align*}
Expanding $F_0(X_j' \hat \gamma) - F_0(X_j'\gamma_0)$, we find that
\begin{align*}
	&\frac{1}{\sqrt{n}}\sum_{i \in N} e_{i,1}\sum_{j \in N_{\mathsf{obs}}(i)} nw_j  \varepsilon_{j,0} \frac{(\hat \mu_{j,0} - \mu_{j,0})(2 \mu_{j,0} -1)}{\sigma_{j,0}^4(1 + o_P(1))} \\ \notag
	&= \left( \frac{1}{\sqrt{n}}\sum_{i \in N} e_{i,1}\sum_{j \in N_{\mathsf{obs}}(i)} nw_j \varepsilon_{j,0} f_{j,0}(\gamma_0) \frac{2 \mu_{j,0} -1}{\sigma_{j,0}^4} X_j'\right) (\hat \gamma - \gamma_0) + o_P(1).
\end{align*}
By Lemma \ref{lemm: Al gamma beta} and Assumption \ref{assump: network formation}, the leading term on the right-hand side is equal to
\begin{align*}
	\left( \frac{1}{n}\sum_{i \in N} e_{i,1}\sum_{j \in N_{\mathsf{obs}}(i)} nw_j \varepsilon_{j,0} f_{j,0}(\gamma_0) \frac{2 \mu_{j,0} -1}{\sigma_{j,0}^4} X_j'\right)\left( - H_0^{-1}(\gamma_0) \frac{1}{\sqrt{n}}\sum_{j \in N} \pi_{j,0} f_{j,0}(\gamma_0) X_j \right) + o_P(1).
\end{align*}
Note that
\begin{align*}
	&\frac{1}{n}\sum_{i \in N} e_{i,1}\sum_{j \in N_{\mathsf{obs}}(i)} nw_j \varepsilon_{j,0} f_{j,0}(\gamma_0) \frac{2 \mu_{j,0} -1}{\sigma_{j,0}^4} X_j'\\
	&=\frac{1}{n}\sum_{i \in N} \mathbf{E}\left[e_{i,1}\sum_{j \in N_{\mathsf{obs}}(i)} nw_j \varepsilon_{j,0} f_{j,0}(\gamma_0) \frac{2 \mu_{j,0} -1}{\sigma_{j,0}^4} X_j' \mid \mathcal{F} \right] + o_P(1)\\
	&=\frac{1}{n}\sum_{i \in N} \mathbf{E}\left[e_{i,1}\sum_{j \in N_{\mathsf{obs}}(i)} \frac{nw_j  f_{j,0}(\gamma_0) X_j'}{\sigma_{j,0}^2} \pi_{j,0}(2 \mu_{j,0} -1) \mid \mathcal{F} \right] + o_P(1),
\end{align*}
using similar arguments as before. We find that
\begin{align*}
	A_{n,1} &= \left( \frac{1}{n}\sum_{i \in N} \mathbf{E}\left[e_{i,1}\sum_{j \in N_{\mathsf{obs}}(i)} \frac{nw_j  f_{j,0}(\gamma_0) X_j'}{\sigma_{j,0}^2} \pi_{j,0}(2 \mu_{j,0} -1) \mid \mathcal{F} \right] \right) \\
	&\qquad \times \left( - H_0^{-1}(\gamma_0) \frac{1}{\sqrt{n}}\sum_{j \in N} \pi_{j,0} f_{j,0}(\gamma_0) X_j \right) + o_P(1).
\end{align*}

Let us deal with $A_{n,2}$. We write it as
\begin{align*}
	\frac{1}{\sqrt{n}}\sum_{i \in N} e_{i,1}\sum_{j \in N_{\mathsf{obs}}(i)} \frac{nw_j (\hat \varepsilon_{j,0} - \varepsilon_{j,0})}{\sigma_{j,0}^2} + o_P(1)
	= - \frac{1}{\sqrt{n}}\sum_{i \in N} e_{i,1}\sum_{j \in N_{\mathsf{obs}}(i)} \frac{nw_j (\hat \mu_{j,0} - \mu_{j,0})}{\sigma_{j,0}^2} + o_P(1).
\end{align*}
Again, expanding $F_0(X_j' \hat \gamma) - F_0(X_j'\gamma_0)$, we can rewrite the last leading term as
\begin{align*}
	&-\left( \frac{1}{\sqrt{n}}\sum_{i \in N} e_{i,1}\sum_{j \in N_{\mathsf{obs}}(i)} \frac{nw_j f_{j,0}(\gamma_0) X_j'}{\sigma_{j,0}^2}\right) (\hat \gamma - \gamma_0) + o_P(1)\\
	&=-\left( \frac{1}{n}\sum_{i \in N} e_{i,1}\sum_{j \in N_{\mathsf{obs}}(i)} \frac{nw_j f_{j,0}(\gamma_0) X_j'}{\sigma_{j,0}^2}\right)   \times \left( - H_0^{-1}(\gamma_0) \frac{1}{\sqrt{n}}\sum_{j \in N} \pi_{j,0} f_{j,0}(\gamma_0) X_j \right) + o_P(1)\\
	&=-\left( \frac{1}{n}\sum_{i \in N} \mathbf{E}\left[e_{i,1}\sum_{j \in N_{\mathsf{obs}}(i)} \frac{nw_j f_{j,0}(\gamma_0) X_j'}{\sigma_{j,0}^2} \mid \mathcal{F}\right] \right) \\ \notag
	&\qquad \qquad \times \left( - H_0^{-1}(\gamma_0) \frac{1}{\sqrt{n}}\sum_{j \in N} \pi_{j,0} f_{j,0}(\gamma_0) X_j \right) + o_P(1).
\end{align*}
Recall the definition of $\psi_j$ in (\ref{psi_i}). Since $H_0(\gamma_0) = - \frac{1}{n}\sum_{i \in N} (f_{i,0}^2 (\gamma_0) /\sigma_{i,0}^2) X_i X_i'$, we find that
\begin{align*}
	\frac{1}{\sqrt{n}}\sum_{i \in N} e_{i,1}(\hat a_i - a_i)  &= \Gamma_n \frac{1}{\sqrt{n}}\sum_{j \in N} \pi_{j,0} f_{j,0}(\gamma_0) X_j  + o_P(1)\\ \notag
	&= \frac{1}{\sqrt{n}}\sum_{j \in N} \left(\psi_j - \mathbf{E}[\psi_j\mid \mathcal{F}]\right)+ o_P(1).
\end{align*}
The condition (\ref{cond psi}) for $\psi_j$ are satisfied under Assumptions \ref{assump: unconfounded}, \ref{assump: non deg} and \ref{assump: ident}.

As for the second statement in Assumption \ref{assump: asymp approx}, we write
\begin{align}
	\label{dev}
	\frac{1}{\sqrt{n}}\sum_{i \in N} (\hat c_i - c_i) a_i =  \frac{1}{n}\sum_{i \in N} f_{i,1}(\beta^*) a_i X_i' \sqrt{n}(\hat \beta - \beta^*) + o_P(1).
\end{align}
By Lemma \ref{lemm: Al gamma beta} and Assumption \ref{assump: network formation}, the leading term on the right-hand side is equal to
\begin{align*}
	&\left(\frac{1}{n}\sum_{i \in N} f_{i,1}(\beta^*) a_i X_i' \right) \left(- H_1^{-1}(\beta^*) \frac{1}{\sqrt{n}}\sum_{i \in N} \pi_{i,1} f_{i,1}(\beta^*) X_i  \right) + o_P(1)\\ \notag
	&= \left(\frac{1}{n}\sum_{i \in N} f_{i,1}(\beta^*) \mathbf{E}[ a_i \mid \mathcal{F}]X_i' \right)
	\left( - H_1^{-1}(\beta^*) \frac{1}{\sqrt{n}}\sum_{i \in N} \left(\pi_{i,1} - \mathbf{E}[\pi_{i,1} \mid \mathcal{F}]\right)f_{i,1}(\beta^*) X_i  \right) + o_P(1).
\end{align*}
 To see the equality, note that
\begin{align*}
	\text{Var} \left(\frac{1}{n}\sum_{i \in N} f_{i,1}(\beta^*) a_i X_i' \mid \mathcal{F} \right) = O_P\left(\frac{d_{av} d_{mx}}{n} \right),
\end{align*}
following the same arguments as in the proof of Lemma \ref{lemm: Var ea}(i), and
\begin{align*}
	\frac{1}{\sqrt{n}}\sum_{i \in N} \pi_{i,1} f_{i,1}(\beta^*) X_i = O_P\left(\sqrt{d_{av} d_{mx}}\right),
\end{align*}
because $\{\pi_{i,1} f_{i,1}(\beta^*) X_i\}_{i \in N}$ has $G^*$ as a conditional dependency graph and $\mathbf{E}[\pi_{i,1} \mid \mathcal{F}] = 0$. Thus the equality below (\ref{dev}) follows by Assumption \ref{assump: ident}(iv) and Assumption \ref{assump: network formation}. However, $\mathbf{E}[ a_i \mid \mathcal{F}] = 0$, and hence we obtain the second condition in Assumption \ref{assump: asymp approx}.

Assumption \ref{assump: rate est funs} follows by Lemma \ref{lemm: unif consist} and  Assumption \ref{assump: network formation}. $\blacksquare$

\section*{D. Further Details on Assumption \ref{assump: non-DCG}}

\setcounter{section}{4}
\setcounter{subsection}{0}
\setcounter{equation}{0}
\setcounter{lemma}{0}

\label{sec: Non-DCG}

In this section, we provide further details that support the discussion after Assumption \ref{assump: non-DCG}. First, let us recall the linear threshold diffusion model considered in the discussion:
\begin{align*}
	\rho_{i,t}\left((A_{j,t-1})_{j \in N_{\mathsf{ctt}}(i)},U_{i,t};G_{\mathsf{ctt}}\right) = 1\left\{ \frac{\beta}{|N_{\mathsf{ctt}}(i)|} \sum_{j \in N_{\mathsf{ctt}}(i)} A_{j,t-1} \ge U_{i,t} \right\},
\end{align*}
where $U_{i,t}$'s are i.i.d. across $t \ge 1$ and $i$'s, following the uniform distribution on $[0,1]$, and conditionally independent of $A_{i,0}$'s given $\mathcal{F}$, and $\beta$ does not depend on $n$. Our primary goal here is to verify the claim that when $\beta>0$, Assumption \ref{assump: non-DCG} follows.

Let us assume that $\beta>0$. This implies that for all $t=1,...,t_1$, $A_{i,t}$ is a non-decreasing function of $A_{j,0}$, $j \in N_{\mathsf{cau}}(i)$, which are conditionally independent of $U_{k,s}$'s, $k \in N$ and $s=1,...,t_1$, given $\mathcal{F}$. Hence, for all $i \in N$ and $j \in N_{\mathsf{cau}}(i)$, and all $t=1,...,t_1$,
\begin{align*}
	\text{Cov}\left(A_{i,t}, A_{j,0} \mid \mathcal{F} \right) \ge 0.
\end{align*}
Thus for $j \in N_{\mathsf{ctt}}(i)$, $\text{Cov}\left( Y_{j,0}, Y_{i,1} \mid \mathcal{F}\right) = \sum_{t=0}^{t_1} \text{Cov}\left( A_{j,0}, A_{i,t} \mid \mathcal{F} \right) \ge \text{Cov}\left(A_{j,0}, A_{i,1} \mid \mathcal{F} \right)$. Note that
\begin{align*}
	\mathbf{E}\left[ A_{i,1} \mid \mathcal{F} \right] &= \mathbf{E}\left[ 1\left\{ \frac{\beta}{|N_{\mathsf{ctt}}(i)|}\sum_{k \in N_{\mathsf{ctt}}(i)} A_{k,0} \ge U_{i,1} \right\} (1 - A_{i,0}) \mid \mathcal{F} \right] \\
	&= \mathbf{E}\left[ \left(\frac{\beta}{|N_{\mathsf{ctt}}(i)|} \sum_{k \in N_{\mathsf{ctt}}(i)} A_{k,0}\right) (1 - A_{i,0}) \mid \mathcal{F} \right] \\
	&= \frac{\beta}{|N_{\mathsf{ctt}}(i)|} \sum_{k \in N_{\mathsf{ctt}}(i)} \mu_{k,0} \left(1 - \mu_{i,0}\right) \equiv \xi(i), \text{ say}.
\end{align*}
The first equality follows by the definition of $A_{i,1}$. The second equality follows because $U_{i,1}$ is independent of $A_{k,0}$'s and $A_{i,0}$, following the uniform distribution on $[0,1]$. The last equality follows because $A_{j,0}$'s are conditionally independent given $\mathcal{F}$ and $\mu_{j,0} = \mathbf{E}[A_{j,0} \mid \mathcal{F}]$.

Write
\begin{align*}
	\text{Cov}\left( A_{j,0}, A_{i,1} \mid \mathcal{F} \right) &= \mathbf{E}\left[A_{j,0} 1\left\{ \frac{\beta}{|N_{\mathsf{ctt}}(i)|} \sum_{k \in N_{\mathsf{ctt}}(i)} A_{k,0} \ge U_{i,1} \right\} (1 - A_{i,0}) \mid \mathcal{F} \right] - \mu_{j,0} \xi(i)\\
	&= \mathbf{E}\left[ A_{j,0} 1\left\{ \frac{\beta}{|N_{\mathsf{ctt}}(i)|} \sum_{k \in N_{\mathsf{ctt}}(i)} A_{k,0} \ge U_{i,1} \right\} \mid \mathcal{F} \right] \left(1 - \mu_{i,0}\right)- \mu_{j,0} \xi(i).
\end{align*}
For the leading term on the right-hand side, since $i \ne j$, we can write it as
\begin{align*}
	\mathbf{E}\left[ A_{j,0} \left( \frac{\beta}{|N_{\mathsf{ctt}}(i)|} \sum_{k \in N_{\mathsf{ctt}}(i)} A_{k,0} \right) \mid \mathcal{F} \right] \left(1 - \mu_{i,0}\right)
	= \frac{\beta \mu_{j,0}}{|N_{\mathsf{ctt}}(i)|} \left( 1 + \sum_{k \in N_{\mathsf{ctt}}(i) \setminus \{j\}} \mu_{k,0} \right) \left(1 - \mu_{i,0}\right).
\end{align*}
The above equality follows because $j \in N_{\mathsf{cau}}(i)$. The last term is equal to
\begin{align*}
	\frac{\beta \mu_{j,0} }{|N_{\mathsf{ctt}}(i)|} \left( 1 +\sum_{k \in N_{\mathsf{ctt}}(i) \setminus \{j\}} \mu_{k,0} \right) \left(1 - \mu_{i,0}\right) = \mu_{j,0} \xi(i) + \frac{\beta \mu_{j,0} (1 -\mu_{j,0})(1 - \mu_{i,0}) }{|N_{\mathsf{ctt}}(i)|}.
\end{align*}
Hence
\begin{align*}
	\text{Cov}\left( A_{j,0}, A_{i,1} \mid \mathcal{F} \right)&= \frac{\beta \mu_{j,0} (1 - \mu_{j,0})(1 - \mu_{i,0}) }{|N_{\mathsf{ctt}}(i)|}
	 \ge \frac{\beta c^3}{d_{mx}},
\end{align*}
by Assumption \ref{assump: unconfounded}(ii). Therefore, we have
\begin{align*}
	\text{Cov}\left( Y_{j,0}, Y_{i,1} \mid \mathcal{F} \right) \ge \frac{\beta c^3}{d_{mx}}.
\end{align*}
By Assumption \ref{assump: network formation}, we find that Assumption \ref{assump: non-DCG} is satisfied.

\putbib[measuring_diffusion]
\end{bibunit}
\end{document}